\def\isarxiv{1} 

\ifdefined\isarxiv
\documentclass[11pt]{article}
\else
\documentclass{article}
\usepackage{xcolor}
\usepackage[accepted]{icml2023}
\fi

\usepackage{amsmath}
\usepackage{amsthm}
\usepackage{amssymb}
\usepackage{algorithm}
\usepackage{subfig}
\usepackage{algpseudocode}
\usepackage{graphicx}
\usepackage{grffile}
\usepackage{wrapfig,epsfig}
\usepackage{url}
\usepackage{xcolor}
\usepackage{epstopdf}

\usepackage{bbm}
\usepackage{dsfont}

\usepackage{tikz}
\usepackage{hyperref}

\definecolor{mydarkblue}{rgb}{0,0.08,0.45}
\hypersetup{colorlinks=true, citecolor=mydarkblue,linkcolor=mydarkblue}

\usetikzlibrary{arrows}
\ifdefined\isarxiv
\usepackage[margin=1in]{geometry}
\else 

\fi
\graphicspath{{./figs/}}

\definecolor{b2}{RGB}{51,153,255}
\definecolor{mygreen}{RGB}{80,180,0}

\newcommand{\Lichen}[1]{\textcolor{red}{[Lichen: #1]}}

\theoremstyle{plain}
\newtheorem{theorem}{Theorem}[section]
\newtheorem{lemma}[theorem]{Lemma}

\newtheorem{fact}[theorem]{Fact}

\newtheorem{corollary}[theorem]{Corollary}

\theoremstyle{definition}
\newtheorem{definition}[theorem]{Definition}

\newtheorem{remark}[theorem]{Remark}

\newcommand{\wh}{\widehat}
\newcommand{\wt}{\widetilde}
\newcommand{\ov}{\overline}
\newcommand{\eps}{\varepsilon}
\renewcommand{\epsilon}{\varepsilon}
\renewcommand{\phi}{\varphi}
\newcommand{\B}{\mathsf{B}}
\newcommand{\A}{\mathsf{A}}
\newcommand{\N}{\mathcal{N}}
\newcommand{\R}{\mathbb{R}}

\renewcommand{\hat}{\wh}

\newcommand{\poly}{\mathrm{poly}}
\newcommand{\tr}{\mathrm{tr}}
\newcommand{\Tmat}{{\cal T}_{\mathrm{mat}}}

\newcommand{\diag}{\mathrm{diag}}
\newcommand{\new}{\mathrm{new}}

\newcommand{\U}{\mathcal{U}}

\DeclareMathOperator*{\E}{\mathbb{E}}

\newcommand{\Var}{\mathbf{Var}}
\newcommand{\vect}{\mathrm{vec}}
\newcommand{\T}{\mathcal{T}}

\newcommand{\pmedian}{\mathsf{pm}}

\begin{document}

\ifdefined\isarxiv

\title{Sketching Meets Differential Privacy: \\
Fast Algorithm for Dynamic Kronecker Projection Maintenance\thanks{A preliminary version of this paper appeared at ICML 2023.}}
\author{
Zhao Song\thanks{\texttt{zsong@adobe.com}. Adobe Research.}
\and 
Xin Yang\thanks{\texttt{yx1992@cs.washington.edu}. University of Washington. Supported in part by NSF grant No. CCF-2006359.}
\and 
Yuanyuan Yang\thanks{\texttt{yyangh@cs.washington.edu}. University of Washington. Supported by NSF grant No. CCF-2045402 and NSF grant No. CCF-2019844.} 
\and 
Lichen Zhang\thanks{\texttt{lichenz@mit.edu}. MIT. Supported by NSF grant No. CCF-1955217 and NSF grant No. DMS-2022448.}
}
\date{}

\else 

\twocolumn[
\icmltitle{Sketching Meets Differential Privacy: 
Fast Algorithm for 
Dynamic Kronecker Projection Maintenance}



\icmlsetsymbol{equal}{*}

\begin{icmlauthorlist}
\icmlauthor{Zhao Song}{adobe}
\icmlauthor{Xin Yang}{uw}
\icmlauthor{Yuanyuan Yang}{uw}
\icmlauthor{Lichen Zhang}{mit}
\end{icmlauthorlist}

\icmlaffiliation{adobe}{Adobe Research}
\icmlaffiliation{uw}{The University of Washington}
\icmlaffiliation{mit}{MIT}

\icmlcorrespondingauthor{Zhao Song}{zsong@adobe.com}
\icmlcorrespondingauthor{Yuanyuan Yang}{yyangh@cs.washington.edu}
\icmlcorrespondingauthor{Lichen Zhang}{lichenz@mit.edu}
\icmlkeywords{}

\vskip 0.3in
]



\printAffiliationsAndNotice{}  

\fi 

\ifdefined\isarxiv
\begin{titlepage}
  \maketitle
  \begin{abstract}
      Projection maintenance is one of the core data structure tasks. Efficient data structures for projection maintenance have led to recent breakthroughs in many convex programming algorithms. In this work, we further extend this framework to the Kronecker product structure. Given a constraint matrix ${\sf A}$ and a positive definite matrix $W\in \R^{n\times n}$, we consider the task of maintaining the projection in the form of ${\sf B}^\top({\sf B}{\sf B}^\top)^{-1}{\sf B}$, where ${\sf B}={\sf A}(W\otimes I)$ or ${\sf B}={\sf A}(W^{1/2}\otimes W^{1/2})$. At each iteration, the weight matrix $W$ receives a low rank change that keeps its eigenbasis fixed, and we receive a new vector $h$. The goal is to maintain the projection matrix and answer the query ${\sf B}^\top({\sf B}{\sf B}^\top)^{-1}{\sf B}h$ with good approximation guarantees. We design a fast dynamic data structure for this task and it is robust against an adaptive adversary. Following the beautiful and pioneering work of [Beimel, Kaplan, Mansour, Nissim, Saranurak and Stemmer, STOC'22], we use tools from differential privacy to reduce the randomness that a set query data structure needs against an adaptive adversary, which improves its preprocessing and update time when the number of iterations exceeds the number of output coordinates.

  \end{abstract}
  \thispagestyle{empty}
\end{titlepage}
\newpage
\else
  \begin{abstract}
      Projection maintenance is one of the core data structure tasks. Efficient data structures for projection maintenance have led to recent breakthroughs in many convex programming algorithms. In this work, we further extend this framework to the Kronecker product structure. Given a constraint matrix ${\sf A}$ and a positive definite matrix $W\in \R^{n\times n}$, we consider the task of maintaining the projection in the form of ${\sf B}^\top({\sf B}{\sf B}^\top)^{-1}{\sf B}$, where ${\sf B}={\sf A}(W\otimes I)$ or ${\sf B}={\sf A}(W^{1/2}\otimes W^{1/2})$. At each iteration, the weight matrix $W$ receives a low rank change that keeps its eigenbasis fixed, and we receive a new vector $h$. The goal is to maintain the projection matrix and answer the query ${\sf B}^\top({\sf B}{\sf B}^\top)^{-1}{\sf B}h$ with good approximation guarantees. We design a fast dynamic data structure for this task and it is robust against an adaptive adversary. Following the beautiful and pioneering work of [Beimel, Kaplan, Mansour, Nissim, Saranurak and Stemmer, STOC'22], we use tools from differential privacy to reduce the randomness that a set query data structure needs against an adaptive adversary, which improves its preprocessing and update time when the number of iterations exceeds the number of output coordinates.

  \end{abstract}
\fi

\ifdefined\isarxiv
\newpage
\else
\fi

\section{Introduction}\label{sec:intro}

Projection maintenance is one of the most important data structure problems in recent years. Many convex optimization algorithms that give the state-of-the-art running time heavily rely on an efficient and robust projection maintenance data structure~\cite{cls19,lsz19,jlsw20,jklps20,hjstz21}. Let $B\in \R^{m\times n}$, consider the projection matrix $P=B (B^\top B)^{-1}B^\top$. The projection maintenance task aims for the design of a data structure with the following guarantees: it can preprocess and compute an initial projection. At each iteration, $B$ receives a low rank or sparse change, and the data structure needs to update $B$ to reflect these changes. It will then be asked to approximately compute the matrix-vector product, between the updated $P$ and an online vector $h$. For example, in linear programming, one sets $B=\sqrt{W}A$, where $A\in \R^{m\times n}$ is the constraint matrix and $W$ is the diagonal slack matrix. Each iteration, $W$ receives relatively small perturbations. Then, the data structure needs to output an approximate vector to $\sqrt{W}A (A^\top WA)^{-1} A^\top \sqrt{W}h$, for an online vector $h\in \R^n$.

In this work, we consider a specific type of projection maintenance problem. Concretely, our matrix $B={\sf A}(W\otimes I)$ or $B={\sf A}(W^{1/2}\otimes W^{1/2})$, where $W\in \R^{n\times n}$ is a positive definite matrix and ${\sf A}\in \R^{m\times n^2}$ is a matrix whose $i$-th row is the vectorization of an $n\times n$ matrix. We call the problem for maintaining such kind of matrices, the \emph{Dynamic Kronecker Product Projection Maintenance Problem}. Maintaining the Kronecker product projection matrix has important implications for solving semi-definite programs using interior point method~\cite{jklps20,hjstz21,hjs+22_quantum,gs22}. Specifically, one has $m$ constraint matrices $A_1,\ldots,A_m\in \R^{n\times n}$, and ${\sf A}$ is constructed by vectorization each of the $A_i$'s as its rows. The matrix $W$ is typically the complementary slack of the program. We study the following setting: the matrices $A_1,\ldots,A_m$ are fixed and linearly independent, and the sequence of update matrices $W^{(0)},W^{(1)},\ldots,W^{(T)}\in \R^{n\times n}$ are positive definite and share the same eigenbasis. In general, the weight matrices of an SDP do not share an eigenbasis. They do share one, for example, when the constraint matrices and the objective matrix are simultaneously diagonalizable~\cite{jl16} by an orthogonal matrix $P$, meaning that $P^\top A_i P$ is diagonal for every $i$ and the objective matrix is diagonalized by $P$ as well. But then all $A_i$ lie in one $n$-dimensional space of matrices, so there are at most $n$ linearly independent constraints. More generally, the slack matrices of the barrier method of Appendix~\ref{sec:app:sdp}, which runs on an $(n+2)\times(n+2)$ embedding of the SDP, stay diagonal in the fixed orthogonal basis $\diag(U,1,1)$ when the objective matrix is diagonal in $U$ and every constraint is either diagonal in $U$, or has zero diagonal in $U$ and right-hand side $0$. Such SDPs can have up to $n(n+1)/2$ linearly independent constraints, but the constraints of the second kind never change the optimal value, and every such SDP has the same optimal value as a linear program in $n$ variables (Theorem~\ref{thm:sdp_fixed_basis}). We present our main result as a data structure for the fixed-eigenbasis setting, and we give an SDP solver for this class in Appendix~\ref{sec:app:sdp}.

The data structure design problem can be decomposed into 2 parts: 1).\ How to update the projection matrix fast, and 2).\ How to answer the query efficiently. For the update portion, we leverage the fact that the updates are relatively small. Hence, by updating the inverse part of projection in a lazy fashion, we can give a fast update algorithm.

For the query portion, it is similar to the Online Matrix Vector Multiplication Problem~\cite{hkns15,lw17,ckl18}, with a changing matrix-to-multiply. To speed up this process, prior works either use importance sampling to sparsify the vector $h^{(t)}$, or using sketching matrices to reduce the dimension. For the latter, the idea is to prepare multiple instances of sketching matrices beforehand, and batch-multiplying them with the initial projection matrix $P^{(0)}$. Let ${\sf R}\in \R^{n^2\times Tb}$ denote the batched sketching matrices, where $b$ is the sketching dimension for each matrix, one prepares the matrix $P {\sf R}$. At each query phase, one only needs to use one sketching matrix, $R^{(t)}$, and compute $(P (R^{(t)})^\top )R^{(t)}h$. By computing this product from right to left, we effectively reduce the running time from $O(n^4)$ to $O(n^2b)$. One significant drawback of this approach is that, if the number of iterations $T$ is large, then the preprocessing and update phase become less efficient due to the sheer number of sketching matrices.

We observe that the fundamental reason that applying one uniform sketch for all iterations will fail is due to the dependence between the current output and all previous inputs: When using the projection maintenance data structure in an iterative process, the new input query is typically formed by a combination of the 
previous outputs. This means that our data structure should be robust against an \emph{adaptive adversary}. Such an adversary can infer the randomness from observing the output of the data structure and design new input to the data structure. Prior works combat this issue by using a uniformly-chosen sketching matrix that won't be used again.

To make our data structure both robust against an adaptive adversary and reduce the number of sketches to use, we adapt a differential privacy framework as in the fantastic work~\cite{bkm+21}. Given a data structure against an oblivious adversary that outputs a real number, the pioneering work~\cite{bkm+21} proves that it is enough to use $\wt O(\sqrt{T})$ data structures instead of $T$ for adaptive adversary, while the runtime is only blew up by a polylogarithmic factor. However, their result is not immediately useful for our applications, since we need an approximate vector with $n^2$ numbers. We generalize their result to $n^2$ dimension by drawing an independent random sample of the data structures for each coordinate and applying the strong composition theorem, which gives rise to $\wt O(n\sqrt T)$ data structures. The price is in the query time: each round queries $\wt O(\min\{n^2, n\sqrt T\})$ of the data structures instead of one, so the query time grows by this factor. While not directly applicable to the SDP problem, we hope the differential privacy framework we develop could be useful for applications when $n<\sqrt T$, i.e., problems require a large number of iterations. Due to the tightness of strong composition~\cite{kov15}, we conjecture our result is essentially tight. If one wants to remove the $n$ dependence in the number of sketches, one might need resort to much more sophisticated machinery such as differentially private mean estimation in nearly-linear time. We further abstract the result as a generic \emph{set query} data structure, with the number of sketches required scaling with the number of coordinates one wants to output.

Nevertheless, our Kronecker product projection maintenance data structure can be run inside a full interior point method through its initialization and update operations, with each step computed exactly from the Hessian approximation given by the update operation, instead of by the sketched query (Remark~\ref{rem:sketch_steps}). Combining it with the robust barrier method of~\cite{hjstz21}, we obtain a solver for the class of SDPs above when the basis $U$ is given as part of the input (Theorem~\ref{thm:sdp_fixed_basis}). For $m=\Theta(n^2)$, if the matrix multiplication exponent $\omega$ were $2$, it would run in $m^{2+(1-a)/4+o(1)}\log(n/\epsilon)$ time, below the $m^{9/4}$ bound of~\cite{hjstz21}. Here $\epsilon$ is the accuracy of the solution, and $a\in(0,1)$ is the cutoff parameter of the data structure in the theorem below. This improvement requires $\omega<2.25$, but $\omega=2.25$~\cite[Theorem~1.1]{OpenAI26}, and for $\omega\geq 2.25$ the solver does not improve on the bound of~\cite{hjstz21}. Even at $\omega=2$, the improvement would not come from the data structure. On this class, the same method with exact Newton steps and no call of \textsc{Update} takes the initialization time plus $O(n^{\omega+1/2}\log(n/\epsilon))$. This bound is at most the bound of the solver above for every $a$ and every $\omega$, and beyond initialization it is smaller by a factor of at least $n^{3/2}$. For $m=\Theta(n^2)$ it is polynomially below the bound of~\cite{hjstz21} whenever $\omega<2.25$, with no condition on $a$, but not at $\omega=2.25$ (Theorem~\ref{thm:sdp_fixed_basis}, parts 3 and 4). When each constraint is given either diagonal in $U$, or with zero diagonal in $U$ and right-hand side $0$, the linear programming algorithm of~\cite{cls19}, applied to the equivalent linear program in $n$ variables, is at least as fast as the solver above up to $n^{o(1)}$ factors, and for every $\omega$ it is below the bound of~\cite{hjstz21} by a factor of at least $n^{1/2-o(1)}$. So this application shows that the data structure can drive an interior point method whose slack matrices share a fixed eigenbasis. On this class, though, it gives no speedup over simpler methods, and the improvement is only over the general SDP bound of~\cite{hjstz21}.



We start with defining notations to simplify further discussions.

\begin{definition}[Time complexity for preprocessing, update and query]\label{def:time_complex}
Let $B^{(0)},B^{(1)},\ldots,B^{(T)} \in \R^{m\times n}$ be an online sequence of matrices and $h^{(1)},\ldots,h^{(T)} \in \R^n$ be an online sequence of vectors.
\begin{itemize}
    \item We define ${\cal T}_{\rm prep}(m, n, s, b)$ as the preprocessing time of a dynamic data structure with input matrix $B\in \R^{m\times n}$, sketching dimension $b$ and the number of sketches $s$.
    \item We define ${\cal T}_{\rm update}(m, n, s, b, \epsilon)$ as the update time of a dynamic data structure with update matrix $B^{(t)}\in \R^{m\times n}$, sketching dimension $b$ and the number of sketches $s$. This operation should maintain the projection of a matrix $\wt B^{(t)}$ that is formed like $B^{(t)}$ but with its weight matrix $W^{(t)}$ (for example, $B^{(t)}=\sqrt{W^{(t)}}A$ in linear programming) replaced by a matrix $\wt W^{(t)}$ that has the same eigenvectors as $W^{(t)}$ and whose eigenvalues are within a factor of $1\pm \epsilon$ of the corresponding eigenvalues of $W^{(t)}$.
    \item We define ${\cal T}_{\rm query}(m, n, b, \epsilon, \delta)$ as the query time of a dynamic data structure with query vector $h \in \R^n$, sketching dimension $b$ and the number of sketches $s$. This operation should output a vector $\wt h^{(t)}\in \R^n$ such that for any $i\in [n]$, $|\wt h^{(t)}_i-(P^{(t)}h^{(t)})_i|\leq \epsilon \|h^{(t)}\|_2$ with probability at least $1-\delta$.
\end{itemize}

\end{definition}




Now, we turn to the harder problem of maintaining a Kronecker product projection matrix. While the update matrices are positive definite instead of diagonal, we note that they still share the same eigenbasis. Essentially, the updates are a sequence of diagonal eigenvalues, and we can leverage techniques from prior works~\cite{cls19,lsz19,sy21} with lazy updates. 

Our data structure also relies on using sketches to speed up the query phase. We present an informal version of our result below.

\begin{theorem}
Let ${\sf A}\in \R^{m\times n^2}$ have full row rank, i.e., $\mathrm{rank}({\sf A})=m\leq n^2$, and let ${\sf B}={\sf A}(W\otimes I)$ or ${\sf B}={\sf A}(W^{1/2}\otimes W^{1/2})$. Let ${\sf R}\in \R^{sb\times n^2}$ be a batch of $s$ sketching matrices, each with dimension $b$. Given a sequence of online positive definite matrices $W^{(1)},\ldots,W^{(T)}\in \R^{n\times n}$ where $W^{(t)}=U\Lambda^{(t)}U^\top$ and online vectors $h^{(1)},\ldots,h^{(T)}\in \R^{n^2}$. The data structure has the following operations:
\begin{itemize}
    \item {\sc Init}: The data structure preprocesses and generates an initial projection matrix in time
    \begin{align*}
    mn^{\omega}+m^\omega+\Tmat(m, m, n^2)+\Tmat(n^2, m, n^2)+\Tmat(n^2, n^2, sb).
    \end{align*}
    \item {\sc Update}$(W)$: the data structure outputs a matrix $\wt V=U\wt\Lambda U^\top$ such that
    \begin{align*}
        (1-\epsilon)\cdot \lambda_i(\wt V)\leq \lambda_i(W)\leq (1+\epsilon)\cdot \lambda_i(\wt V), \quad \forall i\in [n],
    \end{align*}
    where $\lambda_i(\cdot)$ denotes the eigenvalue on the $i$-th column of $U$. Let $\wt P$ denote the projection matrix defined by $\wt V$ in place of $W$. Moreover, if
    \begin{align*}
    \sum_{i=1}^n (\ln \lambda_i(W)-\ln\lambda_i(W^{\rm old}))^2 \leq C^2,
    \end{align*}
    the amortized expected time per call of $\textsc{Update}(W)$ is
    \begin{align*}
        & ~ C/\epsilon^2 \cdot (\max\{n^{F(a)+\omega-2.5}+n^{F(a)-a/2}, \\
        & ~ n^{F(a)+\omega-4.5}sb+n^{F(a)-2-a/2}sb\}).
    \end{align*}
    A single call that changes $n^c$ of the maintained eigenvalues can take time
    \begin{align*}
        \Tmat(n^2,n^{1+c},\max\{n^2,sb\}),
    \end{align*}
    which is $n^{2\omega}$ for $c=1$ and $sb\leq n^2$.
    \item {\sc Query}$(h)$: the data structure outputs $\wt P R_l^\top R_l\cdot h$ in time
    \begin{align*}
        n^{3+a}+n^{2+b}.
    \end{align*}
\end{itemize}
Here, $a$ is a parameter that can be chosen in the range $a\in (0,\min\{\alpha,\allowbreak (3-\omega)/(\omega-1)\}]$, and $F(a):=\max_{c\in [a,1]}f(a,c)$, where $f(a,c)$ is the function defined in Def.~\ref{def:f_a_c}. We have $F(a)\in [\omega+2,2\omega]$, so $F(a)\in [4.25,4.5]$, since $\omega=2.25$~\cite[Theorem~1.1]{OpenAI26}.
\end{theorem}

For the sake of discussion, let us consider the hypothetical parameter setting 
\begin{align*}
    m = \Theta(n^2), b = \wt O(n^{1.5}/\epsilon^2), s = T, \omega=2, F(a) = 4, T = m^{1/4}
\end{align*}
and $C/\epsilon^2=O(1)$. Here we take $m<n^2$, since for $m=n^2$ the full row rank of ${\sf A}$ makes ${\sf B}$ square and invertible, so the projection is the identity. Then $sb=\wt O(m/\epsilon^2)$, and up to polylogarithmic factors the running time simplifies to
\begin{itemize}
    \item Preprocessing in $m^\omega$ time.
    \item Update in $m^{1.75}+m^{2-a/4}$ amortized time per call.
    \item Query in $m^{1.5+a/2}+m^{1.75}$ time.
\end{itemize}
Since there are $m^{1/4}$ iterations in total, the data structure spends $m^\omega$ time on preprocessing and $m^{1/4}\cdot (m^{1.75}+m^{2-a/4}+m^{1.5+a/2})=m^{2}+m^{2.25-a/4}+m^{1.75+a/2}$ time on all updates and queries. As long as we ensure $a<1$, the total time would be $m^\omega+o(m^{2+1/4})$. Hence, after the preprocessing, the amortized cost per iteration would be $o(m^2)$, which is less than the $n^4=\Theta(m^2)$ time of multiplying a vector with the $n^2\times n^2$ projection matrix even when this matrix is given explicitly. This accounting uses $\omega=2$, which does not hold: $\omega=2.25$~\cite[Theorem~1.1]{OpenAI26}. For general $\omega$, the exponent $F(a)+\omega-2.5$ in the amortized update time is at least $2\omega-0.5$, since $F(a)\geq \omega+2$, and it is below $4$ only if $\omega<2.25$. So at $\omega=2.25$ our bound on the amortized update time is not $o(m^2)$. Theorem~\ref{thm:sdp_fixed_basis} turns the preprocessing and update part of this accounting into an SDP solver for the class of SDPs described above, with no sketches and with each step computed exactly from the Hessian approximation given by \textsc{Update}. At $\omega=2$ it would run in $m^{2+(1-a)/4+o(1)}\log(n/\epsilon)$ time, below the $m^{9/4}$ bound of~\cite{hjstz21}, but at $\omega=2.25$ it does not improve on that bound (Theorem~\ref{thm:sdp_fixed_basis}, part 3). The same method with no call of \textsc{Update} would have the smaller bound $m^{2+o(1)}\log(n/\epsilon)$ at $\omega=2$ (Theorem~\ref{thm:sdp_fixed_basis}, parts 3 and 4), so on this class the improvement would not come from the update operation. In this bound $\epsilon$ denotes the accuracy of the SDP solution, not the accuracy parameter of the data structure used above.


\paragraph{Roadmap.}

In Section~\ref{sec:related_work}, we present our related work on sketching, differential privacy and projection maintenance. In Section~\ref{sec:preli}, we provide the basic notations of this paper, preliminaries on 
Kronecker product calculation, and the formulation of the data structure design problem. In Section~\ref{sec:tech}, we provide an overview on techniques we used in this paper.  
In Section~\ref{sec:kron}, we provide the description of our algorithm
, running time analysis and proof of correctness of Kronecker Product Maintenance Data Structure.   
In Section~\ref{sec:set_query}, we present the definition of the set query problem, the set query estimation data structure and its analysis of correctness and approximation guarantee. In Appendix~\ref{sec:app:sdp}, we use the Kronecker product maintenance data structure inside the robust barrier method of~\cite{hjstz21} and obtain an SDP solver for the class of SDPs of Theorem~\ref{thm:sdp_fixed_basis}, on which every slack matrix of the method stays diagonal in a fixed orthogonal basis.

\section{Related Work}\label{sec:related_work}

\paragraph{Sketching.}

Sketching is a fundamental tool and has many applications in machine learning and beyond, such as linear regression, low-rank approximation \cite{cw13,nn13,mm13,bw14,rsw16,swz17,alszz18,swz19_soda}, distributed problems \cite{wz16,bwz16}, reinforcement learning \cite{wzd+20}, projected gradient descent \cite{xss21}, tensor decomposition \cite{swz19_soda}, clustering \cite{emz21}, signal interpolation \cite{sswz22}, distance oracles~\cite{dswz22}, 
generative adversarial networks \cite{xzz18}, 
training neural networks~\cite{lsswy20,bpsw21,syz21,szz21,hswz22}, matrix completion \cite{gsyz23}, matrix sensing \cite{dls23_sensing,qsz23}, attention scheme inspired regression \cite{lsz23,dls23,lsx+23,gsy23_hyper,gms23}, sparsification of attention matrix \cite{dms23}, discrepancy minimization \cite{dsw22}, dynamic tensor regression problem \cite{rsz22}, John Ellipsoid computation \cite{syyz22}, NLP tasks \cite{lsw+20}, total least square regression \cite{dswy19}.

\paragraph{Differential Privacy.}

First introduced in~\cite{dkm+06}, differential privacy has been playing an important role in providing theoretical privacy guarantees for enormous algorithms~\cite{dr14}, for example, robust learning a mixture of Gaussians~\cite{kssu19}, hypothesis selection~\cite{bksw21}, hyperparameter selection~\cite{mshkt22}, convex optimization~\cite{klz22}, first-order method~\cite{svk21} and mean estimation~\cite{ksu20,hkm22}. Techniques from differential privacy are also widely studied and applied in machine learning~\cite{cm08, wm10, je19, tf20}, deep neural networks~\cite{acgm16, bps19}, computer vision~\cite{zycw20, lwaf21, tkp19}, natural language processing~\cite{xmtyhs21, wk18},  large language models \cite{gsy23_dp,ynb+22}, label protection in split learning \cite{ysy+22}, multiple data release \cite{wyy+22}, federated learning \cite{syy+22} and peer review~\cite{dkws22}. Recent works also show that robust statistical estimator implies differential privacy~\cite{hkmn23}.

\paragraph{Projection Maintenance Data Structure.}

The design of efficient projection maintenance data structure is a core step that lies in many optimization problems, such as linear programming~\cite{v89_lp,cls19,song19,lsz19,b20,sy21,jswz21,blss20,y20,dly21,gs22}, cutting plane method \cite{v89_cp,jlsw20}, integral minimization \cite{jlsz23}, empirical risk minimization~\cite{lsz19,qszz23}, semidefinite programming~\cite{jlsw20,jklps20,hjstz21,hjs+22_quantum,gs22}, dynamic least-square regression~\cite{jpw22}, large language models \cite{bsz23}, and sum-of-squares optimization~\cite{jnw22}.  

\section{Preliminaries \& Problem Formulation}\label{sec:preli}
In Section \ref{subsec:notation}, we introduce the basic notations that we will use in the remainder of the paper.  
In Section \ref{subsec:formulation}, we describe the data structure design problem of our paper.

\subsection{Notations and Basic Definitions.}\label{subsec:notation}

For any integer $n>0$, let $[n]$ denote the set $\{1,2,\cdots,n\}$. Let $\Pr[\cdot]$ denote probability and $\E[\cdot]$ denote expectation. We use $\|x\|_2$ to denote the $\ell_2$ norm of a vector $x$. We use ${\cal N}(\mu,\sigma^2)$ to denote the Gaussian distribution with mean $\mu$ and variance $\sigma^2$. We use $\wt O(f(n))$ to denote $O(f(n)\cdot\poly \log(f(n))$.  We use $\Tmat(m,n,k)$ to denote the time for matrix multiplication for matrix with dimension $m \times n$ and matrix with dimension $n \times k$. We denote by $\omega$ the matrix multiplication exponent, i.e., $\Tmat(n,n,n) = n^{\omega+o(1)}$, and by $\alpha$ the dual exponent of matrix multiplication. Over every field, $\omega=2.25$ and $0.5\leq\alpha\leq 2/3$~\cite[Theorem~1.1 and Corollary~1.3]{OpenAI26}. Since $\omega$ is an infimum, a running time written as $n^\omega$ or $O(n^\omega)$ stands for $n^{\omega+o(1)}$, and the same convention applies to $\Tmat$ and to the exponents defined from it.

 We use $\|A\|$ and $\|A\|_F$ to denote the spectral norm and the Frobenius norm of matrix $A$, respectively.  We use $A^\top$ to denote the transpose of matrix $A$. We use $I_m$ to denote the identity matrix of size $m\times m$. For $\alpha$ being a vector or matrix, we use $\|\alpha\|_0$ to denote the number of nonzero entries of $\alpha$. Given a real square matrix $A$, we use $\lambda_{\max}(A)$ and $\lambda_{\min}(A)$ to denote its largest and smallest eigenvalue,  respectively. Given a real matrix $A$, we use $\sigma_{\max}(A)$ and $\sigma_{\min}(A)$ to denote its largest and smallest singular value, respectively. We use $A^{-1}$ to denote the matrix inverse for matrix $A$. For a square matrix $A$, we use $\tr[A]$ to denote the trace of $A$. We use $b$ and $n^b$ interchangeably to denote the sketching dimension, and $b \in [0,2]$ when the sketching dimension is $n^b$.
 
 Given an $n_1 \times d_1$ matrix $A$ and an $n_2 \times d_2$ matrix $B$, we use $\otimes$ to denote their Kronecker product, i.e., $A \otimes B$ is a matrix where its $(i_1+ (i_2-1) \cdot n_1, j_1 + (j_2-1) \cdot d_1 )$-th entry is $A_{i_1,j_1}\cdot B_{i_2,j_2}$. We denote $I_n \in \R ^{n \times n}$ as the $n$ dimensional identity matrix. For matrix $A \in  \R ^{n \times n}$, we denote (column vector) $\vect(A)$ as the vectorization of $A$. We use $\langle \cdot,\cdot \rangle$ to denote the inner product, when applied to two vectors, this denotes the standard dot product between two vectors, and when applied to two matrices, this means $\langle A,B\rangle = \tr[A^\top B]$, i.e., the trace of $A^\top B$.

\subsection{Problem Formulation}\label{subsec:formulation}
In this section, we  introduce our new online Kronecker projection matrix vector multiplication problem. Before that, we will review the standard online matrix vector multiplication problem:
\begin{definition}[Online Matrix Vector Multiplication (OMV),~\cite{hkns15,lw17, ckl18}]\label{def:omv}
Given a fixed matrix $A \in \R^{n \times n}$. The goal is to design a dynamic data structure that maintains matrix $A$ and supports fast matrix-vector multiplication for $A \cdot h$ for any query $h$ with the following operations:
\begin{itemize}
    \item \textsc{Init}$(A \in \R^{n \times n})$: The data structure takes the matrix $A$ as input, and does some preprocessing.
    \item \textsc{Query}$(h \in \R^n)$: The data structure receives a vector $h \in \R^n$, and the goal is to approximate the matrix vector product $A \cdot h$.
\end{itemize}
\end{definition}
It is known that if we are given a list of $h$ (e.g. $T= n$ different vectors $h^{(1)}$, $h^{(2)}, \cdots, h^{(T)}$) at the same time, this can be done in $n^{\omega}$ time. However, if we have to output the answer before we see the next one, it's unclear how to do it in truly sub-quadratic time per query.

Motivated by linear programming, a number of works \cite{cls19,lsz19,sy21,jswz21,dly21} have explicitly studied the following problem:
\begin{definition}[Online Projection Matrix Vector Multiplication (OPMV), \cite{cls19}]\label{def:opmv}
Given a fixed matrix $A \in \R^{m \times n}$ and a diagonal matrix $W \in \R^{m \times m}$ with nonnegative entries. 
The goal is to design a dynamic data structure that maintains $P(W)=\sqrt{W} A ( A^\top W A )^{-1} A^\top \sqrt{W}$ and supports fast multiplication for $P(W) \cdot h$ for any future query $h$ with the following operations:
\begin{itemize}
    \item \textsc{Init}$(A \in \R^{m \times n}, W \in \R^{m \times m})$: The data structure takes the matrix $A$ and  the diagonal matrix $W$ as input, and performs necessary preprocessing.
    \item \textsc{Update}$(W^{\new} \in \R^{m \times m})$: The data structure  takes diagonal matrix $W^{\new}$ and updates $W$ by $W+ W^{\new}$. 
    \item \textsc{Query}$(h \in \R^n)$: The data structure receives a vector $h \in \R^n$, and the goal is to approximate the matrix vector product $P(W) \cdot h$.
\end{itemize}
\end{definition}

In this work, inspired by semidefinite programming \cite{hjstz21}, we introduce the following novel data structure design problem: 
\begin{definition}[Online Kronecker Projection Matrix Vector Multiplication(OKPMV)]\label{def:okpmv}
Suppose we have $A_1\in \R^{n\times n},\ldots, A_m\in \R^{n\times n}$, and let ${\sf A}=\begin{bmatrix}
\vect(A_1) & \vect(A_2) &\cdots & \vect(A_m)
\end{bmatrix}^\top \in \R^{m\times n^2}$. Let $W=U\Lambda U^\top \in \R^{n\times n}$ be positive definite. Define ${\sf B}={\sf A}(W\otimes I_n)\in \R^{m\times n^2}$ or ${\sf B}={\sf A}(W^{1/2}\otimes W^{1/2})\in \R^{m\times n^2}$. The goal is to design a dynamic data structure that maintains the projection ${\sf B}^\top ({\sf B}{\sf B}^\top)^{-1}{\sf B}$ with the following procedures:
\begin{itemize}
    \item {\sc Initialize}: The data structure preprocesses ${\sf B}$ and forms ${\sf B}^\top ({\sf B}{\sf B}^\top)^{-1}{\sf B}$.
    \item {\sc Update}: The data structure receives a matrix $\Delta=U\wt\Lambda U^\top\in \R^{n\times n}$. The goal is to update the matrix ${\sf B}$ to ${\sf A}((W+\Delta)\otimes I_n)$ or ${\sf A}((W+\Delta)^{1/2}\otimes (W+\Delta)^{1/2})$ and the corresponding projection.
    \item {\sc Query}: The data structure receives a vector $h\in \R^{n^2}$, and the goal is to approximate the matrix ${\sf B}$ and forms the matrix vector product ${\sf B}^\top ({\sf B}{\sf B}^\top)^{-1}{\sf B}h$ quickly.  
\end{itemize}
\end{definition}
\begin{remark}
This problem can be viewed as a generalization to the data structure problem posed in~\cite{cls19,lsz19,sy21}. In their settings, the matrix $\mathsf{B}$ 
is not in the Kronecker product form and $W$ is a full rank diagonal matrix. 
 
\end{remark}

\section{Technical Overview}\label{sec:tech}

Our work consists of two independent results, and neither of them is used in the proof of the other.

The first result considers designing an efficient projection maintenance data structure for Kronecker product in the form of ${\sf A}(W\otimes I)$ or ${\sf A}(W^{1/2}\otimes W^{1/2})$. Our main machinery consists of sketching and low rank update for amortization. More concretely, we explicitly maintain the quantity ${\sf B}^\top ({\sf B}{\sf B}^\top)^{-1}{\sf B} {\sf R}^\top$, where ${\sf R}$ is a batch of sketching matrices.  

By using fast rectangular matrix multiplication~\cite{lgu18}, we only update the projection maintenance when necessary and we update matrices related to the batch of sketching matrices. 
To implement the query, we pick one sketching matrix and compute their corresponding vectors ${\sf B}^\top ({\sf B}{\sf B}^\top)^{-1} {\sf B}R^\top Rh$. One of the main challenges in our data structure is to implement this sophisticated step with a Kronecker product-based projection matrix. We show that as long as $W=U\Lambda U^\top$ with only the diagonal matrix $\Lambda$ changing, we can leverage matrix Woodbury identity and still implement this step relatively fast. For query, we note that a naive approach will be just multiplying the $n^2\times n^2$ projection matrix with a vector, which will take $O(n^4)$ time. To break this quadruple barrier, however, is non-trivial. While using sketching seemingly speeds up the matrix-vector product, this is not enough: since we update the projection in a lazy fashion, during query we are required to ``complete'' the low rank update. This needs products of vectors with $U\otimes U$ and with the maintained $n^2\times n^2$ matrix $M$ of Algorithm~\ref{alg:proj_maintain_main}. The orthonormal eigenbasis $U\in \R^{n\times n}$ might be dense, but a product with $U\otimes U$ still takes only $O(n^\omega)$ time through the identity $(U\otimes U)\vect(X)=\vect(UXU^\top)$, so no sparsity assumption on $U$ is needed. The products with $M$ involve at most $2n^{1+a}$ of its rows or columns, which gives the $n^{3+a}$ term of the query time.

The second main result uses techniques from differential privacy to develop robust data structures against an adaptive adversary. The intuition of such data structure is to protect the privacy of internal randomness (i.e., sketching matrices) from the adversary. 

In the inspiring prior work due to~\cite{bkm+21}, they show a generic reduction algorithm that given a data structure that is robust against an oblivious adversary, outputs a data structure that is robust against an adaptive adversary. However, their mechanism has drawbacks --- it requires the data structure to output a real number. Naively adapting their method to higher-dimensional output will lead to significantly more data structures and much slower update and query time. To better characterize the issue caused by high dimensional output, we design a generic data structure for the set query problem. In this problem, we are given a sequence of matrices $P^{(0)},P^{(1)},\ldots,P^{(T)}\in \R^{n\times n}$ and a sequence of vectors $h^{(1)},\ldots,h^{(T)}\in \R^n$. At each iteration $t\in [T]$, we update $P^{(t-1)}$ to $P^{(t)}$ and we are given a set of indices $Q_t\subseteq [n]$ with support size $k\leq n$, and we only need to approximate entries in set $Q_t$. This model has important applications in estimating the heavy-hitter coordinates of a vector~\cite{p11,jlsw20}. To speed up the matrix-vector product, we use batched sketching matrices. Our method departs from the standard approach that uses $T$ sketches for handling adaptive adversary, by using only $\wt O(\sqrt{kT})$ sketches. The main algorithm is to run the private median algorithm for each coordinate on its own independent random sample of the data structures, and use the strong composition theorem~\cite{dr14} over $T$ iterations and $k$ coordinates, which leads to $\wt O(\sqrt{kT})$ data structures. This procedure has the advantage that, as long as $k\leq T$, then it leads to an improvement over the standard approach that uses $T$ sketches. The price is paid in the query time: each round queries $\wt O(\min\{k,\sqrt{kT}\})$ of the data structures instead of one.


\section{Kronecker Product Projection Maintenance Data Structure}\label{sec:kron}

In this section, we provide the main theorem for Kronecker product projection maintenance together with the data structure. 

Before proceeding, we introduce an amortization tool regarding matrix-matrix multiplication that helps us analyze the running time of certain operations:

\begin{definition}[\cite{cls19}]
Given $i\in [r]$, we define the weight function as
\begin{align*}
    g_i = & ~ \begin{cases}
    n^{-a}, & \text{if $i<n^a$}; \\
    i^{\frac{\omega-2}{1-a}-1}n^{-\frac{a(\omega-2)}{1-a}}, & \text{otherwise}.
    \end{cases}
\end{align*}
\end{definition}

Consider multiplying a matrix of size $n\times r$ with a matrix of size $r\times n$, where $a\leq \alpha$. If $r\leq n^a$, then multiplying these matrices takes $O(n^{2+o(1)})$ time, otherwise, it takes $O(n^2+r^{\frac{\omega-2}{1-a}}n^{2-\frac{a(\omega-2)}{1-a}})$ time. Both of these quantities can be captured by $O(rg_r\cdot n^{2+o(1)})$.

\begin{theorem}[Kronecker Product Projection Maintenance. Informal version of Theorem~\ref{thm:app:kronecker_maintain}]\label{thm:kronecker_maintain}
Given a collection of matrices $A_1, \cdots, A_m \in \R^{n \times n}$. We define $B_i = W^{1/2} A_i W^{1/2} \in \R^{n \times n}, \forall i \in [m]$\footnote{Our algorithm also works if $B_i=A_iW$.}. We define $\mathsf{A} \in \R^{m \times n^2}$ to be the matrix where $i$-th row is the vectorization of $A_i \in \R^{n \times n}$ and $\mathsf{B} \in \R^{m \times n^2}$ to be the matrix where $i$-th row is the vectorization of $B_i \in \R^{n \times n}$. Assume that $\mathsf{A}$ has full row rank, i.e., $\mathrm{rank}(\mathsf{A})=m\leq n^2$.
Let $b$ denote the sketching dimension and let $T$ denote the number of iterations. 
Let $R_1, \cdots, R_s \in\R^{b \times n^2}$ denote a list of sketching matrices. Let $\mathsf{R} \in \R^{sb \times n^2}$ denote the batch sketching matrices. Let $\eps_{\mathrm{mp}} \in (0,0.1)$ be a precision parameter. Let $a\in (0,\min\{\alpha,(3-\omega)/(\omega-1)\}]$. There is a dynamic data structure that given a sequence of online matrices
\begin{align*}
    W^{(1)}, \cdots, W^{(T)} \subset \R^{n \times n}; \text{~~~and~~~} h^{(1)}, \cdots, h^{(T)} \in \R^{n^2}
\end{align*}
approximately maintains the projection matrices
\begin{align*}
    \mathsf{B}^\top (\mathsf{B} \mathsf{B}^\top)^{-1} \mathsf{B} 
\end{align*}
for matrices $W^{(k)}=U\Lambda^{(k)}U^\top \in \R^{n \times n}$ where $\Lambda^{(k)}$ is a diagonal matrix with positive entries and $U$ is an orthonormal eigenbasis.

The data structure has the following operations:

\begin{itemize}
    \item \textsc{Init}$(\epsilon_{\mathrm{mp}} \in (0,0.1))$: This step takes 
\begin{align*}
    mn^\omega+m^\omega+\Tmat(m,m,n^2)+\Tmat(n^2,m,n^2)+\Tmat(n^2,n^2,sb)
\end{align*}
time in the worst case.
    \item \textsc{Update}$(W)$: Output a matrix $\wt{V} \in \R^{n \times n}$ such that for all $i \in [n]$  
    \begin{align*}
        (1-\epsilon_{\mathrm{mp}} ) \cdot \lambda_i( \wt{V}) \leq \lambda_i(W) \leq (1+\epsilon_{\mathrm{mp}}) \cdot \lambda_i( \wt{V} )
    \end{align*}
    where $\lambda_i(W)$ denote the $i$-th entry of the $\Lambda$ matrix for $W$.  If this call changes $n^c$ entries of the stored eigenvalue vector, so that the Kronecker matrices receive a change of rank at most $2n^{1+c}$, it takes $O(\Tmat(n^2,n^{1+c},\max\{n^2,sb\}))$ time in the worst case, which is $n^{2\omega}$ for $c=1$ and $sb\leq n^2$. We claim only an amortized bound on the time of \textsc{Update}, which we state in the lemma below in terms of the function $F(a)$ defined in Def.~\ref{def:f_a_c}.
    \item \textsc{Query}$(h)$: Output $\wt{\mathsf{B}}^\top ( \wt{\mathsf{B} } \wt{ \mathsf{B}}^\top )^{-1} \wt{\mathsf{B}} R_{l}^\top R_l \cdot h$ for the $\wt{B}$ defined by positive definite matrix $\wt{V} \in \R^{n \times n}$ outputted by the last call to \textsc{Update}. 
    This operation takes time
\begin{align*}
    n^{3+a+o(1)}+n^{2+b+o(1)}.
\end{align*}
\end{itemize}

\end{theorem}

For simplicity, consider the regime $m=\Theta(n^2)$. Then the initialization takes $O(m^\omega)+\Tmat(m,m,sb)$ time, for $sb\leq m$, this is $O(m^\omega)$. For update, a call that changes $n^c$ of the stored eigenvalues performs an update of rank at most $2n^{1+c}$ of the Kronecker matrices, and the parameter $c$ captures the \emph{auxiliary rank} of the update. Such a call takes $\Tmat(n^2,n^{1+c},\max\{n^2,sb\})$ time, which is $m^\omega$ for $c=1$ and $sb\leq m$, so we only bound the amortized time. For $sb\leq m$, the lemma below gives the amortized expected time
\[
(C_1/\epsilon_{\mathrm{mp}}+C_2/\epsilon_{\mathrm{mp}}^2)\cdot (m^{(F(a)+\omega-2.5)/2+o(1)}+m^{(F(a)-a/2)/2+o(1)})
\]
per call. For constant $C_1$, $C_2$ and $\epsilon_{\mathrm{mp}}$, this amortized time is $o(m^2)$ if
\begin{align*}
    F(a)+\omega-5/2<4 \text{~~~and~~~} F(a)-a/2<4.
\end{align*}
Since $F(a)\geq \omega+2$, the first condition requires $\omega<2.25$, and the second condition requires $a>2(\omega-2)$. Since $\omega=2.25$~\cite[Theorem~1.1]{OpenAI26}, the first condition fails, so our analysis does not show that the amortized update time is $o(m^2)$. If $\omega$ were $2$, then $f(a,c)$ would be $4$ for every $c\in [0,1]$, hence $F(a)$ would be $4$, and both conditions would hold for every $a\in (0,1)$.

For query, the cost is $m^{1.5+a/2+o(1)}+m^{1+b/2+o(1)}$, which is $o(m^2)$ for $a<1$ and $b<2$. So if $\omega$ were $2$, then for $a<1$, $b<2$, $sb\leq m$ and constant $C_1$, $C_2$, $\epsilon_{\mathrm{mp}}$, the amortized cost per iteration of our data structure after the initialization would be $o(m^2)$.

We give an amortized analysis for \textsc{Update} as follows:

\begin{lemma}
Assume the notations are the same as those stated in Theorem~\ref{thm:kronecker_maintain}, in particular $a\in (0,\min\{\alpha,(3-\omega)/(\omega-1)\}]$.
Furthermore, if the initial vector $W^{(0)} \in \R^{n \times n}$ and the (random) update sequence 
\begin{align*}
W^{(1)}, W^{(2)}, \cdots, W^{(T)} \in \R^{n \times n}
\end{align*}
satisfies
\begin{align*}
    \sum_{i=1}^n ( \E[ \ln \lambda_i( W^{(k+1)} ) ] - \ln ( \lambda_i(W^{(k)}) ) )^2 \leq C_1^2
\end{align*}
and 
\begin{align*}    
    \sum_{i=1}^n ( \Var [ \ln \lambda_i(W^{(k+1)}) ] )^2 \leq C_2^2
\end{align*}
with the expectation and variance are conditioned on $\lambda_i(W^{(k)})$ for all $k=0,1,\cdots, T-1$. Then, the amortized expected time\footnote{ When the input is deterministic, the output and the running time of \textsc{update} is also deterministic.} per call of \textsc{Update}$(W)$ is 
\begin{align*}
    & ~ (C_1/\epsilon_{\mathrm{mp}} + C_2/ \epsilon_{\mathrm{mp}}^2 ) \cdot \max\{ n^{F(a)+\omega-5/2+o(1)} + n^{F(a)-a/2+o(1)}, \\
    & ~ n^{F(a)+\omega-9/2+o(1)}sb + n^{F(a)-2-a/2+o(1)}sb \},
\end{align*}
where $F(a)$ is defined in Def.~\ref{def:f_a_c}.
\end{lemma}

\begin{algorithm}[!ht]\caption{An informal version of our projection maintenance data structure}
\label{alg:proj_maintain_main}
\begin{algorithmic}[1]
\Procedure{Init}{${\sf A}\in \R^{m\times n^2},\eps_{\mathrm{mp}}\in (0,0.1), W\in \R^{n\times n}$}
\State Let $W= U\Lambda U^\top$
\State Store ${\sf G}\gets {\sf A}(U\otimes U)$
\State Store $M \gets {\sf G}^\top ({\sf G}(\Lambda\otimes \Lambda){\sf G}^\top)^{-1}{\sf G}$
\State Prepare batched sketching ${\sf R}\gets [R_1,\ldots,R_s]\in \R^{sb\times n^2}$
\State Store $Q\gets M(\Lambda^{1/2}\otimes \Lambda^{1/2})(U^\top\otimes U^\top){\sf R}^\top$
\State Store $P\gets (U\otimes U)(\Lambda^{1/2}\otimes \Lambda^{1/2})Q$
\EndProcedure

\State 
\Procedure{Update}{$W^{\new}$}
\State $y_i\gets \ln \lambda_i^{\new}-\ln \lambda_i$
\State Let $r$ denotes the number such that $|y_i|\geq \epsilon_{\mathrm{mp}}/2$
\If{$r<n^a$}
\State $\wh \lambda\gets \lambda$
\State Keep $M, Q, P$ the same
\Else
\State $\wh \lambda, r \gets \textsc{SoftThreshold}(\lambda, \lambda^{\new},r)$ \Comment{Create a new vector that finds the correct number of entries needs to be updated}
\State Update $M,Q,P$ using matrix Woodbury identity
\EndIf
\State $\wt \lambda_i \gets \begin{cases}
    \wh \lambda_i & \text{if $|\ln \lambda_i^{\new}-\ln \wh \lambda_i|\leq \epsilon_{\mathrm{mp}}/2$} \\
    \lambda^{\new}_i & \text{otherwise}
    \end{cases}$
\State \Return $\wt \lambda$
\EndProcedure
\State 
\Procedure{Query}{$h^{\new}\in \R^{n^2}$}
\State Let $\wt P$ be the projection whose $\lambda$ being updated to $\wt \lambda$
\State Compute $p_g$ as $\wt PR^\top R h$ using matrix Woodbury identity
\State Compute $p_l$ as $(I-\wt P)R^\top Rh$
\State \Return $p_g, p_l$
\EndProcedure
\end{algorithmic}
\end{algorithm}

Let us set $C_1,C_2=1/\log n$, $\epsilon_{\mathrm{mp}}=0.01$, $m=\Theta(n^2)$ and $sb\leq m$ for the sake of discussion, and suppose that our algorithm has $\sqrt n$ iterations. By the lemma above, the initialization and all calls of \textsc{Update} take time at most 
\begin{align*}
n^{2\omega}+n^{F(a)+\omega-2+o(1)}+n^{F(a)-a/2+1/2+o(1)},
\end{align*} 
recall $m=\Theta(n^2)$, in terms of $m$, it becomes 
\begin{align*}
m^\omega+m^{\frac{F(a)+\omega-2}{2}+o(1)}+m^{\frac{F(a)-a/2}{2}+1/4+o(1)}.
\end{align*}
If $F(a)+\omega-5/2<4$ and $F(a)-a/2<4$, then the last two terms are $o(m^{2+1/4})$, so this time is $m^\omega+o(m^{2+1/4})$. As discussed above, these two conditions require $\omega<2.25$, so they fail, since $\omega=2.25$~\cite[Theorem~1.1]{OpenAI26}. They would hold for every $a\in (0,1)$ if $\omega$ were $2$.

\paragraph{Application to SDP.} Appendix~\ref{sec:app:sdp} uses this data structure inside the robust barrier method of~\cite{hjstz21}. For SDPs whose objective is diagonal in a fixed orthogonal basis $U$, given as part of the input, and whose constraints are, after taking invertible linear combinations of them, either diagonal in $U$ or have zero diagonal in $U$ and right-hand side $0$, every slack matrix of the method is diagonal in $\diag(U,1,1)$, and each Newton step can be read off from an $(n+2)\times(n+2)$ block of $M$. Theorem~\ref{thm:sdp_fixed_basis} bounds the total time by the initialization time with no sketching matrices plus $(n^{F(a)+\omega-2}+n^{F(a)-a/2+1/2})n^{o(1)}\log(n/\epsilon)$. For $m=\Theta(n^2)$ this would be below the bound of~\cite{hjstz21} under the two conditions above, for example at $\omega=2$, but these conditions fail at $\omega=2.25$, where it does not improve on~\cite{hjstz21}. Even at $\omega=2$, the improvement would not come from \textsc{Update}. The same method with no call of \textsc{Update} has the smaller bound of the initialization time plus $O(n^{\omega+1/2}\log(n/\epsilon))$. When the constraints are given in the two forms, the linear programming algorithm of~\cite{cls19} is at least as fast up to $n^{o(1)}$ factors (Theorem~\ref{thm:sdp_fixed_basis}, part 4).

We give an overview of our data structure (Algorithm~\ref{alg:proj_maintain_main}). As we have covered in Section~\ref{sec:tech}, we maintain the matrix 
\begin{align*}
{\sf B}^\top ({\sf B}{\sf B}^\top)^{-1}{\sf B}{\sf R},
\end{align*}
where ${\sf R}$ is a batch of $s$ sketching matrices. When receiving an update $W^{\new}$, we utilize the fact that $W^{\new}$ has the form $U\Lambda^{\new} U^\top$ where only the diagonal matrix $\Lambda$ changes. We then perform a lazy update on $\Lambda$, when its entries don't change too much, we defer the update. Otherwise, we compute a threshold on how many entries need to be updated, and update all maintained variables using matrix Woodbury identity.

By using an amortization framework based on fast rectangular matrix multiplication~\cite{lgu18}, we bound the amortized update time. For constant $C_1$, $C_2$ and $\epsilon_{\mathrm{mp}}$, if $sb\leq n^2$ and the two conditions $F(a)+\omega-5/2<4$ and $F(a)-a/2<4$ above held, for example if $\omega$ were $2$, then the amortized update time would be faster than $n^4$, which is the size of the projection matrix. Since $\omega=2.25$~\cite[Theorem~1.1]{OpenAI26}, the first condition fails. For $a<1$ and $b<2$, the query time is also faster than directly multiplying a length $n^2$ vector with an $n^2\times n^2$ matrix.

\section{Set Query Data Structure}\label{sec:set_query}

In this section, we study an abstraction and generalization of the online matrix-vector multiplication problem. Given a projection matrix ${\sf B}^\top ({\sf B}{\sf B}^\top)^{-1}{\sf B}$ and a query vector $h$, we only want to output a subset of entries of the vector ${\sf B}^\top ({\sf B}{\sf B}^\top)^{-1}{\sf B}h$. A prominent example is we know some entries of ${\sf B}^\top ({\sf B}{\sf B}^\top)^{-1}{\sf B}h$ are above some threshold $\tau$ and have already located their indices using sparse recovery tools, then the goal is to output the estimations of values of these entries.

To improve the runtime efficiency and space usage of Monte Carlo data structures, randomness is typically exploited and made internal to the data structure. Examples such as re-using sketching matrices and locality-sensitive hashing~\cite{im98}. To utilize the efficiency brought by internal randomness, these data structures assume the query sequence is chosen \emph{oblivious} to its pre-determined randomness. This assumption, however, is not sufficient when incorporating a data structure in an iterative process, oftentimes the input query is chosen based on the output from the data structure over prior iterations. Since the query is no longer independent of the internal randomness of the data structure, the success probability guaranteed by the Monte Carlo data structure usually fails.

From an adversary model perspective, this means that the adversary is \emph{adaptive}, meaning that it can design input query based on the randomness leaked from the data structure over prior interactions. If we desire to use our projection maintenance data structure (Alg.~\ref{alg:proj_maintain_main}) for efficient query, we need to initialize $T$ different sketching matrices and for each iteration, using a fresh new sketching. This is commonly adapted by prior works, such as~\cite{lsz19,sy21,qszz23}. However, the linear dependence on $T$ becomes troublesome for large number of iterations.

How to reuse the randomness of the data structure while preventing the randomness leakage to an adaptive adversary?~\cite{bkm+21} provides an elegant solution based on differential privacy. Build upon and extend their framework, we show that $\wt O(\sqrt {kT})$ sketches suffice instead of $T$ sketches.

In Section~\ref{subsec:pdefinition}, we present the definition of the set query and estimation problem.  In Section~\ref{subsec:top_k_ds}, we present our main result for the set query problem.
 
\subsection{Problem Definition}\label{subsec:pdefinition}
In this section, we present the definition and the goal of the set query problem. 
\begin{definition}[Set Query]
Let $G\in \R^{n\times n}$ and $h\in \R^n$. Given a set $Q\subseteq [n]$ and $|Q|=k$, the goal is to estimate the coordinates of $Gh$ in set $Q$. Given a precision parameter $\epsilon$, for each $j\in Q$, we want to design a function $f$ such that
\begin{align*}
    f(G, h)_j \in & ~ g_j^\top h\pm \epsilon \|g_j\|_2 \|h\|_2
\end{align*}

where $g_j$ denotes the $j$-th row of $G$.
\end{definition}

\subsection{Robust Set Query Data Structure}\label{subsec:top_k_ds}
In this section, we design a robust set query data structure against an adaptive adversary.

To give an overview, consider estimating only one coordinate. We prepare $\wt O(\sqrt T)$ sketching matrices and initialize them in a batched fashion. During query stage, we sample $\wt O(1)$ sketching matrices, and compute the inner product between the corresponding row of the (sketched) projection matrix and the sketched vector. This gives us $\wt O(1)$ estimators, we then run a \textsc{PrivateMedian} algorithm (Theorem~\ref{thm:app:pm}) to obtain a real-valued output. This makes sure that we \emph{do not reveal the randomness of the sketching matrices we use}. Using a standard composition result in differential privacy (Theorem~\ref{thm:ada_c}), we reduce the required number of sketches from $T$ to $\wt O(\sqrt T)$.

Lifting from a single coordinate estimation to $k$ coordinates, we draw an independent sample of $\wt O(1)$ sketching matrices for each of the $k$ coordinates and adapt the strong composition over $k$ coordinates, leading to a total of $\wt O(\sqrt{kT})$ sketches. Each round then queries $\wt O(\min\{k,\sqrt{kT}\})$ sketching matrices instead of $\wt O(1)$.

\begin{theorem}[Reduction to Adaptive Adversary: Set Query. Informal version of Theorem~\ref{thm:app:top_k}]\label{thm:top_k}
Let $\delta, \alpha >0$ be parameters. Let $f$ be a function that maps elements from domain $G \times H$ to an element in $\mathcal{U}^d$, where
\begin{align*}
    \mathcal{U} := [-U,-\frac{1}{U}] \cup \{0\} \cup [\frac{1}{U}, U]
\end{align*}
for $U>1$. Suppose there is a dynamic algorithm $\cal A$ against an oblivious adversary that, given an initial data point $x_0 \in X$ and $T$ updates, guarantees the following:
\begin{itemize}
    \item The preprocessing time is $\T_{\mathrm{prep}}$.
    \item The per round update time is $\T_{\mathrm{update}}$.
    \item The per round query time is $\T_{\mathrm{query}}$ and given a set $Q_t\subseteq [n]$ with cardinality $k$, with probability $\geq 9/10$, the algorithm outputs $f(G_t, h_t)_j$ where $j \in Q_t$, and each $f(G_t,h_t)_j$ satisfies the following guarantee:
    \begin{align*}
        f(G_t, h_t)_j \geq & ~ g_j^\top h_t - \gamma \|g_j\|_2 \|h_t\|_2  \\
        f(G_t, h_t)_j \leq & ~ g_j^\top h_t + \gamma \|g_j\|_2 \|h_t\|_2 
    \end{align*}
    where $g_j$ denotes the $j$-th row of matrix $G_t$.
\end{itemize}
Then, there exists a dynamic algorithm $\cal B$ against an adaptive adversary, guarantees the following:
\begin{itemize}
    \item The preprocessing time is 
    \begin{align*}
    \wt{O}(\sqrt{kT}\log(\frac{\log U}{\alpha \delta})\T_{\mathrm{prep}}).
    \end{align*}
    \item The per round update time is 
    \begin{align*}
        \wt{O}(\sqrt{kT} \log(\frac{\log U}{\alpha \delta}) \T_{\mathrm{update}}).
    \end{align*}
    \item The per round query time is 
    \begin{align*} 
    \wt{O}(\min\{k,\sqrt{kT}\}\log(\frac{\log U}{\alpha \delta}) \T_{\mathrm{query}} + k \poly\log(\frac{\log U}{\alpha \delta}))
    \end{align*}
    and, with probability $1-\delta$, for every $j \in Q_t$, the answer $u_t$ is an $(\alpha+ \gamma+\alpha \gamma )$-approximation of $g_j^\top h_t$ for all $t$, i.e.
    \begin{align*}
       (u_t)_j \geq & ~ g_j^\top h_t -(\alpha + \gamma + \alpha\gamma) \|g_j\|_2 \|h_t\|_2  \\
        (u_t)_j \leq & ~ g_j^\top h_t +(\alpha + \gamma+ \alpha \gamma) \|g_j\|_2 \|h_t\|_2 
    \end{align*}
    
\end{itemize}
Here $\wt{O}$ hides $\poly\log(kT\log U/(\alpha\delta))$ factors.
\end{theorem}

\section*{Acknowledgement}

The authors would like to thank Jonathan Kelner for many helpful discussions, Shyam Narayanan for discussions about differential privacy and Jamie Morgenstern for continued support and encouragement. The authors would like to thank Ying Feng, George Li and David Woodruff for pointing out an error in the set query data structure for a previous version of the paper. Xin Yang is supported in part by NSF grant No. CCF-2006359. Yuanyuan Yang is supported by NSF grant No. CCF-2045402 and NSF grant No. CCF-2019844. Lichen Zhang is supported by NSF grant No. CCF-1955217 and NSF grant No. DMS-2022448. The corrections in this version were found and written with the help of Claude Opus 5.5. The authors have verified the corrections and take full responsibility for the content of this paper. 

\ifdefined\isarxiv
 \bibliographystyle{alpha}
 \bibliography{ref}
\else
 \bibliography{ref}
\bibliographystyle{icml2023}

\fi

\newpage 
\appendix
\onecolumn

\section*{Appendix}
\paragraph{Roadmap.}

In Section~\ref{sec:app:preli}, we present the preliminaries of this paper.
In Section~\ref{sec:kronecker_ds}, we design a Kronecker product projection maintenance data structure that has fast update and query time. In Section~\ref{sec:dp}, we use differential privacy techniques to design a robust norm estimation data structure. In Section~\ref{sec:top_k}, we extend our DP mechanisms to develop a robust set query data structure.  

\section{Preliminaries}\label{sec:app:preli}

{\bf Notations.} For any integer $n>0$, let $[n]$ denote the set $\{1,2,\cdots,n\}$. Let $\Pr[\cdot]$ denote probability and $\E[\cdot]$ denote expectation. We use $\|x\|_2$ to denote the $\ell_2$ norm of a vector $x$. We use ${\cal N}(\mu,\sigma^2)$ to denote the Gaussian distribution with mean $\mu$ and variance $\sigma^2$. We use $\wt O(f(n))$ to denote $O(f(n)\cdot\poly \log(f(n))$. We denote by $\omega$ the matrix multiplication exponent and by $\alpha$ the dual exponent of matrix multiplication. Over every field, $\omega=2.25$ and $0.5\leq\alpha\leq 2/3$~\cite[Theorem~1.1 and Corollary~1.3]{OpenAI26}. Since $\omega$ is an infimum, a running time written as $n^\omega$ or $O(n^\omega)$ stands for $n^{\omega+o(1)}$, and the same convention applies to $\Tmat$ and to the exponents defined from it.

 We use $\|A\|$ and $\|A\|_F$ to denote the spectral norm and the Frobenius norm of matrix $A$, respectively.  We use $A^\top$ to denote the transpose of matrix $A$. We use $I_m$ to denote the identity matrix of size $m\times m$. For $\alpha$ being a vector or matrix, we use $\|\alpha\|_0$ to denote the number of nonzero entries of $\alpha$. Given a real square matrix $A$, we use $\lambda_{\max}(A)$ and $\lambda_{\min}(A)$ to denote its largest and smallest eigenvalue, respectively. Given a real matrix $A$, we use $\sigma_{\max}(A)$ and $\sigma_{\min}(A)$ to denote its largest and smallest singular value, respectively. We use $A^{-1}$ to denote the matrix inverse for matrix $A$. For a square matrix $A$, we use $\tr[A]$ to denote the trace of $A$. We use $b$ and $n^b$ interchangeably to denote the sketching dimension, and $b \in [0,2]$ when the sketching dimension is $n^b$.
 
 Given an $n_1 \times d_1$ matrix $A$ and an $n_2 \times d_2$ matrix $B$, we use $\otimes$ to denote the Kronecker product, i.e., $A \otimes B$ is a matrix where its $(i_1+ (i_2-1) \cdot n_1, j_1 + (j_2-1) \cdot d_1 )$-th entry is $A_{i_1,j_1}\cdot B_{i_2,j_2}$.  
 For matrix $A \in  \R ^{n \times n}$, we denote $\vect(A)$ as the vectorization of $A$. We use $\langle \cdot,\cdot \rangle$ to denote the inner product, when applied to two vectors, this denotes the standard dot product between two vectors, and when applied to two matrices, this means $\langle A,B\rangle = \sum_{i,j} A_{i,j} B_{i,j}$. Further, $ \langle A, B \rangle = \tr[A^\top B]$.
 
  We denote the data/constraint matrix as $\A \in \R ^{m \times n^2}$, weight matrix as $W \in \R ^{n \times n}$, and the resulting projection matrix as $\B^\top (\B \B^\top)^{-1} \B$, where $\B = \A (W^{1/2} \otimes W^{1/2}) \in \R^{m \times n^2}$. Additionally, we denote matrix $\Delta \in \R^{n \times n }$ as the update matrix of projection maintenance,  $\Delta$ has rank $k$ and has the same eigenbasis as matrix $W$. We denote $A_1, \ldots, A_m \in \R^{n \times n}$ as the collection of data constraint matrices, ${\sf A}=\begin{bmatrix}
\vect(A_1) & \vect(A_2) & \cdots & \vect(A_m)
\end{bmatrix}^\top \in \R^{m\times n^2}$ as the batched constraint matrix, and $\B^\top (\B \B^\top)^{-1} \B$ as the projection matrix, where $m$ is given. We denote $h \in \R ^{n^2}$ as the vector that the data structure receives to be projected. 
 
  We denote $\eps_{\mathsf{mp}} \in [0, 0.1]$ as the tolerance parameter.  
 We denote $T$ as the number of iterations.  
 We denote $\delta$ as the failure probability, and $\alpha$ as the parameter for the dynamic algorithm against an adaptive adversary. We denote $\T_{\mathrm{prep}}, \T_{\mathrm{update}}, \T_{\mathrm{query}}$ as the preprocessing time, update time, and query time for the dynamic algorithm against an oblivious adversary. We denote $U >1$ as the output parameter and $~\U$ as the output range of the above dynamic algorithm, where every coordinate of the output $v$ satisfies $v \in \U = [-U, -\frac{1}{U}] \cup \{0\} \cup [\frac{1}{U}, U] $.

{\bf Probability tools.} We present the probability tools we will use in this paper, and all of them are exponentially decreasing bounds. At first, we present the Chernoff bound, which bounds the probability that the sum of independent random \emph{Boolean} variables deviates from its true mean by a certain amount.  
 \begin{lemma}[Chernoff bound \cite{che52}]
\label{lem:chernoff_bound}
Let $X = \sum_{i =1}^n X_i$, where $X_i = 1$ with probability $p_i$ and $X_i = 0$ with probability $1 - p_i$, and all $X_i$ are independent. Let $\mu = \E[X] = \sum_{i = 1}^n p_i$. Then
\begin{itemize}
    \item $\Pr[X \ge (1+\delta)\mu] \le \exp(-\delta^2\mu/3)$, $\forall \delta > 0$;
    \item $\Pr[X \le (1-\delta)\mu] \le \exp(-\delta^2\mu/2)$, $\forall 0 < \delta < 1$.
\end{itemize}
\end{lemma}

Next, we present the Hoeffding bound, which bounds the probability that the sum of independent random \emph{bounded} variables deviates from its true mean by a certain amount. 

\begin{lemma}[Hoeffding bound \cite{h63}]\label{lem:hoeffding}
Let $X_1, \cdots, X_n$ denote $n$ independent bounded variables in $[a_i,b_i]$. Let $X= \sum_{i=1}^n X_i$, then we have
\begin{align*}
\Pr[ | X - \E[X] | \geq t ] \leq 2\exp \left( - \frac{2t^2}{ \sum_{i=1}^n (b_i - a_i)^2 } \right).
\end{align*}
\end{lemma}

Finally, we present the Bernstein inequality, which bounds the probability that the sum of independent random \emph{bounded zero-mean} variables deviates from its true mean. 
\begin{lemma}[Bernstein inequality \cite{b24}]\label{lem:bernstein}
Let $X_1, \cdots, X_n$ be independent zero-mean random variables. Suppose that $|X_i| \leq M$ almost surely, for all $i$. Then, for all positive $t$,
\begin{align*}
\Pr \left[ \sum_{i=1}^n X_i > t \right] \leq \exp \left( - \frac{ t^2/2 }{ \sum_{j=1}^n \E[X_j^2]  + M t /3 } \right).
\end{align*}
\end{lemma}

\section{Kronecker Product Projection Maintenance Data Structure}

\label{sec:kronecker_ds}

This section is organized as follows: We introduce some basic calculation rules for Kronecker product in Section~\ref{subsec:app:kron_lin_alg}. We give the visualization of OMV, OPMV and OKPMV in Section~\ref{subsec:app:omv},  Section~\ref{subsec:app:opmv}, and Section~\ref{subsec:app:skpmv}, respectively. We introduce the projection matrix and its properties in Section~\ref{subsec:app:kro_pre}. We present our data structure in Section~\ref{subsec:app:kro_data_stru}. We present our main results for Kronecker projection maintenance in Section~\ref{subsec:app:kro_result}. 

\subsection{Basic Linear Algebra for Kronecker Product}\label{subsec:app:kron_lin_alg}
In this section, we state a number of useful facts for Kronecker product:

At first, we present the mixed product property regarding the interchangeability of the conventional matrix product and the Kronecker product. 
\begin{fact}[Mixed Product Property]\label{fact:app:k_mix_product_property}
Given conforming matrices $A, B, C$ and $D$, we have
\begin{align*}
    (A\otimes B)\cdot (C\otimes D) = & ~ (A\cdot C)\otimes (B\cdot D),
\end{align*}
where $\cdot$ denotes matrix multiplication.
\end{fact}

Next, we present the inversion property regarding the calculation on the inverse of Kronecker product of two conforming matrices. 

\begin{fact}[Inversion]\label{fact:app:k_inversion}
Let $A,B$ be full rank square matrices. Then we have
\begin{align*}
    (A\otimes B)^{-1} = & ~ A^{-1}\otimes B^{-1}.
\end{align*}
\end{fact}

Next, we present a fact regarding the vectorization of conventional matrix product of two conforming matrices and their Kronecker product with identity matrix. 

\begin{fact}\label{fact:app:k_vec_AB}
Let $I_m$ denote the $m\times m$ identity matrix, and $A \in \R^{m \times k}, B \in \R^{k \times m}$ be conforming matrices, then: 
\begin{align*}
    \vect{(AB)} = (I_m \otimes A) \vect{(B)} = (B^\top \otimes I_m) \vect{(A)}
\end{align*}
\end{fact}

We present a fact regarding the vectorization of conventional matrix product of three conforming matrices and their Kronecker product.

\begin{fact}\label{fact:app:k_vec_ABCt}
Let $A,B,C$ be conforming matrices, then:
\begin{align*}
    \vect(ABC) = (C^\top \otimes A) \vect(B)
\end{align*}
\end{fact}

We present a fact regarding the trace of the multiplication of two conforming matrices and their vectorization. 

\begin{fact}\label{fact:app:vec_trace}
Let $A,~B$ be conforming matrices, then:
\begin{align*}
    \tr[A^\top B] = \vect(A)^\top \vect(B) = \vect(B)^\top \vect(A)
\end{align*}
\end{fact}

Finally, we present the cyclic property of trace calculation: The calculation of the trace of the conventional matrix product is invariant under cyclic permutation.

\begin{fact}[Cyclic]\label{fact:app:trace_cyclic}
Let $A,B,V$ be conforming matrices, then:
\begin{align*} 
\tr [ABC] = \tr[BCA] = \tr[CAB].
\end{align*}
\end{fact}

\subsection{Online Matrix Vector Multiplication}\label{subsec:app:omv}
In this section, we present the visualization of online matrix vector multiplication.

\begin{figure*}[ht]
\centering

\includegraphics[width = 0.5\textwidth]{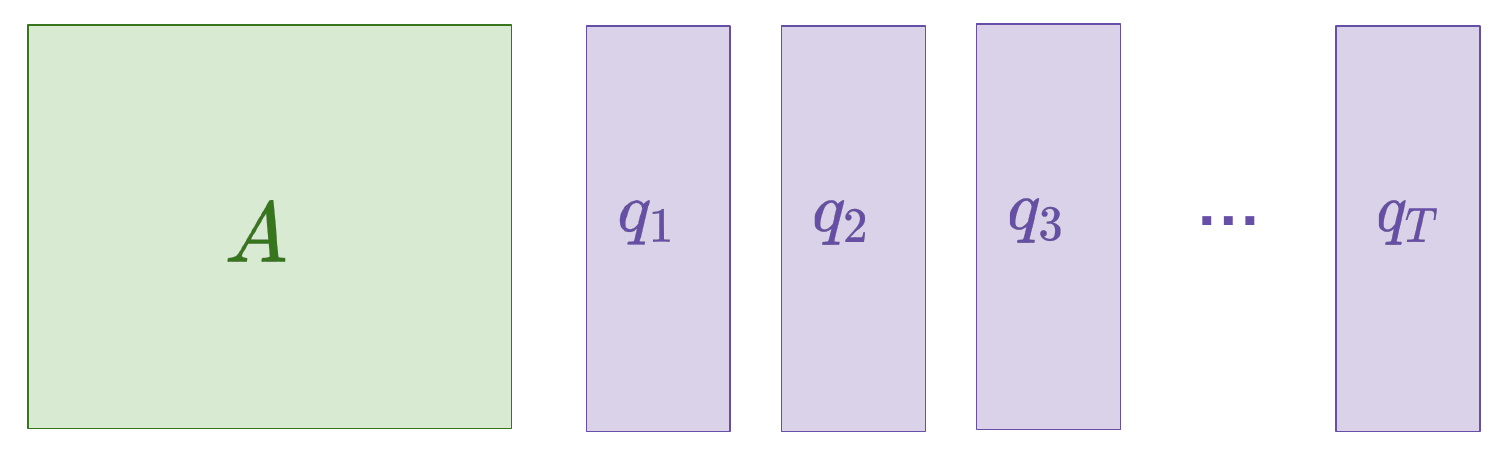}
\caption{Online matrix vector multiplication (Definition~\ref{def:omv}).}
\label{fig:omv}

\end{figure*}

\subsection{Online Projection Matrix Vector Multiplication}\label{subsec:app:opmv}
In this section, we present the visualization of online projection matrix vector multiplication.
\begin{figure*}[ht]
\centering

\includegraphics[width = 1\textwidth]{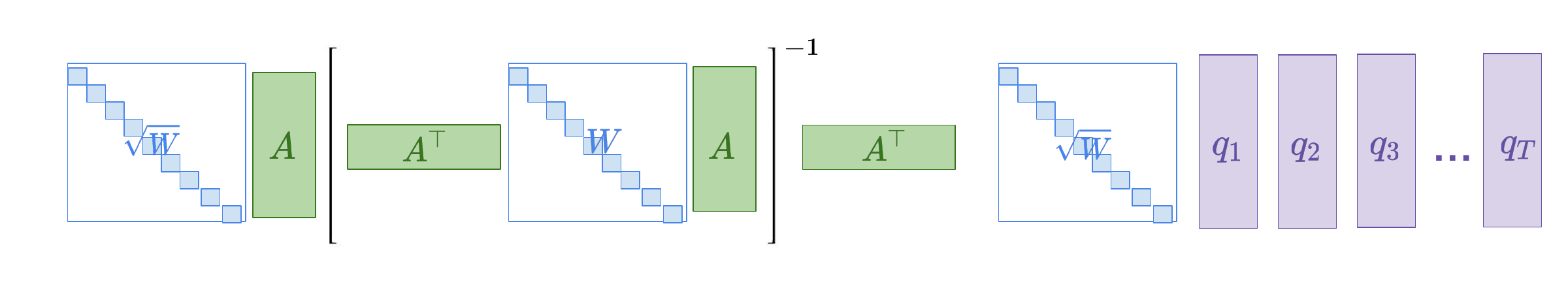}
\caption{Online projection matrix vector multiplication (Definition~\ref{def:opmv}). Usually, we say $\sqrt{W} A (A^\top W A)^{-1} A^\top \sqrt{W}$ is a projection matrix. We say $(A^\top W A)^{-1} A^\top \sqrt{W}$ is a projection matrix without left arm. We say $A (A^\top W A)^{-1} A^\top \sqrt{W}$ is a projection matrix without left hand. Technically, we call $\sqrt{W} A$ arm, and call $\sqrt{W}$ hand.}
\label{fig:opmv}
\end{figure*}

\subsection{Online Kronecker Projection Matrix Vector Multiplication}\label{subsec:app:skpmv}
In this section, we present the visualizations of online Kronecker projection matrix vector multiplication. 

\begin{figure*}[ht]
\centering
\includegraphics[width = 1\textwidth]{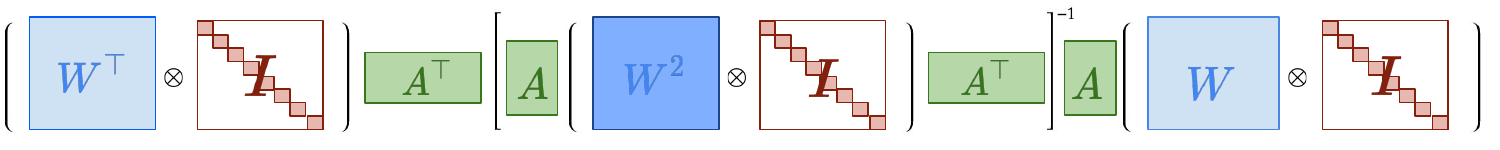}
\caption{Online Kronecker matrix vector multiplication (Definition~\ref{def:okpmv}), where $B_i = A_i W$, and the projection matrix is defined as $\mathsf{B}^\top (\mathsf{B} \mathsf{B}^\top)^{-1} \mathsf{B} = (W^\top \otimes I) \mathsf{A}^\top ( \mathsf{A} (W^2 \otimes I) \mathsf{A}^\top  )^{-1} \mathsf{A} (W \otimes I)$.}
\label{fig:okpmv_identity}
\end{figure*}

\begin{figure*}[ht]
\centering
\includegraphics[width = 1\textwidth]{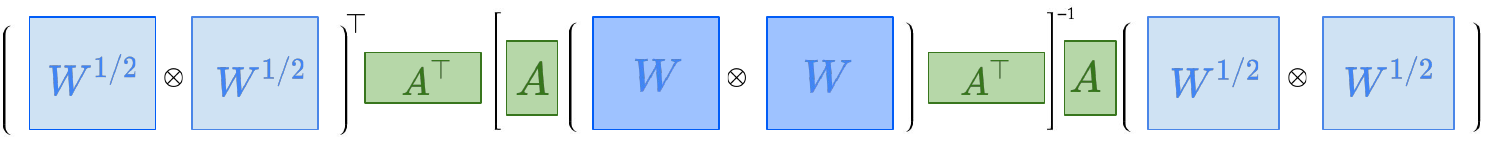}
\caption{Online Kronecker matrix vector multiplication (Definition~\ref{def:okpmv}), where $B_i = W^{1/2}A_i W^{1/2}$, and the projection matrix is defined as $\mathsf{B}^\top (\mathsf{B} \mathsf{B}^\top)^{-1} \mathsf{B} = (W^{1/2} \otimes W^{1/2})^\top \mathsf{A}^\top ( \mathsf{A} (W \otimes W) \mathsf{A}^\top  )^{-1} \mathsf{A} (W^{1/2} \otimes W^{1/2} )$.}
\label{fig:akpmv_sqrt_w}
\end{figure*}

\subsection{Preliminaries}\label{subsec:app:kro_pre}
We present the definitions of the matrices we will be using across the sections.

\begin{definition}
Given a collection of matrices $A_1, \cdots, A_m \in \R^{n \times n}$, we define $\mathsf{A} \in \R^{m \times n^2}$ to be the batched matrix whose $i$-th row is the vectorization of $A_i$, for each $i \in [m]$. Let $W \in \R^{n \times n}$ be a positive definite matrix. We define $\mathsf{B} \in \R^{m \times n^2}$ to be a matrix where each row is the vectorization of $B_i = A_i W$.  
The projection matrix corresponds to ${\sf B}$ is 
\begin{align*}
    {\sf B}^\top ({\sf B}{\sf B}^\top)^{-1}{\sf B} \in \R^{n^2 \times n^2} .
\end{align*}
\end{definition}

Next, we present a fact regarding the batched matrix $\mathsf{B} \in \R^{m \times n^2}$, matrix $W \in \R^{n \times n}$, and the batched matrix $\mathsf{A}$, if $B_i = A_i W$.
\begin{fact}
For $B_i, A_i, W \in \mathbb{R}^{n \times n}$, if $B_i=A_iW$, then we have

\begin{itemize}
    \item $\mathsf{B} = \mathsf{A} (W \otimes I) \in \R^{m \times n^2}$.
    \item $\mathsf{B} \mathsf{B}^\top = \mathsf{A}( W^2 \otimes I) \mathsf{A}^\top \in \R^{m \times m}$.
\end{itemize}
where the $i$-th row of $~\mathsf{B}$, and $\mathsf{A}$ is $\vect(B_i)^\top, \vect(A_i)^\top$, respectively.
\end{fact}

\begin{proof}
We note that each row of $\mathsf{B}$ is in the form of $\vect(B_i)^\top=\vect(A_iW)^\top$, hence,
\begin{align*}
    \vect(B_i) = & ~ \vect(A_i W) \\
    = & ~ \vect(IA_iW) \\
    = & ~ (W^\top \otimes I) \vect(A_i).
\end{align*}
where the last step follows from Fact~\ref{fact:app:k_vec_AB}. 
Therefore, we have: 
\begin{align*}
    \vect(B_i)^\top = & ~ \vect(A_i)^\top (W^\top\otimes I)^\top \\
    = & ~ \vect(A_i)^\top (W\otimes I).
\end{align*}
Hence, we derive that:
\begin{align*}
    \mathsf{B} = & ~ \mathsf{A}(W\otimes I).
\end{align*}
To verify $\mathsf{B} \mathsf{B}^\top = \mathsf{A}( W^2 \otimes I) \mathsf{A}^\top$, we first compute $\mathsf{B}\mathsf{B}^\top$ by the original definition, we have:
\begin{align*}
    (\mathsf{B}\mathsf{B}^\top)_{i,j} = & ~ \vect(A_iW)^\top \vect(A_jW) \\
    = & ~ \tr[WA_i^\top A_jW] \\
    = & ~ \tr[A_jW^2A_i^\top].
\end{align*}
where the second step follows from Fact~\ref{fact:app:vec_trace}, the third step follows from Fact~\ref{fact:app:trace_cyclic}. 

Then, we calculate the $(i,j)$-th coordinate of $\mathsf{A}( W^2 \otimes I) \mathsf{A}^\top$, and we have:
\begin{align*}
      (\mathsf{A}(W\otimes I)(W\otimes I)\mathsf{A}^\top)_{i,j} 
    = & ~ \vect(A_i)^\top (W\otimes I)(W\otimes I)\vect(A_j) \\
    = & ~ \vect(I)^\top(I\otimes A_i^\top)(W\otimes I)(W\otimes I)(I\otimes A_j)\vect(I) \\
    = & ~ \vect(I)^\top (W\otimes A_i^\top)(W\otimes A_j)\vect(I) \\
    = & ~ \vect(A_iW)^\top\vect(A_jW) \\
    = & ~ \tr[WA_i^\top A_jW] \\
    = & ~ \tr[A_jW^2A_i^\top].\\
    = & ~ (\mathsf{B}\mathsf{B}^\top)_{i,j}
\end{align*}
where the first step follows from the definition of $\mathsf{A}$ that $\vect(A_i)^\top$ is the $i$-th row of the matrix $\mathsf{A}$, and $\vect(A_j)$ is the $j$-th column of the matrix $\mathsf{A}^\top$. The second step follows from Fact~\ref{fact:app:k_vec_AB}. The third step follows from Fact~\ref{fact:app:k_mix_product_property}. The fourth step follows from Fact~\ref{fact:app:k_vec_AB}. The fifth step follows from Fact~\ref{fact:app:vec_trace}. The sixth step follows from Fact~\ref{fact:app:trace_cyclic}. The final step follows by the  definition of $\B$. 
\end{proof}

Next, we present a fact regarding the batched matrix $\mathsf{B} \in \R^{m \times n^2}$, matrix $W \in \R^{n \times n}$, and the batched matrix $\mathsf{A} \in \R^{m \times n^2}$, if $B_i = W^{1/2} A_i W^{1/2}$.

\begin{fact} For positive semidefinite matrix $W \in \mathbb{R}^{n \times n}$,
if $B_i = W^{1/2} A_i W^{1/2}$, then we have:
\begin{itemize}
    \item $\mathsf{B} = \mathsf{A} (W^{1/2} \otimes W^{1/2}) \in \R^{m \times n^2}$
    \item $\mathsf{B} \mathsf{B}^\top = \mathsf{A} ( W \otimes W ) \mathsf{A}^\top \in \R^{m \times m}$
\end{itemize}
\end{fact}

\begin{proof}
Suppose $B_i=W^{1/2}A_iW^{1/2}$ whose vectorization is $\vect(W^{1/2}A_iW^{1/2})$, by Fact~\ref{fact:app:k_vec_ABCt}, we have that:
\begin{align*}
    \vect(W^{1/2}A_iW^{1/2}) = & ~ (W^{1/2}\otimes W^{1/2})\vect(A_i).
\end{align*}

Transposing the right hand side gives us:
\begin{align*}
    \vect(A_i)^\top (W^{1/2}\otimes W^{1/2}),
\end{align*}
therefore, we conclude that:
\begin{align*}
    \mathsf{B} = & ~ \mathsf{A} (W^{1/2}\otimes W^{1/2}).
\end{align*}
Consequently, we have:
\begin{align*}
    \mathsf{B}\mathsf{B}^\top = & ~ \mathsf{A} (W^{1/2}\otimes W^{1/2}) (W^{1/2}\otimes W^{1/2})\mathsf{A}^\top \\
    = & ~ \mathsf{A}(W\otimes W)\mathsf{A}^\top.
\end{align*}
where the last step follows from Fact~\ref{fact:app:k_mix_product_property}.
\end{proof}

Next, we present a fact regarding the rank change of the matrix $W^2\otimes I$ and the matrix $W\otimes W$ when $W$ experiences a rank-$k$ change. 
\begin{fact}
Suppose $W\in \R^{n\times n}$ undergoes a rank-$k$ change, i.e., $W\leftarrow W+\Delta$ where $\Delta$ has rank-$k$, then 
\begin{itemize}
    \item The matrix $W^2\otimes I$ undergoes a rank-$3nk$ change.
    \item The matrix $W\otimes W$ undergoes a rank-$(2nk+k^2)$ change.
\end{itemize}
\end{fact}

\begin{proof}
For $W^2\otimes I$, it suffices to understand the rank change on $W^2$. Note that:
\begin{align*}
    (W+\Delta)^2 = & ~ W^2+W\Delta+\Delta W+\Delta^2,
\end{align*}
since $\Delta$ is rank-$k$, we know that $W\Delta$, $\Delta W$ and $\Delta^2$ all have rank at most $k$. Hence, if we let $\wt \Delta$ to denote $W\Delta+\Delta W+\Delta^2$, we have:
\begin{align*}
    \mathrm{rank}(\wt \Delta) \leq & ~ 3k.
\end{align*}
Finally, note that:
\begin{align*}
    (W+\Delta)^2\otimes I = & ~ (W^2+\wt \Delta)\otimes I \\
    = & ~ W^2\otimes I+\wt \Delta\otimes I,
\end{align*}
we have that the rank change of the matrix $W^2\otimes I$ is the same as the rank of the matrix $\wt \Delta\otimes I$ that is at most $3nk$.

We now analyze the rank change of $W\otimes W$. Consider
\begin{align*}
    (W+\Delta)\otimes (W+\Delta) = & ~ W\otimes W+W\otimes \Delta+\Delta\otimes W+\Delta\otimes \Delta,
\end{align*}
the components $W\otimes \Delta$ and $\Delta\otimes W$ both have ranks $nk$, and $\Delta\otimes \Delta$ has rank $k^2$. Hence, if we let $\wt \Delta$ to denote $W\otimes \Delta+\Delta\otimes W+\Delta\otimes \Delta$, then $\mathrm{rank}(\wt \Delta)\leq 2nk+k^2$.
\end{proof}

Next, we present a fact regarding the rank change of the matrix $W^{1/2}$, when $W$ experiences a rank-$k$ change $\Delta$ that has the same eigenbasis of $W$.  
\begin{lemma}
\label{lem:app:sqrt_low_rank}
Suppose $W\in \R^{n\times n}$ undergoes a rank-$k$ change $\Delta$ and $W,\Delta$ have the same eigenbasis, then the matrix $(W+\Delta)^{1/2}$ undergoes a rank-$k$ change, i.e.,
\begin{align*}
    (W+\Delta)^{1/2} = & ~ W^{1/2}+\ov \Delta,
\end{align*}
where $\ov \Delta$ is rank $k$ and shares the same eigenbasis as $W$.
\end{lemma}

\begin{proof}
By spectral theorem, we know that there exists $U, \Lambda,$ and $\wt{\Delta}$ such that, $W=U\Lambda U^\top$ and $\Delta=U\wt \Delta U^\top$, while $\wt \Delta$ has only $k$ nonzero entries. Hence, we notice that $W+\Delta=U (\Lambda+\wt \Lambda) U^\top$ has only $k$ entries being changed. Note that $(W+\Delta)^{1/2}=U(\Lambda+\wt \Lambda)^{1/2}U^\top$, 
which means the diagonal only has $k$ entries being changed. 
We can write it as:
\begin{align*}
    (W+\Delta)^{1/2} = & ~ U(\Lambda+\wt \Lambda)^{1/2}U^\top \\
    = & ~ U\Lambda^{1/2}U^\top+U DU^\top,
\end{align*}
where $D$ is a diagonal matrix with only $k$ nonzeros. Hence, we can write it as $W^{1/2}+\ov \Delta$ where $\ov \Delta$ is rank $k$, as desired.
\end{proof}

Next, we present a fact regarding the rank change of $\mathsf{A}(W\otimes I)$ and $\mathsf{A}(W^{1/2}\otimes W^{1/2})$, if the matrix $W$ experiences a rank-$k$ change. 
\begin{fact}
Suppose $W\in \R^{n\times n}$ undergoes a rank-$k$ change, i.e., $W\leftarrow W+\Delta$ where $\Delta$ has rank-$k$, then
\begin{itemize}
    \item If $\mathsf{B}=\mathsf{A}(W\otimes I)$, then $\mathsf{B}$ undergoes a rank-$nk$ change.
    \item If $\mathsf{B}=\mathsf{A}(W^{1/2}\otimes W^{1/2})$, then $\mathsf{B}$ undergoes a rank-$(2nk+k^2)$ change.
\end{itemize}
\end{fact}

\begin{proof}
We prove item by item.
\begin{itemize}
    \item Suppose $\mathsf{B}=\mathsf{A}(W\otimes I)$, and we have $W+\Delta$, then
    \begin{align*}
        \mathsf{A}((W+\Delta)\otimes I) = & ~ \mathsf{A}(W\otimes I)+\mathsf{A}(\Delta\otimes I),
    \end{align*}
    note that $\Delta\otimes I$ is of rank $nk$, so we conclude that $\mathsf{B}$ experiences a rank-$nk$ change.
    \item Suppose $\mathsf{B}=\mathsf{A}(W^{1/2}\otimes W^{1/2})$, then:
    \begin{align*}
        (W+\Delta)^{1/2}\otimes (W+\Delta)^{1/2} = & ~ (W^{1/2}+ \ov \Delta) \otimes (W^{1/2}+ \ov \Delta) \\
        = & ~ (W^{1/2} \otimes W^{1/2})+W^{1/2}\otimes \ov \Delta+\ov \Delta \otimes W^{1/2}+\ov \Delta \otimes \ov \Delta
    \end{align*}
    where the first step is by Lemma~\ref{lem:app:sqrt_low_rank}. Both $W^{1/2}\otimes \ov \Delta$ and $\ov \Delta \otimes W^{1/2}$ have rank $nk$, and the last term has rank $k^2$. This completes the proof. \qedhere
\end{itemize}
\end{proof}

\begin{remark}
The above results essentially show that if we give a low rank update to $W \in \R^{n \times n}$ and we wish $(W+\Delta)^{1/2}$ is also a low rank update to $W^{1/2}$, then the update $\Delta$ must share the same eigenbasis as $W$.
\end{remark}
 
We define a function which will be heavily used in Section~\ref{subsec:app:kro_result}.
\begin{definition}
\label{def:f_a_c}
Let $\theta$ and $\omega$ be two fixed parameters, which satisfy that $\Tmat(n^2,n,n^2) = n^{\theta}$ and $\Tmat(n,n,n)= n^{\omega}$. 
We define the function $f(a, c)$ as 
\begin{align*}
    f(a, c) := & ~ \frac{c(\theta-\omega-2)+a(2+\theta-c\theta-\omega+2c\omega)-\theta}{a-1}.
\end{align*}
We further define $F(a):=\max_{c\in [a,1]} f(a,c)$. Since $f(a,c)$ is linear in $c$, with $f(a,a)=\theta+a(2\omega-\theta)$ and $f(a,1)=\omega+2$, we have
\begin{align*}
    F(a) = & ~ \max\{\theta+a(2\omega-\theta),\omega+2\}.
\end{align*}
Since $\theta\leq 2\omega$ and $\omega\geq 2$, we have $F(a)\in [\omega+2,2\omega]$ for every $a\in (0,1)$.
\end{definition}

\subsection{Our Data Structure}\label{subsec:app:kro_data_stru}
In this section, we present our data structure for initialization (Algorithm~\ref{alg:app:init}), update (Algorithm~\ref{alg:app:update}) and query (Algorithm~\ref{alg:app:query}). 
\begin{algorithm}[!ht]\caption{Initialization and members}\label{alg:app:init}
\begin{algorithmic}[1]
\State {\bf data structure} \textsc{KrockerProjMaintain} \Comment{Theorem~\ref{thm:app:kronecker_maintain}}
\State {\bf members}
\State \hspace{4mm} ${W} \in \R^{n \times n}$
\State \hspace{4mm} $\mathsf{A} \in \R^{m\times n^2}$ \Comment{Fixed data matrix}
\State \hspace{4mm} $\mathsf{G} \in \R^{m\times n^2}$ \Comment{Data matrix with eigenbasis}
\State \hspace{4mm} $M\in \R^{n^2\times n^2}$ \Comment{Inverse Hessian}
\State \hspace{4mm} $\lambda, \wt \lambda\in \R^n$ \Comment{Eigenvalues and its approximation}
 
\State \hspace{4mm}$Q\in \R^{n^2\times sb}$
\State \hspace{4mm}$P\in \R^{n^2\times sb}$

\State \hspace{4mm} $\epsilon_{\mathrm{mp}} \in (0,0.1)$ \Comment{Accuracy parameter}
\State \hspace{4mm} $a \in (0,\min\{\alpha,(3-\omega)/(\omega-1)\}]$ \Comment{Cutoff threshold}
\State {\bf end members}
\State 
\Procedure{Init}{$\mathsf{A} \in \R^{m \times n^2}, W \in \R^{n \times n}$} \Comment{Lemma~\ref{lem:app:init}}
    \State  $\mathsf{A}\gets \mathsf{A}$
    \State  $W\gets W$
    \State Let $W = U \Lambda U^\top$ \Comment{Compute the spectral decomposition for $W$}
   
    \State ${\sf G}\gets {\sf A}(U\otimes U)$
    \State Generate $R_{1,*},\ldots,R_{s,*}\in \R^{b\times n^2}$ to be sketching matrices
    \State $\mathsf{R}\gets [R_{1,*},\ldots,R_{s,*}] \in \R^{bs\times n^2}$
    \State $\lambda\gets \lambda$
    \State $\wt \lambda\gets \lambda$
    \State $M\gets {\sf G}^\top (\mathsf{G}(\Lambda\otimes \Lambda)\mathsf{G}^\top)^{-1} {\sf G}$ 
    \State $Q\gets M (\Lambda^{1/2} \otimes \Lambda^{1/2})(U^\top \otimes U^\top){\sf R}^\top$
    \State $P\gets (U\otimes U)(\Lambda^{1/2}\otimes \Lambda^{1/2})Q$
\EndProcedure
\State 
\State {\bf private:}
\Procedure{SoftThreshold}{$\lambda\in \R^n, \lambda^{\new}\in \R^n, r\in \mathbb{N}_+$}
\State $y_i\gets \ln \lambda_i^{\new}-\ln \lambda_i$
\State Let $\pi:[n]\rightarrow [n]$ be a sorting permutation such that $|y_{\pi(i)}|\geq |y_{\pi(i+1)}|$
        \While{$1.5\cdot r<n$ and $|y_{\pi(\lceil{1.5\cdot r\rceil})}|\geq (1-1/\log n)|y_{\pi(r)}|$}
            \State $r\gets \min(\lceil{1.5\cdot r\rceil},n)$
        \EndWhile
        \State $\wh \lambda_{\pi(i)}\gets \begin{cases}
        \lambda^{\new}_{\pi(i)} & i\in \{1,2,\ldots,r\} \\
        \lambda_{\pi(i)} & i \in \{r+1,\ldots,n \}
        \end{cases}$
    \State \Return $\wh \lambda, r$
\EndProcedure
\State {\bf end data structure}
\end{algorithmic}
\end{algorithm}

\begin{algorithm}[!ht]\caption{ \textsc{Update}  
part of our data structure.  
}\label{alg:app:update}
\begin{algorithmic}[1]
\State {\bf data structure} \textsc{KrockerProjMaintain}
\Procedure{Update}{$W^{\new}$} \Comment{Lemma~\ref{lem:app:correct}}

    \State \Comment{$W^{\new}=U\diag(\lambda^{\new})U^\top$}
    \State $y_i\gets \ln \lambda^{\new}_i-\ln  \lambda_i$
    \State $r\gets \text{the number of indices $i$ such that $|y_i|\geq \epsilon_{\mathrm{mp}}/2$}$
    \If{$r<n^{a}$} \Comment{No update}
        \State $\wh \lambda\gets \lambda$
        \State $V^{\new}\gets W$ 
        \State $M^{\new}\gets M$
        \State $Q^{\new}\gets Q$
        \State $P^{\new}\gets P$
        
    \Else
        \State $\wh \lambda, r \gets \textsc{SoftThreshold}(\lambda, \lambda^{\new}, r)$
        \State $C \gets \wh \lambda - \lambda$ \Comment{Entries updated by $\lambda^{\new}$}
        \State $\Delta \gets \Lambda \otimes C+C\otimes \Lambda+C\otimes C$ \Comment{$\Delta\in \R^{n^2\times n^2}$ is diagonal, has at most $2nr$ nonzero entries}
        \State $\wt S\gets \mathrm{supp}(\Delta)$, the set of indices of the nonzero diagonal entries of $\Delta$ \Comment{$|\wt S|\leq 2nr$}
        \State Let ${M}_{\wt S}\in \R^{n^2\times |\wt S|}$ be the columns corresponding to $\wt S$
        \State Let $M_{\wt S,\wt S}$ be the rows and columns corresponding to $\wt S$
        \State Let $\Delta_{\wt S,\wt S}$ be the diagonal matrix of the nonzero entries of $\Delta$
        \State $M^{\new}\gets M-M_{\wt S}\cdot (\Delta_{\wt S,\wt S}^{-1}+M_{\wt S, \wt S})^{-1} M_{\wt S}^\top$
        \State $\Gamma \gets (\Lambda+C)^{1/2} \otimes (\Lambda+C)^{1/2}-\Lambda^{1/2}\otimes \Lambda^{1/2}$
        \State $Q^{\new} \leftarrow Q + ( M^{\new} \cdot \Gamma) \cdot (U^\top\otimes U^\top)\cdot \mathsf{R}^\top + (M^{\new} - M) \cdot (\Lambda^{1/2}\otimes \Lambda^{1/2}) \cdot (U^\top\otimes U^\top)\cdot \mathsf{R}^\top$ \label{line:app:update_Q}
        \State $P^{\new}\gets P+(U\otimes U)\cdot \Gamma \cdot Q^{\new}+(U\otimes U)\cdot (\Lambda^{1/2}\otimes \Lambda^{1/2})\cdot (Q^{\new}-Q)$\label{line:app:update_P}
        \State $V^{\new}\gets U\diag(\wh \lambda) U^\top$
    \EndIf
    \State $\lambda\gets \wh \lambda$
    \State $Q \gets Q^{\new}$
    \State $P \gets P^{\new}$
    \State $M\gets M^{\new}$ \label{line:app:update_M}
    \State $W\gets V^{\new}$
    \State $\wt \lambda_i \gets \begin{cases}
    \wh \lambda_i & \text{if $|\ln \lambda_i^{\new}-\ln \wh \lambda_i|\leq \epsilon_{\mathrm{mp}}/2$} \\
    \lambda^{\new}_i & \text{otherwise}
    \end{cases}$
    \State \Return $U\diag(\wt \lambda )U^\top$
\EndProcedure
\State {\bf end data structure}
\end{algorithmic}
\end{algorithm}

\begin{algorithm}[!ht]\caption{Query}\label{alg:app:query}
\begin{algorithmic}[1]
\Procedure{Query}{$h^{\new}$} \Comment{Lemma~\ref{lem:app:correct} } 
    \State Let $S$ denote the set of indices $i$ such that $\wt \lambda_i\neq \lambda_i$ \Comment{$|S|<n^a$}
    \State $\wt C\gets \wt \lambda-\lambda$
    \State $\wt \Delta\gets \Lambda\otimes \wt C+\wt C\otimes \Lambda+\wt C\otimes \wt C$
    \State $\wt S\gets \mathrm{supp}(\wt \Delta)$, the set of indices of the nonzero diagonal entries of $\wt \Delta$ \Comment{$|\wt S|\leq 2n|S|$}
    \State $\wt \Gamma\gets (\Lambda+\wt C)^{1/2}\otimes (\Lambda+\wt C)^{1/2}-\Lambda^{1/2} \otimes \Lambda^{1/2}$

    \If{$S=\emptyset$}
    \State $p_g\gets 0$
    \Else
    \State $p_g\gets (U\otimes U)\cdot(\wt \Lambda^{1/2}\otimes \wt \Lambda^{1/2})\cdot (M_{*,\wt S})\cdot (\wt \Delta_{\wt S, \wt S}^{-1}+M_{\wt S, \wt S})^{-1}\cdot (Q_{\wt S, l}+M_{\wt S, *}\cdot \wt \Gamma \cdot (U^\top\otimes U^\top)\cdot R_{*,l}^\top)\cdot R_{*,l}h^{\new}$
    \EndIf
    \State $p_l\gets (U\otimes U)\cdot(\wt\Lambda^{1/2}\otimes \wt\Lambda^{1/2})\cdot (Q_{*,l}+M\cdot \wt \Gamma\cdot (U^\top\otimes U^\top)\cdot R^\top_{*,l})R_{*,l}h^{\new}-p_g$
    \State \Return $p_l$
\EndProcedure
\end{algorithmic}
\end{algorithm}

\subsection{Main Results}\label{subsec:app:kro_result}

The goal of this section is to prove Theorem~\ref{thm:app:kronecker_maintain} that, given a sequence of online matrices and queries, there exists a data structure that approximately maintains the projection matrix and the requested matrix-vector product. 
\begin{theorem}[Formal verison of Theorem~\ref{thm:kronecker_maintain}]\label{thm:app:kronecker_maintain}
Given a collection of matrices $A_1, \cdots, A_m \in \R^{n \times n}$. We define $B_i = W^{1/2} A_i W^{1/2} \in \R^{n \times n}, \forall i \in [m]$. We define $\mathsf{A} \in \R^{m \times n^2}$ to be the matrix whose $i$-th row is the vectorization of $A_i \in \R^{n \times n}$ and $\mathsf{B} \in \R^{m \times n^2}$ to be the matrix whose $i$-th row is the vectorization of $B_i \in \R^{n \times n}$. Assume that $\mathsf{A}$ has full row rank, i.e., $\mathrm{rank}(\mathsf{A})=m\leq n^2$.
Let $b$ denote the sketching dimension and let $T$ denote the number of iterations. Let $\epsilon_{\mathrm{mp}} \in (0, 0.1)$ and $a \in (0,\min\{\alpha,(3-\omega)/(\omega-1)\}]$ be parameters.
Let $R_1, \cdots, R_s \in\R^{b \times n^2}$ denote a list of $s$ sketching matrices, and let $\mathsf{R} \in \R^{sb \times n^2}$ denote the batched matrix of these matrices. Then, there is a dynamic maintenance data structure (\textsc{KroneckerProjMaintain}) that given a sequence of online matrices
\begin{align*}
    W^{(1)}, \cdots, W^{(T)} \subset \R^{n \times n}; \text{~~~and~~~} h^{(1)}, \cdots, h^{(T)} \in \R^{n^2}
\end{align*}
that approximately maintains the projection matrices
\begin{align*}
    \mathsf{B}^\top (\mathsf{B} \mathsf{B}^\top)^{-1} \mathsf{B} 
\end{align*}
for positive definite matrices $W^{(k)}=U\Lambda^{(k)}U^\top \in \R^{n \times n}$ that share a fixed orthonormal eigenbasis $U$, through the following two operations:

\begin{itemize}
    \item \textsc{Update}$(W)$: Output a positive definite matrix $\wt{V} \in \R^{n \times n}$ such that for all $i \in [n]$
    \begin{align*}
        (1-\epsilon_{\mathrm{mp}} ) \cdot \lambda_i( \wt{V}) \leq \lambda_i(W) \leq (1+\epsilon_{\mathrm{mp}}) \cdot \lambda_i( \wt{V} )
    \end{align*}
    where $\lambda_i(W)$ denote the eigenvalue of $W \in \R^{n \times n}$ on the $i$-th column of $U$. If this call changes $n^c$ entries of the stored eigenvalue vector, so that the Kronecker matrices receive a change of rank at most $2n^{1+c}$, it takes $O(\Tmat(n^2,n^{1+c},\max\{n^2,sb\}))$ time in the worst case, which is $n^{2\omega}$ for $c=1$ and $sb\leq n^2$. We claim only the amortized bound below for \textsc{Update}.
    \item \textsc{Query}$(h)$: Output $\wt{\mathsf{B}}^\top ( \wt{\mathsf{B} } \wt{ \mathsf{B}}^\top )^{-1} \wt{\mathsf{B}} R_{l}^\top R_l \cdot h$ for the $\wt{B}$ defined by positive definite matrix $\wt{V} \in \R^{n \times n}$ outputted by the last call to \textsc{Update}.
\end{itemize}
The data structure takes $mn^\omega+m^\omega+\Tmat(m,m,n^2)+\Tmat(n^2,m,n^2)+\Tmat(n^2,n^2,sb)$
time to initialize, and each call of \textsc{Query} takes time
\begin{align*}
    n^{2+b+o(1)} + n^{3+a+o(1)}.
\end{align*}
Furthermore, if the initial matrix $W^{(0)} \in \R^{n \times n}$ and the (random) update sequence $W^{(1)}, W^{(2)}, \cdots, W^{(T)} \in \R^{n \times n}$ satisfies
\begin{align*}
    \sum_{i=1}^n ( \E[ \ln \lambda_i( W^{(k+1)} ) ] - \ln ( \lambda_i(W^{(k)}) ) )^2 \leq C_1^2
\end{align*}
and 
\begin{align*}    
    \sum_{i=1}^n ( \Var [ \ln \lambda_i(W^{(k+1)}) ] )^2 \leq C_2^2
\end{align*}
with the expectation and variance is conditioned on $\lambda_i(W^{(k)})$ for all $k=0,1,\cdots, T-1$. Then, the amortized expected time per call of \textsc{Update}$(W)$ is 
\begin{align*}
    (C_1/\epsilon_{\mathrm{mp}} + C_2/ \epsilon_{\mathrm{mp}}^2 ) \cdot \max\{{\cal T}_1, {\cal T}_2 \}.
\end{align*}
Here, ${\cal T}_1=n^{\omega-2.5+F(a)+o(1)}+n^{F(a)-a/2+o(1)}$ and ${\cal T}_2=n^{\omega+F(a)-4.5+o(1)}sb+n^{F(a)-(4+a)/2+o(1)}sb$, where $F(a)$ is defined in Definition~\ref{def:f_a_c}.
\end{theorem}

\begin{proof}
The correctness of update matrices and queries follows from Lemma~\ref{lem:app:correct}. The runtime of initialization follows from~Lemma~\ref{lem:app:init}. The runtime of $\textsc{Query}$ follows from Lemma~\ref{lem:app:query}.

For the runtime of update, we note that by Lemma~\ref{lem:app:update}, a call at step $t$ that changes $r_t=n^{c_t}\geq n^a$ stored eigenvalues costs $O(n^{c_t}g_{n^{c_t}}\cdot n^{f(a,c_t)}\cdot\max\{1,sb/n^2\})$ time. Since $c_t\in [a,1]$, we have $f(a,c_t)\leq F(a)$, so this cost is at most $O(r_tg_{r_t}\cdot n^{F(a)}\cdot \max\{1,sb/n^2\})$. Using a potential analysis similar to~\cite{cls19}, which applies since $a\leq \alpha$, we have that

\begin{align*}
    \sum_{t=1}^T r_t g_{r_t} = O(T\cdot (C_1/\epsilon_{mp}+C_2/\epsilon_{mp}^2)\cdot \log^{1.5}n\cdot (n^{\omega-2.5}+n^{-a/2})),
\end{align*}

where the sum is over the calls that change $M$. Multiplying by $n^{F(a)}\max\{1,sb/n^2\}$ gives the bound $(C_1/\epsilon_{\mathrm{mp}}+C_2/\epsilon_{\mathrm{mp}}^2)\cdot\max\{{\cal T}_1,{\cal T}_2\}$ per call, and this concludes the proof of the amortized running time of update.

Regarding the guarantees of eigenvalues, we can reuse a similar analysis of~\cite{cls19}, Section 5.4 and 5.5.
\end{proof}

Next, we present the runtime analysis of the initialization of our data structure. 
\begin{lemma}[Initialization Time]\label{lem:app:init}
Let $s$ denote the number of sketches. Let $b$ denote the size of each sketch. 
The $\textsc{Init}$ takes time
\begin{align*}
    mn^\omega+m^\omega+\Tmat(m,m,n^2)+\Tmat(n^2,m,n^2)+\Tmat(n^2,n^2,sb).
\end{align*}
Suppose $m=n^2$, then the time becomes
\begin{align*}
    m^\omega+\Tmat(m,m,sb).
\end{align*}
\end{lemma}
\begin{proof}

The initialization contains the following computations:
\begin{itemize}
    \item Compute the spectral decomposition for $W\in \R^{n\times n}$, takes $O(n^\omega)$ time.
    \item Compute matrix ${\sf G}={\sf A}(U\otimes U)$. We note that ${\sf A}$ can be viewed as $m$ different $n\times n$ matrices, and we can use the identity $\vect(A_i)(U\otimes U)=\vect(U^\top A_i U)$, hence, it takes $O(mn^\omega)$ time to compute ${\sf G}$. Note that the naive computation of ${\sf G}$ takes $\Tmat(m,n^2,n^2)$ time which is about $mn^{2(\omega-1)}$ time, and it's worse than the time by using the Kronecker product identity.
    \item Compute $M={\sf G}^\top ({\sf G}(\Lambda\otimes \Lambda){\sf G}^\top)^{-1}{\sf G}$. We split this computation into several parts:
    \begin{itemize}
        \item Compute ${\sf G}(\Lambda\otimes \Lambda)$. Since $\Lambda$ is diagonal, this takes $O(mn^2)$ time.
        \item Computing ${\sf G}(\Lambda\otimes \Lambda){\sf G}^\top$ takes $\Tmat(m,n^2,m)$ time.
        \item Computing the inverse $({\sf G}(\Lambda\otimes \Lambda){\sf G}^\top)^{-1}$ takes $O(m^\omega)$ time.
        \item Finally, computing $M$ as ${\sf G}^\top \cdot (({\sf G}(\Lambda\otimes \Lambda){\sf G}^\top)^{-1}{\sf G})$ takes $\Tmat(m, m, n^2)+\Tmat(n^2,m,n^2)$ time, where the second term accounts for the $n^2\times n^2$ output.
    \end{itemize}
    Since $\Tmat(m,n^2,m)=O(\Tmat(m,m,n^2))$, computing $M$ takes time
    \begin{align*}
        O(\Tmat(m,m,n^2)+\Tmat(n^2,m,n^2)+m^\omega).
    \end{align*}
    \item Computing $Q$ takes two steps: Appling sketching ${\sf R}^\top$ takes $\Tmat(n^2, n^2, sb)$ time, and computing the product between $M$ and $(\Lambda^{1/2}U^\top \otimes \Lambda^{1/2}U^\top){\sf R}^\top$ takes $\Tmat(n^2, n^2, sb)$ time.
    \item Computing $P$ takes $\Tmat(n^2, n^2, sb)$ time.
\end{itemize}
Thus, the total running time is
\begin{align*}
    mn^\omega+m^\omega+\Tmat(m,m,n^2)+\Tmat(n^2,m,n^2)+\Tmat(n^2,n^2,sb).
\end{align*}
Specifically, for $m=n^2$, we have $mn^\omega\leq n^{2\omega}=m^\omega$ and $\Tmat(m,m,n^2)=\Tmat(n^2,m,n^2)=m^\omega$, so the time becomes $O(m^\omega+\Tmat(m,m,sb))$.
\end{proof}

Next, we present the runtime analysis of the \textsc{Query} of our data structure. 

\begin{lemma}[Query Time]\label{lem:app:query}
Let $b$ denote the size of sketch and let $a\leq (3-\omega)/(\omega-1)$, then \textsc{Query} takes time
\begin{align*}
    O(n^{3+a}+n^{2+b}).
\end{align*}
\end{lemma}

\begin{proof}
First, observe that $|S|<n^a$. Before the first call of \textsc{Update}, $\wt\lambda=\lambda$, so $S=\emptyset$. After a call of \textsc{Update} that changes $M$, every index with $|y_i|\geq \epsilon_{\mathrm{mp}}/2$ is among the $r$ updated indices, so $\wt \lambda=\lambda$ and $S=\emptyset$. After a call that does not change $M$, $S$ consists of the indices with $|y_i|>\epsilon_{\mathrm{mp}}/2$, and there are fewer than $n^a$ of them. A diagonal entry of $\wt \Delta$ or of $\wt \Gamma$ can be nonzero only if at least one of its two Kronecker coordinates lies in $S$. Therefore $|\wt S|\leq 2n|S|\leq 2n^{1+a}$, and $\wt \Gamma$ has at most $2n^{1+a}$ nonzero entries. If $S=\emptyset$, then $p_g=0$ and only the steps for $p_l$ below are performed.

We also use that, for any $X\in \R^{n\times n}$, Fact~\ref{fact:app:k_vec_ABCt} gives $(U\otimes U)\vect(X)=\vect(UXU^\top)$ and $(U^\top\otimes U^\top)\vect(X)=\vect(U^\top XU)$, so multiplying a vector of length $n^2$ by $U\otimes U$ or by $U^\top\otimes U^\top$ takes $O(n^\omega)$ time for any $U$.

We compute time for each term as follows:

\begin{itemize}
    \item Computing $\wt \Delta$ takes $O(n^{1+a})$ time.
    \item Compute $\wt \Gamma$. Note that this is a diagonal matrix with at most $2n^{1+a}$ nonzero entries, so it takes $O(n^{1+a})$ time to compute.
    \item Computing $p_g$, involving the following steps:
    \begin{itemize}
        \item Computing $R_{*,l}h$, takes $O(n^{2+b})$ time.
        \item Computing $R^\top_{*,l} (R_{*,l}h)$ takes $O(n^{2+b})$ time.
        \item Compute $(U^\top\otimes U^\top)R^\top_{*,l} (R_{*,l}h)$ takes $O(n^{\omega})$ time by the identity above.
        \item Computing $\wt \Gamma\cdot (U^\top\otimes U^\top)\cdot R^\top_{*,l} (R_{*,l}h)$ takes $O(n^{1+a})$ time since $\wt \Gamma$ has $O(n^{1+a})$ nonzero entries on the diagonal.
        \item Computing $M_{\wt S, *}\cdot \wt \Gamma \cdot(U^\top\otimes U^\top)\cdot R^\top_{*,l} (R_{*,l}h)$ takes $O(n^{3+a})$ time.
        \item Similarly, computing $Q_{\wt S, l}R_{*,l}h^{\new}$ takes $O(n^{2+b}+n^{1+a+b})$ time.
        \item Computing the inverse $(\wt \Delta_{\wt S, \wt S}^{-1}+M_{\wt S,\wt S})^{-1}$ takes $O(n^{(1+a)\omega})$ time, which is $O(n^{3+a})$ because $a\leq (3-\omega)/(\omega-1)$ is equivalent to $(1+a)\omega\leq 3+a$.
        \item Computing $(\wt \Delta_{\wt S, \wt S}^{-1}+M_{\wt S,\wt S})^{-1}M_{\wt S, *}\cdot \wt \Gamma\cdot R^\top_{*,l} (R_{*,l}h)$ takes $O(n^{2+2a})$ time.
        \item Computing $M_{*, \wt S}(\wt \Delta_{\wt S, \wt S}^{-1}+M_{\wt S,\wt S})^{-1}M_{\wt S, *}\cdot \wt \Gamma\cdot (U^\top\otimes U^\top)\cdot R^\top_{*,l} (R_{*,l}h)$ takes $O(n^{3+a})$ time.
        \item Computing $(\wt\Lambda^{1/2}\otimes \wt\Lambda^{1/2})M_{*, \wt S}(\wt \Delta_{\wt S, \wt S}^{-1}+M_{\wt S,\wt S})^{-1}M_{\wt S, *}\cdot \wt \Gamma\cdot R^\top_{*,l} (R_{*,l}h)$, since $\Lambda$ is a diagonal matrix, it takes $O(n^2)$ time to form this matrix, and multiplying it with a vector of length $n^2$ takes $O(n^2)$ time.
        \item Computing $(U\otimes U)(\wt\Lambda^{1/2}\otimes \wt\Lambda^{1/2})M_{*, \wt S}(\wt \Delta_{\wt S, \wt S}^{-1}+M_{\wt S,\wt S})^{-1}M_{\wt S, *}\cdot \wt \Gamma\cdot R^\top_{*,l} (R_{*,l}h)$ takes $O(n^{\omega})$ time by the identity above.
    \end{itemize}
    \item Compute $p_l$. Note that the product $R_{*,l}^\top R_{*,l}h^{\new}$ takes $O(n^{2+b})$ time, $(U^\top\otimes U^\top)R_{*,l}^\top R_{*,l}h^{\new}$ takes $O(n^{\omega})$ time, and $\wt \Gamma (U^\top\otimes U^\top)(R_{*,l}^\top R_{*,l}h^{\new})$ takes $O(n^{1+a})$ time due to the sparsity of $\wt \Gamma$ and the resulting vector contains at most $2n^{1+a}$ nonzero entries. Therefore, computing $M(\wt \Gamma (U^\top\otimes U^\top) R_{*,l}^\top R_{*,l}h^{\new})$ takes $O(n^{3+a})$ time.

    Similarly, computing $Q_{*,l} R_{*,l} h^{\new}$ takes $O(n^{2+b})$ time.

    Finally, computing the product between an $n^2\times n^2$ diagonal matrix and a length $n^2$ vector takes $O(n^2)$ time, and multiplying the vector with $(U\otimes U)$ takes $O(n^{\omega})$ time.
\end{itemize}

Overall, since $\omega\leq 3$, it takes
\begin{align*}
    O(n^{3+a}+n^{2+b}+n^{(1+a)\omega}+n^\omega)=O(n^{3+a}+n^{2+b})
\end{align*}
time to realize this step.
\end{proof}

To adapt an amortized analysis for \textsc{Update}, we introduce several definitions.

\begin{definition}
Given $i\in [r]$, we define the weight function as
\begin{align*}
    g_i = & ~ \begin{cases}
    n^{-a} & \text{if $i<n^a$} \\
    i^{\frac{\omega-2}{1-a}-1}n^{-\frac{a(\omega-2)}{1-a}} & \text{otherwise}
    \end{cases}
\end{align*}
\end{definition}

This is a well-known weight function used in~\cite{cls19,lsz19} and many subsequent works that use rank-aware amortization for matrix multiplication.

\begin{definition}
Let $\theta\in [4,5]$ be the value such that 
\begin{align*}
    \Tmat(n^2,n,n^2)=n^\theta.
\end{align*}
\end{definition}

Note that when $\omega=2$, we have that $\theta=4$.

\begin{lemma}
Let $c\in [0,1]$, then we have
\begin{align*}
    \Tmat(n^2, n^{1+c}, n^2) = & ~ n^{c (2\omega-\theta)+\theta}
\end{align*}
\end{lemma}

Recall Definition~\ref{def:f_a_c}, we define the function $f(a,c)$ as follows: 
\begin{align*}
    f(a, c) :=  \frac{c(\theta-\omega-2)+a(2+\theta-c\theta-\omega+2c\omega)-\theta}{a-1}.
\end{align*}
We will use this function $f$ to simplify our amortization.

\begin{corollary}
For every $c\in [a,1]$, we have that
\begin{align*}
    \Tmat(n^2, n^{1+c}, n^2) = & ~ n^c g_{n^c} \cdot n^{f(a,c)}
\end{align*}
\end{corollary}

\begin{proof}
By basic algebraic manipulation, we have
\begin{align*}
    c(2\omega-\theta)+\theta = & ~ \frac{(c-a)(\omega-2)}{1-a}+\frac{c(\theta-\omega-2)+a(2+\theta-c\theta-\omega+2c\omega)-\theta}{a-1} \\
    = & ~ \frac{(c-a)(\omega-2)}{1-a}+f(a,c),
\end{align*}
where the second step is by definition of $f(a,c)$.
\end{proof}

\begin{lemma}[Property of $f(a,c)$]
If $\omega=2$ and $\theta=4$, then
\begin{align*}
    f(a,c) = & ~ 4,
\end{align*}
for any $a\in (0,1)$ and $c\in (0,1)$.
\end{lemma}

\begin{proof}
Suppose $\omega=2$ and $\theta=4$, We can simplify $f(a,c)$ as 
\begin{align*}
    f(a,c) = & ~ \frac{4+a(4c-4+2-4c-2)}{1-a} \\
    = & ~ \frac{4-4a}{1-a} \\
    = & ~ 4 \qedhere
\end{align*}
\end{proof}

\begin{remark}
Our proof shows that if $\omega=2$, and therefore $\theta=4$, then the term $f(a,c)$ is always 4. In that case, as we will show below, for $sb\leq n^2$, the running time of a call of update that changes $r$ stored eigenvalues would be
\begin{align*}
    O(rg_r n^4),
\end{align*}
and, using a result proved in~\cite{cls19}, the amortized running time would be 
\begin{align*}
    \wt O(n^{\omega+1.5}+n^{4-a/2}).
\end{align*}

This means the amortized time of update would be subquadruple if $\omega$ were $2$. For general $\omega$, the exponent $\omega-2.5+F(a)$ in Theorem~\ref{thm:app:kronecker_maintain} is at least $2\omega-0.5$, since $F(a)\geq \omega+2$, so it is below $4$ only if $\omega<2.25$, which fails, since $\omega=2.25$~\cite[Theorem~1.1]{OpenAI26}. Theorem~\ref{thm:sdp_fixed_basis} runs this data structure inside an SDP solver for the class of SDPs of that theorem. For $m=\Theta(n^2)$ and $\omega=2$ that solver would be below the bound of~\cite{hjstz21}, but at $\omega=2.25$ it is not. By part 4 of that theorem, the same solver with no call of \textsc{Update} has a smaller bound, so even at $\omega=2$ this improvement would not come from the data structure.
\end{remark}

\begin{lemma}[Update Time]
\label{lem:app:update}
Let $r=n^c\geq n^a$ be the number of stored eigenvalues that a call of \textsc{Update} changes, so that the Kronecker matrices receive a change of rank at most $2n^{1+c}$. Then this call takes time $O(\Tmat(n^2,n^{1+c},\max\{n^2,sb\}))$, which is $O(rg_r\cdot n^{f(a,c)})$ if $sb\leq n^2$ and $O(rg_r\cdot n^{f(a,c)-2}sb)$ otherwise.

\end{lemma}

\begin{proof}
We note that if the number of indices $i$ with $|y_i|\geq \epsilon_{\mathrm{mp}}/2$ is less than $n^a$, then we simply update some variables in the data structure.

In the other case, we perform the following operations. Let $r=n^c$ be the value returned by \textsc{SoftThreshold}, so that $nr=n^{1+c}$ and $|\wt S|\leq 2nr$. Then
\begin{itemize}
    \item Forming $\Delta$ and $\wt S$ takes $O(n^{2})$ time.
    \item Adding two $|\wt S|\times |\wt S|$ matrices takes $O(n^{2+2c})$ time.
    \item Inverting a $|\wt S|\times |\wt S|$ matrix takes $O(n^{(1+c)\omega})$ time.
    \item Multiplying an $n^2\times |\wt S|$ matrix with an $|\wt S|\times n^2$ matrix takes $O(\Tmat(n^2,nr,n^2))=O(rg_r\cdot n^{f(a, c)})$ time by the corollary above, since $c\geq a$.
\end{itemize}
The first three costs are at most $O(\Tmat(n^2,nr,n^2))$, since $nr\leq n^2$.

To compute $Q^{\new}$, note that $(U^\top \otimes U^\top) {\sf R}^\top$ can be pre-computed and stored, yielding a matrix of size $n^2\times sb$. 
\begin{itemize}
    \item Computing $(\Lambda^{1/2}\otimes \Lambda^{1/2})\cdot (U^\top \otimes U^\top)\cdot {\sf R}^\top$ takes $O(sbn^2)$ time since $\Lambda$ is diagonal.
    \item Computing $(M^{\new}-M)\cdot (\Lambda^{1/2}\otimes \Lambda^{1/2})\cdot (U^\top \otimes U^\top)\cdot {\sf R}^\top$ can be viewed as a product of four matrices:
    \begin{align*}
        n^2 \times nr \rightarrow nr\times nr\rightarrow nr\times n^2\rightarrow n^2\times sb,
    \end{align*}
    the time is thus dominated by $\Tmat(n^2, nr, \max\{n^2, sb\})$.
    \item Computing $(M^{\new}\cdot \Gamma)\cdot (U^\top \otimes U^\top)\cdot {\sf R}^\top$. Note that $\Gamma$ is a diagonal matrix with at most $2nr$ nonzero entries, therefore $M^{\new}\cdot \Gamma$ can be viewed as selecting and scaling at most $2nr$ columns of $M^{\new}$, which gives a matrix of size $n^2\times 2nr$. Multiplying with the corresponding rows of $(U^\top \otimes U^\top){\sf R}^\top$ then takes $\Tmat(n^2, nr, sb)$ time.
\end{itemize}
To compute $P^{\new}$, we use that a diagonal entry of $\Gamma$ is nonzero exactly when the corresponding entry of $\Delta$ is nonzero, so $\Gamma$ is supported on $\wt S$. Let $K=(\Delta_{\wt S,\wt S}^{-1}+M_{\wt S,\wt S})^{-1}$ and $Y=(U^\top \otimes U^\top){\sf R}^\top$.
\begin{itemize}
    \item The product $(U\otimes U)\cdot \Gamma\cdot Q^{\new}$ only involves $|\wt S|\leq 2nr$ columns of $U\otimes U$ and the same number of rows of $Q^{\new}$, so it takes $\Tmat(n^2,nr,sb)$ time.
    \item Since $M^{\new}=M-M_{\wt S}KM_{\wt S}^\top$, the nonzero columns of $M^{\new}\cdot\Gamma$ form the matrix $M_{\wt S}(I-KM_{\wt S,\wt S})\Gamma_{\wt S,\wt S}$, and $M^{\new}-M=-M_{\wt S}KM_{\wt S}^\top$. Hence $Q^{\new}-Q=M_{\wt S}\cdot Z$ with
    \begin{align*}
        Z=(I-KM_{\wt S,\wt S})\Gamma_{\wt S,\wt S}Y_{\wt S,*}-KM_{\wt S}^\top(\Lambda^{1/2}\otimes \Lambda^{1/2})Y\in \R^{|\wt S|\times sb},
    \end{align*}
    which takes $O(\Tmat(nr,n^2,sb)+\Tmat(nr,nr,sb)+n^{(1+c)\omega})$ time to compute. Then $(U\otimes U)\cdot(\Lambda^{1/2}\otimes \Lambda^{1/2})\cdot (Q^{\new}-Q)=((U\otimes U)(\Lambda^{1/2}\otimes \Lambda^{1/2})M_{\wt S})\cdot Z$ takes $\Tmat(n^2,n^2,nr)+\Tmat(n^2,nr,sb)$ time.
\end{itemize}
Since $\Tmat$ is invariant under permuting its three arguments up to a constant factor, each of the above costs, and hence the total running time, is $O(\Tmat(n^2,\allowbreak nr,\allowbreak \max\{n^2,sb\}))$. This is $O(rg_r\cdot n^{f(a, c)})$ if $sb\leq n^2$. Otherwise, splitting the $sb$ columns into $\lceil sb/n^2\rceil$ blocks of at most $n^2$ columns shows that the running time is $O(\Tmat(n^2, nr, sb))=O(rg_r\cdot n^{f(a,c)-2}sb)$.

\end{proof}

Next, we present the matrix Woodbury Identity regarding the calculation of the inverse of the matrix $(A+UCV)^{-1}$. 
\begin{lemma}[Matrix Woodbury Identity]\label{lem:woodbury}
Let $A\in \R^{n\times n}$, $U\in \R^{n\times k},C\in \R^{k\times k}$ and $V\in \R^{k\times n}$ and both $A$ and $C$ are non-singular. Then we have
\begin{align*}
    (A+UCV)^{-1} = & ~ A^{-1}-A^{-1}U(C^{-1}+VA^{-1}U)^{-1}VA^{-1}.
\end{align*}
\end{lemma}

Next, we present the proof of the correctness of our \textsc{Update} and \textsc{Query} procedure in our data structure.
 
\begin{lemma}[The correctness of Update and Query]\label{lem:app:correct}
 
By line~\ref{line:app:update_M} of \textsc{Update}$(W^{\new})$ (Algorithm~\ref{alg:app:update}), the variables satisfy
\begin{align*}
    M^{\new} = & ~(U^\top \otimes U^\top)\mathsf{A}^\top (\mathsf{A} ( V^{\new} \otimes V^{\new} ) \mathsf{A}^\top )^{-1} \mathsf{A}(U\otimes U) \\
    Q^{\new} = & ~ M^{\new}  ( \wh\Lambda^{1/2}U^\top \otimes \wh\Lambda^{1/2}U^\top ) \mathsf{R}^\top \\
    P^{\new} = & ~ (U\wh\Lambda^{1/2}\otimes U\wh\Lambda^{1/2}) M^{\new}  ( \wh\Lambda^{1/2}U^\top \otimes \wh\Lambda^{1/2}U^\top ) \mathsf{R}^\top
\end{align*}
where $\wh\Lambda=\diag(\wh\lambda)$ and $V^{\new}=U\wh\Lambda U^\top$ is the matrix stored by \textsc{Update}. After a call that does not change $M$, $V^{\new}$ is the previously stored matrix.
Additionally, the output of \textsc{Query}$(h^{\new})$ satisfies
\begin{align*}
     p_l^{\new} = & ~ \wt P\cdot R_{*,l}^\top R_{*,l}\cdot h^{\new}
\end{align*}
where $\wt{P} = \wt{\B}^\top ( \wt{\B} \wt{\B}^\top )^{-1} \wt{\B}$ and $\wt{\B} $ is defined based on $\wt{V}$ which is outputted by \textsc{Update}$(W)$.
\end{lemma}

\begin{remark}
We generalize the Lemma E.3 in \cite{sy21} from the diagonal $W$ case to the case of positive definite $W$ with a fixed eigenbasis.
\end{remark}

\begin{proof}

We proceed by induction over the calls of \textsc{Update}. After \textsc{Init}, the three identities hold with $\wh\Lambda=\Lambda$ and $V^{\new}=W$. A call that does not change $M$ keeps $M$, $Q$, $P$, $\lambda$ and the stored matrix, so the identities continue to hold. It remains to consider a call that changes $M$, where $\wh\Lambda=\Lambda+C$.

{\bf Correctness for $M$.} Since $(U^\top \otimes U^\top)\mathsf{A}^\top=\mathsf{G}^\top$ and $\mathsf{A} ( V^{\new} \otimes V^{\new} ) \mathsf{A}^\top=\mathsf{G}(\wh\Lambda\otimes\wh\Lambda)\mathsf{G}^\top$, the correctness follows from Lemma~\ref{cor:app:update_eigen}.

{\bf Correctness for $Q$.}

We have
\begin{align*}
     & ~ Q^{\new} \\
    = & ~ Q + (M^{\new}\cdot \Gamma)\cdot (U^\top\otimes U^\top)\cdot {\sf R}^\top+(M^{\new}-M)\cdot (\Lambda^{1/2}U^\top\otimes \Lambda^{1/2}U^\top)\cdot {\sf R}^\top \\
    = & ~ M(\Lambda^{1/2}U^\top \otimes \Lambda^{1/2}U^\top) {\sf R}^\top+M^{\new}((\Lambda+C)^{1/2} U^\top \otimes (\Lambda+C)^{1/2} U^\top) {\sf R}^\top-M^{\new}(\Lambda^{1/2}U^\top \otimes \Lambda^{1/2} U^\top) {\sf R}^\top\\
    + & ~ (M^{\new}-M)\cdot (\Lambda^{1/2}U^\top\otimes \Lambda^{1/2}U^\top)\cdot {\sf R}^\top\\
    = & ~ M(\Lambda^{1/2}U^\top \otimes \Lambda^{1/2}U^\top) {\sf R}^\top + M^{\new}( (\Lambda+C)^{1/2}U^\top \otimes (\Lambda+C)^{1/2} U^\top  ) \mathsf{R}^\top - M(\Lambda^{1/2}U^\top \otimes \Lambda^{1/2}U^\top) {\sf R}^\top \\
    = & ~ M^{\new} ( (\Lambda+C)^{1/2}U^\top \otimes (\Lambda+C)^{1/2} U^\top  ) \mathsf{R}^\top
\end{align*}
 
where the first step follows from line~\ref{line:app:update_Q} in Algorithm~\ref{alg:app:update}, the second step follows from definition of $Q$ and of $\Gamma$, the third step follows from re-organizing terms, and the last step follows from cancelling the first term and the last term.

{\bf Correctness for $P$.}
\begin{align*}
    P^{\new}
    = & ~ P + (U\otimes U)\cdot \Gamma \cdot Q^{\new} + (U\Lambda^{1/2}\otimes U\Lambda^{1/2})\cdot (Q^{\new}-Q) \\
    = & ~ ( U\Lambda^{1/2} \otimes U\Lambda^{1/2} ) Q  + (U(\Lambda+C)^{1/2}\otimes U(\Lambda+C)^{1/2}) \cdot Q^{\new}-(U\Lambda^{1/2}\otimes U\Lambda^{1/2})Q^{\new} \\
    + & ~ (U\Lambda^{1/2}\otimes U\Lambda^{1/2})\cdot(Q^{\new}-Q) \\
    = & ~ (U(\Lambda+C)^{1/2}\otimes U(\Lambda+C)^{1/2}) Q^{\new}
\end{align*}

where the first step follows from line~\ref{line:app:update_P} in Algorithm~\ref{alg:app:update}, the second step follows from definition of $P$ and of $\Gamma$, and the last step follows from merging the terms.

{\bf Correctness of \textsc{Query}.}

Here $\Lambda=\diag(\lambda)$ for the stored vector $\lambda$, $\wt\Lambda=\Lambda+\wt C=\diag(\wt\lambda)$, and $\wt \Gamma=\wt\Lambda^{1/2}\otimes\wt\Lambda^{1/2}-\Lambda^{1/2}\otimes\Lambda^{1/2}$. If $S=\emptyset$, then $\wt\Lambda=\Lambda$, $\wt\Gamma=0$ and $p_g=0$, so $p_l^{\new}=(U\otimes U)(\Lambda^{1/2}\otimes \Lambda^{1/2})Q_{*,l}R_{*,l}h^{\new}=\wt P R_{*,l}^\top R_{*,l}h^{\new}$ by the identity for $Q$. Assume now that $S\neq\emptyset$, so that $\wt S=\mathrm{supp}(\wt\Delta)$ is nonempty.

We first unravel $p_g^{\new}$:
\begin{align*}
    & ~ (U\otimes U)(\wt \Lambda^{1/2}\otimes \wt \Lambda^{1/2})(M_{*,\wt S})(\wt \Delta^{-1}_{\wt S,\wt S}+M_{\wt S,\wt S})^{-1} (Q_{\wt S,l}+M_{\wt S,*} \wt \Gamma (U^\top\otimes U^\top) R_{*,l}^\top) \\
    = & ~ (U\otimes U)(\wt \Lambda^{1/2}\otimes \wt \Lambda^{1/2})(M_{*,\wt S})(\wt \Delta^{-1}_{\wt S,\wt S}+M_{\wt S,\wt S})^{-1} (M_{\wt S,*}(\Lambda^{1/2}\otimes \Lambda^{1/2})(U^\top\otimes U^\top) R_{*,l}^\top+M_{\wt S,*} \wt \Gamma (U^\top\otimes U^\top) R_{*,l}^\top) \\
    = & ~ (U\otimes U)(\wt \Lambda^{1/2}\otimes \wt \Lambda^{1/2})(M_{*,\wt S})(\wt \Delta^{-1}_{\wt S,\wt S}+M_{\wt S,\wt S})^{-1} (M_{\wt S,*})(\wt \Lambda^{1/2}\otimes \wt \Lambda^{1/2})(U^\top\otimes U^\top) R_{*,l}^\top.
\end{align*}
Hence, 
\begin{align*}
    & ~ p_g^{\new}\\
    = & ~ (U\otimes U)(\wt \Lambda^{1/2}\otimes \wt \Lambda^{1/2})(M_{*,\wt S})(\wt \Delta^{-1}_{\wt S,\wt S}+M_{\wt S,\wt S})^{-1} (Q_{\wt S,l}+M_{\wt S,*} \wt \Gamma (U^\top\otimes U^\top) R_{*,l}^\top) R_{*,l}h^{\new} \\
     = & ~ (U\otimes U)(\wt \Lambda^{1/2}\otimes \wt \Lambda^{1/2})(M_{*,\wt S})(\wt \Delta^{-1}_{\wt S,\wt S}+M_{\wt S,\wt S})^{-1} (M_{\wt S,*})(\wt \Lambda^{1/2}\otimes \wt \Lambda^{1/2})(U^\top\otimes U^\top) R_{*,l}^\top R_{*,l}h^{\new}.
\end{align*}

To see $p_l^{\new}$, it suffices to show the following:
\begin{align*}
    & ~ (U\otimes U)(\wt \Lambda^{1/2}\otimes \wt \Lambda^{1/2})(Q_{*,l}+M\wt \Gamma (U^\top\otimes U^\top) R_{*,l}^\top) \\
    = & ~ (U\otimes U)(\wt \Lambda^{1/2}\otimes \wt \Lambda^{1/2})(M (\Lambda^{1/2}\otimes \Lambda^{1/2})(U^\top\otimes U^\top) R_{*,l}^\top+M\wt \Gamma (U^\top\otimes U^\top) R_{*,l}^\top) \\
    = & ~ (U\otimes U)(\wt \Lambda^{1/2}\otimes \wt \Lambda^{1/2})M (\wt\Lambda^{1/2}\otimes \wt\Lambda^{1/2})(U^\top\otimes U^\top) R_{*,l}^\top.
\end{align*}

To stitch everything together, we notice that
\begin{align*}
    & ~ M-(M_{*,\wt S})(\wt \Delta^{-1}_{\wt S,\wt S}+M_{\wt S,\wt S})^{-1} (M_{\wt S,*})\\
    = & ~ {\sf G}^\top ({\sf G}(\Lambda\otimes \Lambda){\sf G}^\top)^{-1}{\sf G}-{\sf G}^\top ({\sf G}(\Lambda\otimes \Lambda){\sf G}^\top)^{-1}{\sf G}_{*,\wt S}(\wt \Delta^{-1}_{\wt S,\wt S}+M_{\wt S,\wt S})^{-1} {\sf G}^\top_{\wt S,*} ({\sf G}(\Lambda\otimes \Lambda){\sf G}^\top)^{-1}{\sf G} \\
    = & ~ {\sf G}^\top ({\sf G}(\Lambda\otimes \Lambda){\sf G}^\top)^{-1}{\sf G}-{\sf G}^\top ({\sf G}(\Lambda\otimes \Lambda){\sf G}^\top)^{-1}{\sf G}_{*,\wt S}(\wt \Delta^{-1}_{\wt S,\wt S}+{\sf G}^\top_{\wt S,*} ({\sf G}(\Lambda\otimes \Lambda){\sf G}^\top)^{-1}{\sf G}_{*, \wt S})^{-1} {\sf G}^\top_{\wt S,*} ({\sf G}(\Lambda\otimes \Lambda){\sf G}^\top)^{-1}{\sf G} \\
    = & ~ {\sf G}^\top (({\sf G}(\Lambda\otimes \Lambda){\sf G}^\top)^{-1}- ({\sf G}(\Lambda\otimes \Lambda){\sf G}^\top)^{-1}{\sf G}_{*,\wt S}(\wt \Delta^{-1}_{\wt S,\wt S}+{\sf G}^\top_{\wt S,*} ({\sf G}(\Lambda\otimes \Lambda){\sf G}^\top)^{-1}{\sf G}_{*, \wt S})^{-1} {\sf G}^\top_{\wt S,*} ({\sf G}(\Lambda\otimes \Lambda){\sf G}^\top)^{-1}){\sf G} \\
    = & ~ {\sf G}^\top ({\sf G}(\wt \Lambda\otimes \wt \Lambda){\sf G}^\top)^{-1} {\sf G},
\end{align*}
where the last step is Lemma~\ref{cor:app:update_eigen} with $\wt\Lambda=\Lambda+\wt C$ in place of $\Lambda^{\new}$ and with $\wt\Delta$ and $\wt S=\mathrm{supp}(\wt\Delta)$ in place of $\Delta$ and $\wt S$.

Therefore,
\begin{align*}
     & ~ p_l^{\new} \\
    = & ~ (U\otimes U)(\wt \Lambda^{1/2}\otimes \wt \Lambda^{1/2})M (\wt\Lambda^{1/2}\otimes \wt\Lambda^{1/2})(U^\top\otimes U^\top) R_{*,l}^\top R_{*,l}h^{\new}-p_g^{\new} \\
    = & ~ (U\otimes U)(\wt \Lambda^{1/2}\otimes \wt \Lambda^{1/2})(M-(M_{*,\wt S})(\wt \Delta^{-1}_{\wt S,\wt S}+M_{\wt S,\wt S})^{-1} (M_{\wt S,*})) (\wt\Lambda^{1/2}\otimes \wt\Lambda^{1/2})(U^\top\otimes U^\top) R_{*,l}^\top R_{*,l}h^{\new} \\
    = & ~ (U\otimes U)(\wt \Lambda^{1/2}\otimes \wt \Lambda^{1/2})(M-(M_{*,\wt S})(\wt \Delta^{-1}_{\wt S,\wt S}+M_{\wt S,\wt S})^{-1} (M_{\wt S,*})) (\wt\Lambda^{1/2}\otimes \wt\Lambda^{1/2})(U^\top\otimes U^\top) R_{*,l}^\top R_{*,l}h^{\new} \\
    = & ~ (U\otimes U)(\wt \Lambda^{1/2}\otimes \wt \Lambda^{1/2}){\sf G}^\top ({\sf G}(\wt \Lambda\otimes \wt \Lambda){\sf G}^\top)^{-1} {\sf G}(\wt\Lambda^{1/2}\otimes \wt\Lambda^{1/2})(U^\top\otimes U^\top) R_{*,l}^\top R_{*,l}h^{\new} \\
    = & ~ \wt P R_{*,l}^\top R_{*,l}h^{\new}
\end{align*}

as desired.
\end{proof}

The following corollary uses the assumption that all $W$ we received share the same eigenspace, and presents the formula of matrix $M^{\new}$ as a function of $\mathsf{G}$ and $\Lambda^{\new}$.

\begin{lemma}\label{cor:app:update_eigen}
Let $\mathsf{G} = \mathsf{A} (U \otimes U)$, where $\mathsf{A}$ has full row rank.
Suppose that $W=U\Lambda U^\top$ and we receive $V^{\new}=W+UCU^\top=U\Lambda^{\new}U^\top$, where $\Lambda$ and $\Lambda^{\new}=\Lambda+C$ are diagonal with positive diagonal entries and $C\in \R^{n\times n}$ only has $k$ nonzero entries. Let $\Delta:=C\otimes \Lambda+\Lambda\otimes C+C\otimes C$, let $\wt S=\mathrm{supp}(\Delta)$ be the set of indices of the nonzero diagonal entries of $\Delta$, and let $M:=\mathsf{G}^\top ( \mathsf{G} (  \Lambda  \otimes \Lambda ) \mathsf{G}^\top )^{-1} \mathsf{G}$.

Then we have
\begin{align*}
   \mathsf{G}^\top ( \mathsf{G} (  \Lambda^{\new}  \otimes \Lambda^{\new} ) \mathsf{G}^\top )^{-1} \mathsf{G} = M-M_{*, \wt S}(\Delta_{\wt S,\wt S}^{-1}+ M_{\wt S, \wt S})^{-1}M_{*, \wt S}^\top.
\end{align*}
In particular, the right-hand side is the matrix $M^{\new}$ computed by \textsc{Update} (Algorithm~\ref{alg:app:update}) with $\Lambda^{\new}=\diag(\wh\lambda)$.
\end{lemma}

\begin{proof}
Let $K:=\mathsf{G} (  \Lambda  \otimes \Lambda ) \mathsf{G}^\top$. Since $\mathsf{G}$ has full row rank (because $\mathsf{A}$ does and $U\otimes U$ is orthogonal) and $\Lambda\otimes\Lambda$ and $\Lambda^{\new}\otimes\Lambda^{\new}$ are diagonal with positive diagonal entries, both $K$ and $\mathsf{G} (  \Lambda^{\new}  \otimes \Lambda^{\new} ) \mathsf{G}^\top$ are positive definite. The diagonal matrix $\Delta_{\wt S,\wt S}$ is invertible by the choice of $\wt S$, and since $\Delta$ vanishes outside $\wt S$, we have $\mathsf{G}\Delta\mathsf{G}^\top={\sf G}_{*,\wt S}\Delta_{\wt S,\wt S}{\sf G}_{*,\wt S}^\top$. Moreover, $\det(K+{\sf G}_{*,\wt S}\Delta_{\wt S,\wt S}{\sf G}_{*,\wt S}^\top)=\det(K)\det(\Delta_{\wt S,\wt S})\det(\Delta_{\wt S,\wt S}^{-1}+M_{\wt S,\wt S})$, so $\Delta_{\wt S,\wt S}^{-1}+M_{\wt S,\wt S}$ is invertible.
We prove via matrix Woodbury identity:
\begin{align*}
    & ~ \mathsf{G}^\top ( \mathsf{G} (  \Lambda^{\new}  \otimes \Lambda^{\new} ) \mathsf{G}^\top )^{-1} \mathsf{G} \\
    = & ~ \mathsf{G}^\top ( \mathsf{G} (  (\Lambda+C)  \otimes (\Lambda+C) ) \mathsf{G}^\top )^{-1} \mathsf{G} \\
    = & ~ \mathsf{G}^\top ( K+{\sf G}_{*,\wt S}\Delta_{\wt S,\wt S}{\sf G}_{*,\wt S}^\top )^{-1} \mathsf{G} \\
    = & ~ \mathsf{G}^\top K^{-1} \mathsf{G} - {\sf G}^\top K^{-1}{\sf G}_{*,\wt S}(\Delta_{\wt S,\wt S}^{-1}+{\sf G}_{*,\wt S}^\top K^{-1} {\sf G}_{*, \wt S})^{-1}{\sf G}_{*,\wt S}^\top K^{-1} {\sf G} \\
    = & ~ M-M_{*, \wt S}(\Delta_{\wt S,\wt S}^{-1}+ M_{\wt S, \wt S})^{-1}M_{*, \wt S}^\top,
\end{align*}
where the first step follows from the definition of $\Lambda^{\new} = \Lambda + C$, the second step follows from the linearity of Kronecker product calculation, the definition of $\Delta$ and the support of $\Delta$, the third step follows from matrix Woodbury identity (Lemma~\ref{lem:woodbury}), and the last step follows from plugging in the definition of $M$. The last claim follows from the definition of $M^{\new}$ in the algorithm.
\end{proof}

\section{Application: SDPs Whose Dual Slack Stays Diagonal in a Fixed Basis}\label{sec:app:sdp}

\paragraph{Notation.} We write $\mathbb{S}^n$ for the real symmetric $n\times n$ matrices with $\langle X,Y\rangle=\tr[XY]$, $\|\cdot\|$ for the spectral norm and $\|\cdot\|_1$ for the Schatten-$1$ norm. For $A_1,\dots,A_m\in\mathbb{S}^n$ let $\mathcal{A}(X):=(\langle A_i,X\rangle)_{i\in[m]}={\sf A}\vect(X)$ and $\mathcal{A}^*(y):=\sum_{i=1}^m y_iA_i$. For an orthogonal matrix $U=[u_1,\dots,u_n]$ let
\begin{align*}
\mathcal{D}_U:= & ~ \{U\diag(x)U^\top:x\in\R^n\},\\
\mathcal{D}_U^\perp:= & ~ \{X\in\mathbb{S}^n:u_j^\top Xu_j=0~~\forall j\in[n]\},\\
\Phi_U(X):= & ~ U\diag(u_1^\top Xu_1,\dots,u_n^\top Xu_n)U^\top,
\end{align*}
and let $\wh D\in\R^{m\times n}$ and $\wh c\in\R^n$ be given by $\wh D_{ij}:=u_j^\top A_iu_j$ and $\wh c_j:=u_j^\top Cu_j$. We write $\alpha$ for the dual exponent of matrix multiplication, the largest number such that $\Tmat(n,n^{\alpha},n)=n^{2+o(1)}$. Numbered results of~\cite{hjstz21} and~\cite{cls19} refer to their full versions, arXiv:2101.08208v2 and arXiv:1810.07896v2.

\begin{theorem}[SDPs whose dual slack stays diagonal in a fixed basis]\label{thm:sdp_fixed_basis}
Let $C,A_1,\dots,A_m\in\mathbb{S}^n$ with $A_1,\dots,A_m$ linearly independent, let $b\in\R^m$, and let $\mathrm{OPT}:=\max\{\langle C,X\rangle:\mathcal{A}(X)=b,\ X\succeq0\}$, where the SDP is assumed feasible. Let $R,L>0$ be such that $\|X\|\leq R$ for every feasible $X$ and $\|C\|\leq L$. Let $U$ be an orthogonal matrix, given as part of the input, such that
\begin{itemize}
\item[(H1)] $C\in\mathcal{D}_U$,
\item[(H2)] $\Phi_U(A_i)\in\mathrm{span}\{A_1,\dots,A_m\}$ for every $i\in[m]$,
\item[(H3)] $b\in\mathcal{A}(\mathcal{D}_U)$.
\end{itemize}
Let $a\in(0,\min\{\alpha,(3-\omega)/(\omega-1)\}]$, let $F(a)$ be as in Definition~\ref{def:f_a_c}, and let $\epsilon\in(0,1/2)$. Then the following hold.
\begin{enumerate}
\item (Shape of the class.) (H2) and (H3) hold if and only if, for some invertible $N\in\R^{m\times m}$, every constraint of the equivalent system $N\mathcal{A}(X)=Nb$ has the form $\langle D,X\rangle=\beta$ with $D\in\mathcal{D}_U$, or the form $\langle O,X\rangle=0$ with $O\in\mathcal{D}_U^\perp$. At most $n$ constraints have the first form, and every $X\in\mathcal{D}_U$ satisfies all constraints of the second form. Removing the constraints of the second form does not change the optimal value. Moreover, if $P^\top A_iP$ is diagonal for every $i$ and some invertible $P$, then $m\leq n$, and ${\sf B}{\sf B}^\top={\sf A}(W\otimes W){\sf A}^\top$ is singular whenever more than $n$ matrices that are diagonal in one basis are used. For every $n\geq1$ and every $1\leq m\leq n(n+1)/2$ there are instances that satisfy all the hypotheses.
\item (Solver.) Algorithm~\ref{alg:sdp_fixed_basis} runs the robust log-barrier method of~\cite{hjstz21} with Hessian approximations taken from the data structure of Theorem~\ref{thm:app:kronecker_maintain}. Every matrix that it passes to \textsc{Update} is positive definite and diagonal in the fixed basis $\diag(U,1,1)$. It outputs $X\in\mathcal{D}_U$ with $X\succeq0$ and
\begin{align*}
\langle C,X\rangle\geq\mathrm{OPT}-\epsilon LR\quad\text{and}\quad\sum_{i=1}^m|\langle A_i,X\rangle-b_i|\leq4n\epsilon\Big(R\sum_{i=1}^m\|A_i\|_1+\|b\|_1\Big),
\end{align*}
which is the guarantee of~\cite[Theorem 8.1]{hjstz21} when $L=\|C\|$. Its running time is
\begin{align*}
& ~ O\big(mn^\omega+m^\omega+\Tmat(m,m,n^2)+\Tmat(n^2,m,n^2)\big)\\
& ~ +\big(n^{F(a)+\omega-2}+n^{F(a)-a/2+1/2}\big)\cdot n^{o(1)}\log(n/\epsilon).
\end{align*}
\item (Comparison with~\cite{hjstz21} for $m=\Theta(n^2)$.) For $m=\Theta(n^2)$ the time in part 2 is
\begin{align*}
O(m^\omega)+\big(m^{(F(a)+\omega-2)/2}+m^{(F(a)-a/2)/2+1/4}\big)\cdot n^{o(1)}\log(n/\epsilon),
\end{align*}
while~\cite[Theorem 8.1]{hjstz21} gives $O^*((m^{9/4}+m^\omega)\log(1/\epsilon))$. If $F(a)+\omega<13/2$ and $F(a)-a/2<4$, the time in part 2 is $m^{9/4-\Omega(1)}\log(n/\epsilon)$, which is polynomially below the bound of~\cite{hjstz21}. Since $F(a)\geq\omega+2$, the first condition requires $\omega<2.25$. If $\omega=2$, both conditions hold for every $a\in(0,1)$, and the time is $m^{2+(1-a)/4+o(1)}\log(n/\epsilon)$. If $\omega\geq2.25$, the time in part 2 is at least $m^{\omega-o(1)}\log(n/\epsilon)$, so it does not improve on~\cite{hjstz21}. By part 4, the variant without \textsc{Update} takes $O(m^\omega)+O(m^{\omega/2+1/4}\log(n/\epsilon))$ time for $m=\Theta(n^2)$, which is below the bound of~\cite{hjstz21} whenever $\omega<2.25$, for every $a$. Since $\omega=2.25$~\cite[Theorem~1.1]{OpenAI26}, the case $\omega\geq2.25$ applies, and neither variant is polynomially faster than~\cite{hjstz21}. The linear programming algorithm of part 4 is polynomially faster, by a factor of at least $n^{1/2-o(1)}$ up to polylogarithmic factors.
\item (The class is a linear program.) $\mathrm{OPT}=\max\{\wh c^\top x:\wh Dx=b,\ x\geq0\}$, and $x\mapsto U\diag(x)U^\top$ maps every feasible point of this linear program to a feasible point of the SDP with the same objective value. If the constraints are given in the form of part 1, that is, each $A_i$ lies in $\mathcal{D}_U$, or lies in $\mathcal{D}_U^\perp$ and has $b_i=0$, then the linear programming algorithm of~\cite[Theorem 2.1]{cls19}, applied to the constraints of the first form, outputs $X\in\mathcal{D}_U$ with the guarantee of part 2 in expected time
\begin{align*}
O(mn^2+n^{\omega+1})+\big(n^{\omega+o(1)}+n^{2.5-\alpha/2+o(1)}+n^{2+1/6+o(1)}\big)\log(n/\epsilon).
\end{align*}
This is at most the time in part 2 up to $n^{o(1)}$ factors, and for every $\omega$ and every $m$ it is below the bound of~\cite[Theorem 8.1]{hjstz21} by a factor of at least $n^{1/2-o(1)}$, up to polylogarithmic factors. Algorithm~\ref{alg:sdp_fixed_basis} also keeps the guarantee of part 2 if every iteration of its loop updates $\sigma$ with $\textsc{DiagBlock}(1/\sigma)$ in place of $(1-\epsilon_{\mathrm{mp}}^2)^{-1}\textsc{DiagBlock}(\wt\lambda)$ and \textsc{Update} is never called. Then all iterations together take $O(n^{\omega+1/2}\log(n/\epsilon))$ time after initialization.
\item (A fixed eigenbasis forces a linear program.) Drop (H1), (H2) and (H3). Let $y^{(0)},\dots,y^{(T)}$ be dual feasible points, that is, $S(y^{(t)}):=\mathcal{A}^*(y^{(t)})-C\succeq0$, such that $S(y^{(t)})\in\mathcal{D}_U$ for all $t$. Then $y^{(t)}-y^{(0)}\in K_U:=\{z\in\R^m:\mathcal{A}^*(z)\in\mathcal{D}_U\}$ and $\dim K_U\leq n$. On $y^{(0)}+K_U$ we have $S(y)=U\diag(s(y))U^\top$ with $s(y)$ affine in $y$, so $S(y)\succeq0$ if and only if $s(y)\geq0$, and $-\log\det S(y)=-\sum_j\log s_j(y)$. If moreover $b^\top y^{(T)}\leq\mathrm{OPT}+\epsilon''$, then the linear program $\min\{b^\top y:y\in y^{(0)}+K_U,\ s(y)\geq0\}$, which has at most $n$ free variables and $n$ inequalities, has optimal value in $[\mathrm{OPT},\mathrm{OPT}+\epsilon'']$.
\end{enumerate}
\end{theorem}

\begin{remark}[The sketched query cannot keep the slack in $\mathcal{D}_U$]\label{rem:sketch_steps}
Let $A_1,\dots,A_m$ be linearly independent, let $S(y)\in\mathcal{D}_U$, let $\wt V\succ0$ and $\wt{\sf B}:={\sf A}(\wt V^{1/2}\otimes\wt V^{1/2})$, let $h\in\R^{n^2}\setminus\{0\}$ and $\kappa\neq0$, and let $R_l\in\R^{k\times n^2}$ ($k\geq1$) have i.i.d.\ ${\cal N}(0,1)$ entries. Let $z:=(\wt{\sf B}\wt{\sf B}^\top)^{-1}\wt{\sf B}R_l^\top R_lh$ be the dual direction encoded by the sketched output $\wt{\sf B}^\top z=\wt{\sf B}^\top(\wt{\sf B}\wt{\sf B}^\top)^{-1}\wt{\sf B}R_l^\top R_lh$ of \textsc{Query}. If $\Pr[S(y+\kappa z)\in\mathcal{D}_U]>0$, then $A_1,\dots,A_m,C\in\mathcal{D}_U$, and hence $m\leq n$. So, with Gaussian sketching matrices and $m>n$, the step encoded by the output of \textsc{Query} almost surely moves the slack out of $\mathcal{D}_U$. It therefore cannot serve as a step of the solver of Theorem~\ref{thm:sdp_fixed_basis}, whose inputs to \textsc{Update} must stay diagonal in one basis. Algorithm~\ref{alg:sdp_fixed_basis} instead computes each step without sketching, exactly for the Hessian approximation given by \textsc{Update}, from an $(n+2)\times(n+2)$ block of the matrix $M$ stored by the data structure. The argument uses that the Gaussian law has a density. It does not cover the SRHT or AMS sketches of Section~\ref{subsec:coord_embed_sketch}, and it does not exclude other inexact steps.
\end{remark}

\begin{algorithm}[!ht]\caption{Solver for Theorem~\ref{thm:sdp_fixed_basis}}\label{alg:sdp_fixed_basis}
\begin{algorithmic}[1]
\Procedure{FixedBasisSDP}{$C,\{A_i\}_{i=1}^m,b,U,R,L,\epsilon$}
\State $\nu\gets n+2$, $\epsilon_N\gets10^{-2}$, $\epsilon_{\mathrm{mp}}\gets4\cdot10^{-5}$, $\epsilon'\gets\min\{\epsilon,\epsilon_N/\sqrt{\nu}\}$
\State Form the instance $(\ov C,\{\ov A_i\}_{i=1}^{m+1},\ov b)$ of~\cite[Lemma A.1]{hjstz21} with $\epsilon'$ in place of $\epsilon$, and let $\ov U\gets\diag(U,1,1)=[\ov u_1,\dots,\ov u_\nu]$
\State $\sigma\gets\big(u_1^\top(I-\tfrac{\epsilon'}{L}C)u_1,\dots,u_n^\top(I-\tfrac{\epsilon'}{L}C)u_n,1,1\big)$ \Comment{$\ov U\diag(\sigma)\ov U^\top$ is the slack of $\ov y^{(0)}=(0_m,1)$}
\State \textsc{Init}$(\ov{\sf A},\ov U\diag(1/\sigma)\ov U^\top)$ of Theorem~\ref{thm:app:kronecker_maintain}, with no sketching matrices and with the eigenbasis $\ov U$ supplied
\State $\wh D\gets{\sf G}J$, where ${\sf G}=\ov{\sf A}(\ov U\otimes\ov U)$ is formed by \textsc{Init} and $Je_j=e_{(j,j)}$ \Comment{$\wh D_{ij}=\ov u_j^\top\ov A_i\ov u_j$}
\State $\eta\gets1$, $T\gets\lceil(1+20\sqrt{\nu}/\epsilon_N)\ln(2\nu/\epsilon'^2)\rceil$
\For{$t=1\to T$}
  \State $\eta\gets\eta\cdot(1+\epsilon_N/(20\sqrt{\nu}))$
  \State $\wt\lambda\gets$ the eigenvalues of $\textsc{Update}(\ov U\diag(1/\sigma)\ov U^\top)$ \Comment{$\wt V=\ov U\diag(\wt\lambda)\ov U^\top$}
  \State $\sigma\gets\sigma-(1-\epsilon_{\mathrm{mp}}^2)^{-1}\cdot\textsc{DiagBlock}(\wt\lambda)\cdot(\eta\mathbf{1}-1/\sigma)$ \Comment{$\ov S\gets\ov S-\ov{\mathcal A}^*(\wt H^{-1}g(\ov y,\eta))$}
\EndFor
\State $q\gets\textsc{DiagBlock}(1/\sigma)\cdot(\eta\mathbf{1}-1/\sigma)$ \Comment{exact Newton step}
\State $\ov X\gets\eta^{-1}\,\ov U\diag(1/\sigma+q/\sigma^2)\ov U^\top$
\State \Return $X\gets R\cdot\ov X_{[n]\times[n]}$
\EndProcedure
\Procedure{DiagBlock}{$\mu\in\R^{\nu}_{>0}$}
\State $M_{DD}\gets J^\top MJ$, where $M$ and the stored eigenvalue vector $\lambda$ are the current members of the data structure
\State $\mathcal I\gets\{j:\mu_j\neq\lambda_j\}$, $E\gets\diag((\mu_j^2-\lambda_j^2)_{j\in\mathcal I})$
\State \Return $M_{DD}-(M_{DD})_{*,\mathcal I}\,E\,(I+(M_{DD})_{\mathcal I,\mathcal I}E)^{-1}(M_{DD})_{\mathcal I,*}$
\EndProcedure
\end{algorithmic}
\end{algorithm}
(All operations on vectors such as $1/\sigma$ and $q/\sigma^2$ are entrywise.)

\begin{proof}[Proof of Theorem~\ref{thm:sdp_fixed_basis}]
Let $\mathcal V:=\mathrm{span}\{A_1,\dots,A_m\}$ and $\Phi:=\Phi_U$. Since the $A_i$ are linearly independent, $\mathcal A^*$ is injective. For $W\succ0$ let $H(W):={\sf A}(W\otimes W){\sf A}^\top$. Then $H(W)_{ik}=\tr[A_iWA_kW]$ and $H(W)z=\mathcal A(W\mathcal A^*(z)W)$. For $S=\mathcal A^*(y)-C\succ0$, $H(S^{-1})$ is the Hessian of $\phi(y)=-\log\det S$, and the gradient of $\eta b^\top y+\phi(y)$ is $g(y,\eta)=\eta b-\mathcal A(S^{-1})$. We use the same notation for the modified instance of Step 3.

\paragraph{Step 0: two facts.}
(Q) If $S\succ0$ and $g=\mathcal A(Y)$, then $g^\top H(S^{-1})^{-1}g\leq\|S^{1/2}YS^{1/2}\|_F^2$. Indeed, let $\mathcal B(z):=S^{-1/2}\mathcal A^*(z)S^{-1/2}$. Then $H(S^{-1})=\mathcal B^*\mathcal B$ and $g=\mathcal B^*(Z)$ with $Z:=S^{1/2}YS^{1/2}$, so $g^\top H(S^{-1})^{-1}g=\langle Z,\mathcal B(\mathcal B^*\mathcal B)^{-1}\mathcal B^*Z\rangle\leq\|Z\|_F^2$, because $\mathcal B(\mathcal B^*\mathcal B)^{-1}\mathcal B^*$ is an orthogonal projection.

(P) The map $\Phi$ is self-adjoint, since $\langle\Phi(X),Y\rangle=\sum_j(u_j^\top Xu_j)(u_j^\top Yu_j)$, and idempotent, with range $\mathcal D_U$ and kernel $\mathcal D_U^\perp$. So $\mathbb S^n=\mathcal D_U\oplus\mathcal D_U^\perp$ is an orthogonal sum. If $X\succeq0$ then $\Phi(X)\succeq0$ and $\tr\Phi(X)=\tr X$. The space $\mathcal D_U$ is closed under products and inverses. Finally, $u_j^\top\mathcal A^*(z)u_j=(\wh D^\top z)_j$ and $\mathcal A(U\diag(x)U^\top)=\wh Dx$, and an element of $\mathcal D_U$ is determined by its $U$-diagonal.

\paragraph{Step 1: part 1.}
As $\ker\mathcal A=\mathcal V^\perp$ and $\Phi$ is self-adjoint, (H2), that is $\Phi(\mathcal V)\subseteq\mathcal V$, is equivalent to $\Phi(\ker\mathcal A)\subseteq\ker\mathcal A$. Assume (H2). Every $Y\in\mathcal V$ is $\Phi(Y)+(Y-\Phi(Y))$ with both parts in $\mathcal V$, so $\mathcal V=\mathcal V_D\oplus\mathcal V_O$ with $\mathcal V_D:=\mathcal V\cap\mathcal D_U$ and $\mathcal V_O:=\mathcal V\cap\mathcal D_U^\perp$, and $\dim\mathcal V_D\leq n$. Take bases $\{D_k\}$ of $\mathcal V_D$ and $\{O_l\}$ of $\mathcal V_O$. Since $\mathcal A^*$ is injective, some invertible $N$ maps the constraint matrices onto this basis. By (H3) there is $X_b\in\mathcal D_U$ with $\mathcal A(X_b)=b$, so the new right-hand sides are $\langle D_k,X_b\rangle$ and $\langle O_l,X_b\rangle=0$. Conversely, suppose the constraints have the two forms. Then $\Phi$ fixes each $D_k$ and kills each $O_l$, which gives (H2). The $D_k$ are linearly independent elements of $\mathcal D_U$, so their $U$-diagonals $d_k$ are linearly independent, and $d_k^\top x=\beta_k$ has a solution $x$. The matrix $U\diag(x)U^\top$ satisfies every constraint, because the constraints of the second form vanish on $\mathcal D_U$. This gives (H3).

Under (H2) and (H3), $\mathcal A(X)=b$ implies $\mathcal A(\Phi(X))=b$: write $X=X_b+N_0$ with $N_0\in\ker\mathcal A$, so $\Phi(X)=X_b+\Phi(N_0)$ and $\Phi(N_0)\in\ker\mathcal A$. Removing the constraints of the second form does not change the optimal value. Indeed, removing constraints can only enlarge the feasible set. Conversely, if $X\succeq0$ satisfies the constraints of the first form, then so does $\Phi(X)\succeq0$, since $\langle D,\Phi(X)\rangle=\langle\Phi(D),X\rangle=\langle D,X\rangle$ for $D\in\mathcal D_U$. Also $\Phi(X)\in\mathcal D_U$ satisfies the constraints of the second form, and $\langle C,\Phi(X)\rangle=\langle\Phi(C),X\rangle=\langle C,X\rangle$ by (H1).

If every $P^\top A_iP$ is diagonal, then $\mathcal V\subseteq\{P^{-\top}\Lambda P^{-1}:\Lambda\text{ diagonal}\}$, a space of dimension $n$, so $m\leq n$. For more than $n$ such matrices, $\mathrm{rank}({\sf A}(W\otimes W){\sf A}^\top)\leq\mathrm{rank}({\sf A})\leq n<m$.

For the examples take $U=I$ and $C$ diagonal. The constraints of the first form are $\tr X=1$ and $X_{ll}-X_{l+1,l+1}=0$ for $l<r$. The constraints of the second form are $\langle e_je_k^\top+e_ke_j^\top,X\rangle=0$ for $p$ pairs $j<k$. Here $1\leq r\leq n$ and $0\leq p\leq n(n-1)/2$. These matrices are linearly independent, and $m=r+p$ takes every value in $[1,n(n+1)/2]$. (H1) holds, and (H2), (H3) hold by the equivalence above. The point $X=I/n$ is feasible, and $\tr X=1$ gives $R=1$.

\paragraph{Step 2: steps stay in $\mathcal D_U$.}
\emph{Claim.} Assume (H2). Let $\wt W\in\mathcal D_U$ with $\wt W\succ0$, and let $g\in\mathcal A(\mathcal D_U)$. Then $H(\wt W)\succ0$ and $\mathcal A^*(H(\wt W)^{-1}g)\in\mathcal D_U$.

\emph{Proof.} For $z\neq0$, $z^\top H(\wt W)z=\|\wt W^{1/2}\mathcal A^*(z)\wt W^{1/2}\|_F^2>0$. Let $z:=H(\wt W)^{-1}g$ and write $g=\mathcal A(Y)$ with $Y\in\mathcal D_U$. Then $\mathcal A(\wt W\mathcal A^*(z)\wt W-Y)=0$, so $\wt W\mathcal A^*(z)\wt W-Y\in\mathcal V^\perp$. By Step 1, $\mathcal A^*(z)=Z_D+Z_O$ with $Z_D\in\mathcal V_D$ and $Z_O\in\mathcal V_O\subseteq\mathcal V$. Taking the inner product with $Z_O$ gives
\[
\langle Z_O,\wt WZ_D\wt W\rangle+\|\wt W^{1/2}Z_O\wt W^{1/2}\|_F^2-\langle Z_O,Y\rangle=0.
\]
The first and the last term vanish, because $\wt WZ_D\wt W$ and $Y$ lie in $\mathcal D_U$ and $Z_O\in\mathcal D_U^\perp$. Hence $Z_O=0$ and $\mathcal A^*(z)=Z_D\in\mathcal D_U$.

\paragraph{Step 3: the modified instance.}
The algorithm uses the instance of~\cite[Lemma A.1]{hjstz21} with $\epsilon'<1/2$ in place of $\epsilon$. It has dimension $\nu=n+2$, matrices $\ov A_i=\diag(A_i,0,b_i/R-\tr A_i)$ for $i\in[m]$ and $\ov A_{m+1}=\diag(I_n,1,0)$, right-hand side $\ov b=(b/R,n+1)$ and objective $\ov C=\diag(\frac{\epsilon'}{L}C,0,-1)$. Let $\ov U:=\diag(U,1,1)$. Then $\Phi_{\ov U}(\diag(Y,\gamma,\gamma'))=\diag(\Phi(Y),\gamma,\gamma')$, and the modified instance satisfies the hypotheses of the theorem with $\ov U$:
\begin{itemize}
\item The $\ov A_i$ are linearly independent, since $\ov A_{m+1}$ is the only one with a nonzero $(n+1,n+1)$ entry and the $A_i$ are linearly independent.
\item (H1): $\ov C\in\mathcal D_{\ov U}$ because $C\in\mathcal D_U$.
\item (H3): $\ov b=\ov{\mathcal A}(I_\nu)$, since $\langle\ov A_i,I\rangle=\tr A_i+b_i/R-\tr A_i=b_i/R$ and $\langle\ov A_{m+1},I\rangle=n+1$, and $I_\nu\in\mathcal D_{\ov U}$.
\item (H2): let $(z,z')\in\R^m\times\R$. Then $\ov{\mathcal A}^*(z,z')=\diag(\mathcal A^*(z)+z'I_n,\ z',\ \sum_iz_i(b_i/R-\tr A_i))$, and its pinching is $\diag(\Phi(\mathcal A^*(z))+z'I_n,\ z',\ \sum_iz_i(b_i/R-\tr A_i))$. By (H2) there is $w$ with $\mathcal A^*(w)=\Phi(\mathcal A^*(z))$. Then $b^\top w=\langle X_b,\Phi(\mathcal A^*(z))\rangle=\langle\Phi(X_b),\mathcal A^*(z)\rangle=b^\top z$ and $\sum_iw_i\tr A_i=\tr\Phi(\mathcal A^*(z))=\tr\mathcal A^*(z)$, so the pinching equals $\ov{\mathcal A}^*(w,z')$.
\end{itemize}
The slack of $\ov y^{(0)}=(0_m,1)$ is $\ov S^{(0)}=\diag(I_n-\frac{\epsilon'}{L}C,1,1)\in\mathcal D_{\ov U}$, and $\ov S^{(0)}\succeq(1-\epsilon')I\succ0$.

\paragraph{Step 4: correctness.}
Let $\ov y^{(t)}$ be the dual point after $t$ iterations. The algorithm stores only its slack $\ov S^{(t)}=\ov U\diag(\sigma)\ov U^\top$, which is all the method needs. Let $\eta_t:=(1+\epsilon_N/(20\sqrt\nu))^t$ and $H_t:=H((\ov S^{(t)})^{-1})$. In iteration $t$ the algorithm takes the step $\wt\delta:=-\wt H^{-1}g(\ov y^{(t-1)},\eta_t)$ with $\wt H:=(1-\epsilon_{\mathrm{mp}}^2)H(\wt V)$. Step 5 shows that the update of $\sigma$ computes exactly $\ov S^{(t-1)}+\ov{\mathcal A}^*(\wt\delta)$.

(a) Every $\ov S^{(t)}$ lies in $\mathcal D_{\ov U}$. Suppose $\ov S:=\ov S^{(t-1)}\in\mathcal D_{\ov U}$. Then $W:=\ov S^{-1}\in\mathcal D_{\ov U}$, \textsc{Update} returns $\wt V=\ov U\diag(\wt\lambda)\ov U^\top\in\mathcal D_{\ov U}$, and $g(\ov y,\eta)=\ov{\mathcal A}(\eta I-\ov S^{-1})\in\ov{\mathcal A}(\mathcal D_{\ov U})$. By Step 2, $\ov{\mathcal A}^*(\wt\delta)\in\mathcal D_{\ov U}$, so $\ov S^{(t)}\in\mathcal D_{\ov U}$.

(b) The conditions of~\cite[Lemma 11.10]{hjstz21} hold, with $\epsilon_N=10^{-2}$ and barrier parameter $\nu$.
\begin{itemize}
\item Condition 0 at $\eta_0=1$: $g(\ov y^{(0)},1)=\ov{\mathcal A}(I-(\ov S^{(0)})^{-1})$. By (Q), $g^\top H_0^{-1}g\leq\|\ov S^{(0)}-I\|_F^2=(\epsilon'/L)^2\|C\|_F^2\leq n\epsilon'^2\leq\epsilon_N^2$.\footnote{Unlike~\cite[Algorithm 5]{hjstz21}, which starts at a smaller value of $\eta$, we start at $\eta=1$. Condition 0 holds there by the bound above, which is where we use $\epsilon'\leq\epsilon_N/\sqrt\nu$, and $T$ is chosen for this starting value.}
\item Condition 1': the guarantee of \textsc{Update} is $(1-\epsilon_{\mathrm{mp}})\wt\lambda_j\leq\lambda_j(W)\leq(1+\epsilon_{\mathrm{mp}})\wt\lambda_j$. The matrices $W\otimes W$ and $\wt V\otimes\wt V$ are diagonal in $\ov U\otimes\ov U$ with entries $\lambda_i(W)\lambda_j(W)$ and $\wt\lambda_i\wt\lambda_j$. So $(1-\epsilon_{\mathrm{mp}})^2\wt V\otimes\wt V\preceq W\otimes W\preceq(1+\epsilon_{\mathrm{mp}})^2\wt V\otimes\wt V$, and congruence by $\ov{\sf A}$ gives $(1-\epsilon_{\mathrm{mp}})^2H(\wt V)\preceq H(W)\preceq(1+\epsilon_{\mathrm{mp}})^2H(\wt V)$. Hence $\alpha_H^{-1}H(W)\preceq\wt H\preceq\alpha_HH(W)$ with $\alpha_H=(1+\epsilon_{\mathrm{mp}})/(1-\epsilon_{\mathrm{mp}})\leq1+10^{-4}$.
\item Conditions 2 and 3 hold with $\epsilon_g=\epsilon_\delta=0$, since the gradient and the step are exact.
\end{itemize}
By induction,~\cite[Lemma 11.10]{hjstz21} gives $g(\ov y^{(t)},\eta_t)^\top H_t^{-1}g(\ov y^{(t)},\eta_t)\leq\epsilon_N^2$ for all $t\leq T$.

(c) Output. Let $\ov y:=\ov y^{(T)}$, $\ov S:=\ov S^{(T)}$, $\eta:=\eta_T$, $g:=g(\ov y,\eta)$ and $\ov\delta:=H(\ov S^{-1})^{-1}g$, and set $\ov X:=\eta^{-1}(\ov S^{-1}+\ov S^{-1}\ov{\mathcal A}^*(\ov\delta)\ov S^{-1})$. Then
\[
\ov{\mathcal A}(\ov X)=\eta^{-1}\big(\ov{\mathcal A}(\ov S^{-1})+H(\ov S^{-1})\ov\delta\big)=\eta^{-1}\big(\ov{\mathcal A}(\ov S^{-1})+g\big)=\ov b.
\]
Let $Z:=\ov S^{-1/2}\ov{\mathcal A}^*(\ov\delta)\ov S^{-1/2}$. Then $\|Z\|_F^2=\ov\delta^\top H(\ov S^{-1})\ov\delta=g^\top H(\ov S^{-1})^{-1}g\leq\epsilon_N^2<1$, so $\ov X=\eta^{-1}\ov S^{-1/2}(I+Z)\ov S^{-1/2}\succ0$. The duality gap is $\langle\ov X,\ov S\rangle=\eta^{-1}\tr[I+Z]\leq\eta^{-1}\nu(1+\epsilon_N)$. Let $x:=\epsilon_N/(20\sqrt\nu)$. Since $\ln(1+x)\geq x/(1+x)$, the choice of $T$ gives $\eta_T\geq\exp(Tx/(1+x))\geq2\nu/\epsilon'^2\geq\nu(1+\epsilon_N)/\epsilon'^2$, so the gap is at most $\epsilon'^2$. By~\cite[Lemma A.1(2)]{hjstz21} applied with $\epsilon'$, the matrix $X=R\cdot\ov X_{[n]\times[n]}$ is PSD and satisfies the guarantee in part 2 with $\epsilon'\leq\epsilon$, and hence with $\epsilon$. By Step 2, applied with $\wt W=\ov S^{-1}$ and $g\in\ov{\mathcal A}(\mathcal D_{\ov U})$, $\ov{\mathcal A}^*(\ov\delta)\in\mathcal D_{\ov U}$. So $\ov X\in\mathcal D_{\ov U}$ and $X\in\mathcal D_U$.

\paragraph{Step 5: the steps come from an $(n+2)\times(n+2)$ block of $M$.}
Let ${\sf G}:=\ov{\sf A}(\ov U\otimes\ov U)$. By $\vect(XYZ)=(Z^\top\otimes X)\vect(Y)$, row $i$ of ${\sf G}$ is $\vect(\ov U^\top\ov A_i\ov U)^\top$. So ${\sf G}J=\wh D$, the matrix of Step 0 for the modified instance, and $H(\ov U\diag(\mu)\ov U^\top)={\sf G}(\diag(\mu)\otimes\diag(\mu)){\sf G}^\top$. By \textsc{Init} and Lemma~\ref{lem:app:correct}, the member $M$ equals ${\sf G}^\top H(\ov U\diag(\lambda)\ov U^\top)^{-1}{\sf G}$ for the stored eigenvalue vector $\lambda$.

\emph{Claim.} For $\mu\in\R^\nu_{>0}$, $\textsc{DiagBlock}(\mu)=\wh D^\top H(\ov U\diag(\mu)\ov U^\top)^{-1}\wh D$, and the call takes $O(n^\omega)$ time.

\emph{Proof.} Both sides are unchanged if the constraint system is replaced by $N\ov{\mathcal A}$ for an invertible $N$, since ${\sf G}\mapsto N{\sf G}$, $\wh D\mapsto N\wh D$ and $H\mapsto NHN^\top$. By Step 1 for the modified instance, choose $N$ that gives the two forms. Then each row of $N{\sf G}$ is supported on the coordinates $\{(j,j)\}$, namely $(Jd_k)^\top$ for a constraint of the first form, or vanishes on them, for a constraint of the second form. For a positive diagonal $\Lambda'$, the matrix $(N{\sf G})(\Lambda'\otimes\Lambda')(N{\sf G})^\top$ is therefore block diagonal, and its first-form block is $\wh D_D\Lambda'^2\wh D_D^\top$, where $\wh D_D$ has rows $d_k^\top$. Also $N\wh D=\begin{bmatrix}\wh D_D\\0\end{bmatrix}$. Hence
\[
\wh D^\top H(\ov U\Lambda'\ov U^\top)^{-1}\wh D=\wh D_D^\top(\wh D_D\Lambda'^2\wh D_D^\top)^{-1}\wh D_D.
\]
For $\Lambda'=\diag(\lambda)$ this is $J^\top MJ=M_{DD}$. For $\Lambda'=\diag(\mu)$ we have $\diag(\mu)^2=\diag(\lambda)^2+P_{\mathcal I}EP_{\mathcal I}^\top$, where $P_{\mathcal I}$ embeds the coordinates in $\mathcal I$, and the Woodbury identity gives the returned matrix. The matrix $I+(M_{DD})_{\mathcal I,\mathcal I}E$ is invertible, because its determinant equals
\[
\det(\wh D_D\diag(\mu)^2\wh D_D^\top)/\det(\wh D_D\diag(\lambda)^2\wh D_D^\top)\neq0.
\] Reading $J^\top MJ$ costs $O(\nu^2)$, and the products cost $O(\nu^\omega)$ because $|\mathcal I|\leq\nu$.

Now $\ov b=\ov{\mathcal A}(I)=\wh D\mathbf 1$ and $\ov{\mathcal A}(\ov S^{-1})=\wh D(1/\sigma)$, so $g(\ov y,\eta)=\wh D(\eta\mathbf 1-1/\sigma)$. Hence the $\ov U$-diagonal of $\ov{\mathcal A}^*(\wt\delta)$ is
\[
\wh D^\top\wt\delta=-(1-\epsilon_{\mathrm{mp}}^2)^{-1}\wh D^\top H(\wt V)^{-1}\wh D(\eta\mathbf 1-1/\sigma)=-(1-\epsilon_{\mathrm{mp}}^2)^{-1}\textsc{DiagBlock}(\wt\lambda)(\eta\mathbf 1-1/\sigma).
\]
Since $\ov{\mathcal A}^*(\wt\delta)\in\mathcal D_{\ov U}$ by Step 4(a), and an element of $\mathcal D_{\ov U}$ is determined by its $\ov U$-diagonal, the update of $\sigma$ sets $\ov S\gets\ov S+\ov{\mathcal A}^*(\wt\delta)$. In the same way, $q=\wh D^\top\ov\delta$, so $\ov{\mathcal A}^*(\ov\delta)=\ov U\diag(q)\ov U^\top$, and the algorithm forms the $\ov X$ of Step 4(c).

\paragraph{Step 6: running time.}
Forming the modified instance takes $O(mn^2)$ time. \textsc{Init}, with no sketching matrices and with dimensions $n+2$ and $m+1$, takes $O(mn^\omega+m^\omega+\Tmat(m,m,n^2)+\Tmat(n^2,m,n^2))$ time by Theorem~\ref{thm:app:kronecker_maintain}. Supplying $\ov U$ replaces only the spectral decomposition in \textsc{Init}. It is needed because the first input has repeated eigenvalues (the last two entries of $\sigma$ are both $1$), so a spectral decomposition of it need not use the basis $\ov U$, while every later input is diagonal in $\ov U$. The proofs of Lemma~\ref{lem:app:correct} and Theorem~\ref{thm:app:kronecker_maintain} use only that every input is diagonal in the basis that \textsc{Init} uses to form ${\sf G}$.

Drift of the inputs to \textsc{Update}. Let $W=\ov S^{-1}$ and $W'=\ov S'^{-1}$ be consecutive inputs. The numbers $x_j:=\sigma'_j/\sigma_j-1$ are the eigenvalues of $\ov S^{-1/2}\ov{\mathcal A}^*(\wt\delta)\ov S^{-1/2}$, so
\[
\|x\|_2^2=\wt\delta^\top H(W)\wt\delta=\|\wt H^{-1}g\|_{H(W)}^2\leq\alpha_H^2\|g\|_{H(W)^{-1}}^2,
\]
using $H(W)^{1/2}\wt H^{-1}H(W)^{1/2}\preceq\alpha_HI$, where $g=g(\ov y,\eta_t)$. By (Q) with $Y=\ov S^{-1}$, $\|\ov{\mathcal A}(\ov S^{-1})\|_{H(W)^{-1}}\leq\sqrt\nu$, so $\eta_{t-1}\|\ov b\|_{H(W)^{-1}}\leq\epsilon_N+\sqrt\nu$. Therefore
\[
\|g\|_{H(W)^{-1}}\leq\epsilon_N+\frac{\epsilon_N}{20\sqrt\nu}(\epsilon_N+\sqrt\nu)\leq1.06\,\epsilon_N.
\]
So $\|x\|_2\leq1.07\,\epsilon_N<1$, which also shows $\sigma'>0$. Since $|\ln(1+x)|\leq|x|/(1-|x|)$, we get $\sum_j(\ln\lambda_j(W')-\ln\lambda_j(W))^2=\sum_j\ln^2(1+x_j)\leq(1.1\,\epsilon_N)^2$. The input sequence is deterministic, so the amortized hypothesis of Theorem~\ref{thm:app:kronecker_maintain} holds with $C_1=1.1\,\epsilon_N$ and $C_2=0$. Since $\epsilon_{\mathrm{mp}}$ is an absolute constant and $sb=0$, the amortized time per call of \textsc{Update} is $(n^{F(a)+\omega-5/2}+n^{F(a)-a/2})n^{o(1)}$.

Number of iterations. $\ln(2\nu/\epsilon'^2)\leq\ln(2\nu)+2\ln(1/\epsilon)+\ln(\nu/\epsilon_N^2)$, so $T=O(\sqrt n\log(n/\epsilon))$. So all calls of \textsc{Update} take $(n^{F(a)+\omega-2}+n^{F(a)-a/2+1/2})n^{o(1)}\log(n/\epsilon)$ time.

The rest of each iteration is one call of \textsc{DiagBlock} and $O(n^\omega)$ further work, such as forming $\ov U\diag(1/\sigma)\ov U^\top$, so all iterations together take $O(n^{\omega+1/2}\log(n/\epsilon))$ time beyond the calls of \textsc{Update}. Since $F(a)\geq\omega+2$, this is at most $n^{F(a)+\omega-2}\log(n/\epsilon)$. The output step takes $O(n^\omega)$ time. This proves part 2.

\paragraph{Step 7: part 3.}
For $m=\Theta(n^2)$ we have $mn^\omega=O(n^{2\omega})$, and $\Tmat(m,m,n^2)$ and $\Tmat(n^2,m,n^2)$ are $O(\Tmat(n^2,n^2,n^2))=O(n^{2\omega})=O(m^\omega)$. Also $n^{F(a)+\omega-2}=m^{(F(a)+\omega-2)/2}$ and $n^{F(a)-a/2+1/2}=m^{(F(a)-a/2)/2+1/4}$. Both exponents are below $9/4$ exactly when $F(a)+\omega<13/2$ and $F(a)-a/2<4$. In that case $2\omega\leq F(a)+\omega-2<9/2$ as well, so the time is $m^{9/4-\Omega(1)}\log(n/\epsilon)$.

For $m=\Theta(n^2)$, the bound of~\cite[Theorem 8.1]{hjstz21} is
\[
O^*((\sqrt n(m^2+n^4)+m^\omega+n^{2\omega})\log(1/\epsilon))=O^*((m^{9/4}+m^\omega)\log(1/\epsilon)).
\]

By Definition~\ref{def:f_a_c}, $F(a)\geq f(a,1)=\omega+2$. If $\omega=2$, then $\theta=4$ and $\alpha=1$, so $F(a)=4$ for every $a\in(0,1)$, and the two exponents are $2$ and $2+(1-a)/4$. If $\omega\geq9/4$, then $n^{F(a)+\omega-2}\geq n^{2\omega}=\Theta(m^\omega)$, so the time is at least $m^{\omega-o(1)}\log(n/\epsilon)$.

For the variant without \textsc{Update}, the initialization takes $O(m^\omega)$ time by the first paragraph of this step, and by the last paragraph of Step 8 the iterations take $O(n^{\omega+1/2}\log(n/\epsilon))$ time, which is $O(m^{\omega/2+1/4}\log(n/\epsilon))$. Since $\omega/2+1/4\leq7/4$, this total is $m^{9/4-\Omega(1)}\log(n/\epsilon)$ whenever $\omega<9/4$, and it does not depend on $a$.

\paragraph{Step 8: part 4.}
Let $X$ be feasible for the SDP and set $x:=(u_j^\top Xu_j)_j\geq0$. Then $\Phi(X)=U\diag(x)U^\top$. By Step 1, $\wh Dx=\mathcal A(\Phi(X))=b$, and by (H1), $\wh c^\top x=\langle C,\Phi(X)\rangle=\langle\Phi(C),X\rangle=\langle C,X\rangle$. Conversely, if $x\geq0$ and $\wh Dx=b$, then $U\diag(x)U^\top\succeq0$, $\mathcal A(U\diag(x)U^\top)=\wh Dx=b$, and $\langle C,U\diag(x)U^\top\rangle=\wh c^\top x$. So the two optimal values are equal.

Now suppose the constraints have the two forms of part 1, with first-form constraints $\langle D_k,X\rangle=\beta_k$ for $k\in[r]$, $r\leq n$, and let $d_k$ be the $U$-diagonal of $D_k$. The rows of $\wh D$ for the second form are zero with right-hand side $0$. So the linear program has the same feasible set as $\min\{-\wh c^\top x:\wh D_Dx=\beta,\ x\geq0\}$, whose rows $d_k^\top$ are linearly independent. For a feasible $x$, the matrix $U\diag(x)U^\top$ is feasible for the SDP, so $\|x\|_1=\tr[U\diag(x)U^\top]\leq nR$. Apply~\cite[Theorem 2.1]{cls19} with $\ell_1$ diameter $nR$ and accuracy $\delta=\epsilon/n$. Using $\|\wh c\|_\infty\leq\|C\|\leq L$ and $\sum_j|d_{kj}|=\|D_k\|_1$ (the $d_{kj}$ are the eigenvalues of $D_k$), it outputs $x\geq0$ with $\wh c^\top x\geq\mathrm{OPT}-\epsilon LR$ and
\[
\sum_k|d_k^\top x-\beta_k|\leq\epsilon\Big(R\sum_k\|D_k\|_1+\|\beta\|_1/n\Big)\leq4n\epsilon\Big(R\sum_i\|A_i\|_1+\|b\|_1\Big).
\]
The constraints of the second form have zero residual at $X=U\diag(x)U^\top\in\mathcal D_U$.

Time: reading the input takes $O(mn^2)$. To find the constraints of the first form, draw $\xi\sim\mathcal N(0,I_n)$, form $U\diag(\xi)U^\top$ in $O(n^\omega)$ time, and compute $\langle A_i,U\diag(\xi)U^\top\rangle=\sum_j\xi_ju_j^\top A_iu_j$ for all $i$ in $O(mn^2)$ time. This number is $0$ for a constraint of the second form. For a constraint of the first form, $A_i$ is a nonzero element of $\mathcal D_U$, so its $U$-diagonal is nonzero and the number is nonzero with probability $1$. Computing $\wh c$ and the $d_k$ takes at most $n+1$ products $U^\top(\cdot)U$, that is $O(n^{\omega+1})$. The linear program takes $(n^{\omega+o(1)}+n^{2.5-\alpha/2+o(1)}+n^{2+1/6+o(1)})\log(n^2/\epsilon)$. Forming $X$ takes $O(n^\omega)$.

Comparison with part 2: $mn^2\leq mn^\omega$, $n^{\omega+1}\leq n^{2\omega}\leq n^{F(a)+\omega-2}$, and $n^{2.5}\leq n^{2\omega}$. Comparison with~\cite{hjstz21}: its bound contains $(\sqrt n\cdot n^4+n^{2\omega})\log(1/\epsilon)$ for every $m$. Since $m\leq n^2$, $mn^2\leq n^4$, and $n^{\omega+1}\leq n^4$ because $\omega\leq3$. The remaining terms are at most $n^{3+o(1)}\log(n/\epsilon)$. So the ratio is at least $n^{1/2}$, up to $n^{o(1)}$ and polylogarithmic factors.

Finally, suppose that no iteration calls \textsc{Update}, and that every iteration updates $\sigma$ with the matrix $\textsc{DiagBlock}(1/\sigma)$ in place of $(1-\epsilon_{\mathrm{mp}}^2)^{-1}\textsc{DiagBlock}(\wt\lambda)$. The claim of Step 5 holds for every positive vector. With $\mu=1/\sigma$ and $W=\ov S^{-1}=\ov U\diag(1/\sigma)\ov U^\top$, it shows as in Step 5 that the new $\sigma$ is the $\ov U$-diagonal of $\ov S-\ov{\mathcal A}^*(H(W)^{-1}g(\ov y,\eta))$, so the method takes exact Newton steps. Steps 4 and 5 go through with $\wt H=H(W)$ in place of $(1-\epsilon_{\mathrm{mp}}^2)H(\wt V)$, and Condition 1' holds with $\alpha_H=1$. Each iteration takes $O(n^\omega)$ time for the call of \textsc{DiagBlock} and $O(n^2)$ time for the rest, so the $T=O(\sqrt n\log(n/\epsilon))$ iterations take $O(n^{\omega+1/2}\log(n/\epsilon))$ time after initialization.

\paragraph{Step 9: part 5.}
For each $t$, $\mathcal A^*(y^{(t)}-y^{(0)})=S(y^{(t)})-S(y^{(0)})\in\mathcal D_U$, so $y^{(t)}-y^{(0)}\in K_U$. Since $\mathcal A^*$ is injective and maps $K_U$ into the $n$-dimensional space $\mathcal D_U$, $\dim K_U\leq n$. For $y\in y^{(0)}+K_U$, $S(y)=S(y^{(0)})+\mathcal A^*(y-y^{(0)})\in\mathcal D_U$, so $S(y)=U\diag(s(y))U^\top$ with $s(y)=(u_j^\top S(y)u_j)_j$ affine in $y$. The eigenvalues of $S(y)$ are the entries of $s(y)$, which gives the two equivalences. Every dual feasible $y$ has $b^\top y\geq\mathrm{OPT}$, since $b^\top y-\langle C,X\rangle=\langle S(y),X\rangle\geq0$ for every primal feasible $X$. The point $y^{(T)}$ is feasible for the linear program and has $b^\top y^{(T)}\leq\mathrm{OPT}+\epsilon''$.
\end{proof}

\begin{proof}[Proof of Remark~\ref{rem:sketch_steps}]
Let $K:=\wt{\sf B}\wt{\sf B}^\top$, which is invertible because ${\sf A}$ has full row rank and $\wt V^{1/2}\otimes\wt V^{1/2}$ is invertible, and let $\ell(w):=\mathcal A^*(w)-\Phi_U(\mathcal A^*(w))$. Since $S(y)\in\mathcal D_U$ and $\kappa\neq0$, $S(y+\kappa z)\in\mathcal D_U$ holds if and only if $\ell(z(R_l))=0$. Each entry of $\ell(z(R_l))$ is a polynomial of degree $2$ in the entries of $R_l$. A nonzero polynomial vanishes only on a Lebesgue-null set, and the Gaussian law has a density. So positive probability forces $\ell(z(R_l))=0$ for every $R_l$.

Take $R_l$ with first row $r^\top$ and all other rows zero. Then $R_l^\top R_lh=(r^\top h)r$, and with the linear map $\mathcal L:=\ell\circ K^{-1}\wt{\sf B}$ we get $(r^\top h)\mathcal L(r)=0$ for all $r\in\R^{n^2}$. Polarizing gives $(r^\top h)\mathcal L(s)+(s^\top h)\mathcal L(r)=0$ for all $r,s$. With $r=s=h$ this gives $\mathcal L(h)=0$, and then $s=h$ gives $\|h\|_2^2\mathcal L(r)=0$, so $\mathcal L=0$. As $\wt{\sf B}$ has full row rank, $K^{-1}\wt{\sf B}$ maps onto $\R^m$, so $\ell=0$ on $\R^m$, that is, $A_i=\mathcal A^*(e_i)\in\mathcal D_U$ for all $i$. Finally $C=\mathcal A^*(y)-S(y)\in\mathcal D_U$, and $m\leq n$ by part 1 of Theorem~\ref{thm:sdp_fixed_basis}.
\end{proof}

\section{Differential Privacy}
\label{sec:dp}
This section is organized as follows: We present the preliminaries on coordinate-wise embedding in Section~\ref{subsec:app:coor_embed}. We present the preliminaries on differential privacy in Section~\ref{subsec:app:dp_know}. We present the formal results on the data structure with norm guarantee in Section~\ref{subsec:app:norm_result}. 

\subsection{Coordinate-wise Embedding}\label{subsec:app:coor_embed}

\subsubsection{Definition and Results}

First, we state the definition of coordinate-wise embedding:
\begin{definition}[Coordinate-wise embedding \cite{sy21}]\label{def:coordinate_embed}
We say a random matrix $R\in \R^{b\times n}$ from a family $\Pi$ satisfies $(\alpha,\beta,\delta)$-coordinatewise embedding(CE) property if for any two fixed vector $g,h\in \R^n$, we have the following:
\begin{enumerate}
    \item $\E_{R\sim \Pi}[g^\top R^\top Rh]=g^\top h$.
    \item $\E_{R\sim \Pi}[(g^\top R^\top Rh)^2] \leq (g^\top h)^2+\frac{\alpha}{b}\|g\|_2^2\|h\|_2^2$.
    \item $\Pr_{R\sim \Pi}[|g^\top R^\top Rh-g^\top h|\geq \frac{\beta}{\sqrt{b}}\|g\|_2\|h\|_2]\leq\delta.$
\end{enumerate}
\end{definition}

In \cite{sy21}, they had proved that for certain choices of $\alpha,\beta,\delta$, the coordinate-wise embedding properties are existing. Additionally, we give the $(\alpha,\beta,\delta)$-guarantee for some commonly used sketching matrices in Section~\ref{subsec:coord_embed_sketch}. 

Next, we present the data structure whose output satisfied coordinate-wise embedding property.  

\begin{lemma}[Simple coordinate-wise embedding data structure]\label{lem:app:coor_wise_data_struc}
There exists a randomized data structure such that, for any oblivious sequence $\{g_0,\ldots,g_{T-1}\}\in (\R^n)^T$ and $\{h_0,\ldots,h_{T-1}\}\in (\R^n)^T$ and parameters $\alpha,\beta,\delta$, with probability at least $1-T\delta$, we have for any $t\in \{0,\ldots,T-1\}$, each pair of vectors $(g_t,h_t)$,  
satisfies $(\alpha,\beta,\delta)$-coordinatewise embedding property (Def.~\ref{def:coordinate_embed}). 
\end{lemma}

\begin{proof}
 
The algorithm is simply picking an $(\alpha,\beta,\delta)$-coordinate-wise embedding matrix $R$ and apply it to $g_t$ and $h_t$.
\end{proof}

Note that $(\alpha,\beta,\delta)$-CE gives three guarantees: expectation, variance and high probability. For our applications, we focus on the high probability part and parameters $\beta$, when coupled with the sketching dimension $b$, gives us the approximation factor $\gamma$. 

At first, we present the approximation factor $\gamma$ of given \emph{vectors} $g$ and $h$. 

 \begin{lemma} 
\label{lem:one_row_approx}
Let $R\in \R^{b\times n}$ satisfies $(\alpha,\beta,\delta)$-coordinate-wise embedding property with $\beta \leq \sqrt{b}$, then given vectors $g, h \in \R^n$, then we have, with probability at least $1-\delta$,

\begin{align*}
    |\langle Rg, Rh\rangle|  = & ~ |\langle g,h\rangle| \pm \gamma \|g\|_2 \|h\|_2, \\
    \langle Rg, Rh\rangle^2 = & ~ \langle g,h\rangle^2 \pm 3\gamma \|g\|_2^2 \|h\|_2^2.
\end{align*}
where $\gamma=\frac{\beta}{\sqrt b}$.
\end{lemma}

\begin{proof}
By property 3 of coordinate-wise embedding (Def.~\ref{def:coordinate_embed}), we have that, with probability $1-\delta$,
\begin{align*}
    |\langle Rg, Rh\rangle-\langle g,h\rangle | \leq \frac{\beta}{\sqrt b} \|g\|_2 \|h\|_2.
\end{align*}
Note that for any two real numbers $a$ and $b$, we have

\begin{align*}
    | |a| - |b| | \leq & ~ | a - b |,
\end{align*}

therefore, 

\begin{align*}
    | | \langle Rg, Rh\rangle | - |\langle g, h\rangle | | \leq & ~ |\langle Rg, Rh\rangle-\langle g,h\rangle | \\
    \leq & ~ \frac{\beta}{\sqrt b} \|g\|_2 \|h\|_2.
\end{align*}

Suppose $|\langle Rg, Rh\rangle | \geq |\langle g,h\rangle|$, then we have:
\begin{align*}
    |\langle Rg, Rh\rangle| \leq & ~ |\langle g,h\rangle| +\frac{\beta}{\sqrt b} \|g\|_2 \|h\|_2.
\end{align*}

Square both sides of the inequality yields:
\begin{align*}
    \langle Rg, Rh\rangle^2 \leq & ~ \langle g,h\rangle^2 + \frac{\beta^2}{b} \|g\|_2^2\|h\|_2^2+\frac{2\beta}{\sqrt b} | \langle g,h\rangle| \|g\|_2 \|h\|_2 \\
    \leq & ~ \langle g,h\rangle^2 + \frac{3\beta}{\sqrt b} \|g\|_2^2 \|h\|_2^2.
\end{align*}
where the last step follows from Cauchy-Schwartz inequality that $|\langle g,h \rangle| \leq \|g\|_2\|h\|_2$ and the assumption $\beta / \sqrt{b} \leq 1$.

Suppose $|\langle Rg, Rh\rangle| \leq |\langle g,h\rangle|$, then we have:
\begin{align*}
    |\langle Rg, Rh\rangle| \geq & ~ |\langle g,h\rangle| -\frac{\beta}{\sqrt b} \|g\|_2\|h\|_2.
\end{align*}

Again, we square both sides:

\begin{align*}
    \langle Rg, Rh\rangle^2 \geq & ~ \langle g,h\rangle^2 + \frac{\beta^2}{b}\|g\|_2^2\|h\|_2^2-\frac{2\beta}{\sqrt b} |\langle g,h\rangle| \|g\|_2\|h\|_2 \\
    \geq & ~ \langle g,h\rangle^2 + \frac{\beta^2}{b}\|g\|_2^2\|h\|_2^2-\frac{2\beta}{\sqrt b}  \|g\|_2^2\|h\|_2^2 \\
    \geq & ~ \langle g,h\rangle^2 - \frac{2\beta}{\sqrt b} \|g\|_2^2\|h\|_2^2 \\
    \geq & ~\langle g,h\rangle^2 - \frac{3\beta}{\sqrt b} \|g\|_2^2\|h\|_2^2,
\end{align*}
where the second step follows from Cauchy-Schwartz property that $|\langle g,h \rangle| \leq \|g\|_2\|h\|_2$, and the third step follows from dropping the nonnegative term $\frac{\beta^2}{b}\|g\|_2^2\|h\|_2^2$. Squaring both sides is valid when $|\langle g,h\rangle| \geq \frac{\beta}{\sqrt b}\|g\|_2\|h\|_2$. If instead $|\langle g,h\rangle| < \frac{\beta}{\sqrt b}\|g\|_2\|h\|_2$, then $\langle g,h\rangle^2 < \frac{\beta^2}{b}\|g\|_2^2\|h\|_2^2 \leq \frac{3\beta}{\sqrt b}\|g\|_2^2\|h\|_2^2$ by the assumption $\beta/\sqrt{b} \leq 1$, so the lower bound $\langle Rg, Rh\rangle^2 \geq 0 \geq \langle g,h\rangle^2 - \frac{3\beta}{\sqrt b}\|g\|_2^2\|h\|_2^2$ holds as well.

Combining the two cases gives the desired result with $\gamma = \frac{\beta}{\sqrt b}$.
\end{proof}

Then, we present the approximation factor $\gamma$ of given matrix $G$ and vector $h$. 

\begin{corollary}\label{cor:app:coor_wise_embed_norm}
Let $R\in \R^{b\times n}$ satisfies $(\alpha,\beta,\delta)$-coordinate-wise embedding property with $\beta \leq \sqrt{b}$, then given matrix $G\in \R^{n\times n}$ and $h \in \R^n$, then we have, with probability at least $1-n\delta$,

\begin{align*}
    \|GR^\top Rh\|_2^2 = & ~ \|Gh\|_2^2 \pm\frac{3\beta}{\sqrt b} \|G\|_F^2 \|h\|_2^2.
\end{align*}
\end{corollary}

\begin{proof}
We apply Lemma~\ref{lem:one_row_approx} to each row $i$ of $G$. Each application fails with probability at most $\delta$, so by a union bound over the $n$ rows, with probability at least $1-n\delta$ all of them hold. Use $g_i$ to denote $i$-th row of $G$, we have that:
\begin{align*}
    \langle g_i,h\rangle^2 - \frac{3\beta}{\sqrt b} \|g_i\|_2^2 \|h\|_2^2 \leq \langle Rg_i, Rh\rangle^2 \leq  \langle g_i,h\rangle^2 + \frac{3\beta}{\sqrt b} \|g_i\|_2^2 \|h\|_2^2,
\end{align*}
Observe that we have the following properties:
\begin{align*}
    \sum_{i=1}^n \langle g_i,h\rangle^2 = & ~ \|Gh\|_2^2, \\
    \sum_{i=1}^n \|g_i\|_2^2 \|h\|_2^2 = & ~ \|G\|_F^2 \|h\|_2^2, \\
    \sum_{i=1}^n \langle Rg_i, Rh\rangle^2 = & ~ \|GR^\top Rh\|_2^2.
\end{align*}
Then, summing over all $i\in [n]$ concludes the proof.
\end{proof}

\begin{remark}
The above two results show that, given that the data structure satisfies $(\alpha,\beta,\delta)$-coordinate-wise embedding property, the same data structure has a $\gamma=\frac{3\beta}{\sqrt b}$-approximation guarantee when $\beta \leq \sqrt{b}$.
\end{remark}

\subsubsection{Guarantee on Several Well-known Sketching Matrices}\label{subsec:coord_embed_sketch}
 
In this section, we present the definitions of several commonly used sketching matrices, and their parameters $\alpha, \beta, \delta$ when acting as the matrices for coordinate-wise embedding.  

\begin{table}[!ht]
    \centering
    \begin{tabular}{|l|l|l|l|l|l|} \hline
        \textbf{sketching matrix} & $\alpha$ & $\beta$ \\ \hline
        Random Gaussian (Definition~\ref{def:gaussian}) & $O(1)$ & $O(\log^{1.5} (n /\delta) )$ \\ \hline
        SRHT (Definition~~\ref{def:srht}) & $O(1)$ & $O(\log^{1.5} (n /\delta) )$ \\ \hline
        AMS (Definition~\ref{def:ams}) & $O(1)$ & $O(\log^{1.5} (n /\delta) )$ \\ \hline
        Count-Sketch (Definition~\ref{def:cs}) & $O(1)$ & $O(\sqrt{b} \log (1/\delta))$ or $O(1 /\sqrt{\delta})$  \\ \hline
        Sparse Embedding (Definition~\ref{def:sparse_2}) & $O(1)$ & $O(\sqrt{b/s} \log^{1.5} (n/\delta))$ \\ \hline
    \end{tabular}
    \caption{Summary for different sketching matrices. (Table 1 in~\cite{sy21})}
    \label{tab:my_label}
\end{table}

We give definitions of the sketching matrices below, starting with the definition of random Gaussian matrix.

\begin{definition}[Random Gaussian Matrix, folklore]\label{def:gaussian}
Let $R\in\R^{b\times n}$ denote a random Gaussian matrix such that all entries are i.i.d. sampled from $\N(0,1/b)$.
\end{definition}

Next, we present the definition of subsampled randomized Hadamard/Fourier transform(SRHT) matrix, which can be applied efficiently via fast Fourier transform (FFT).

\begin{definition}[Subsampled Randomized Hadamard/Fourier Transform(SRHT) Matrix \cite{ldfu13}]\label{def:srht}
We use $R\in\R^{b\times n}$ to denote a subsampled randomized Hadamard transform matrix\footnote{In this case, we require $\log{n}$ to be an integer.}. Then $R$ has the form 
\begin{align*} 
R = \sqrt{ \frac{n}{b} } \cdot  SHD,
\end{align*}
where $S\in\R^{b\times n}$ is a random matrix whose rows are $b$ uniform samples (without replacement) from the standard basis of $\R^n$, $H\in\R^{n\times n}$ is a normalized Walsh-Hadamard matrix and $D\in\R^{n\times n}$ is a diagonal matrix whose diagonal elements are i.i.d. $\{-1,+1\}$ random variables.
\end{definition}

Next, let us present the definition of AMS sketch matrix which is generated by 4-wise hash functions.

\begin{definition}[AMS Sketch Matrix \cite{ams99}]\label{def:ams}
Suppose that $g_1, g_2, \cdots, g_b$ be $b$ random hash functions picking from a 4-wise independent hash family 
\begin{align*} 
\mathcal{G} = \Big\{ g : [n]\to\{-\frac{1}{\sqrt{b}},+\frac{1}{\sqrt{b}}\} \Big\}.
\end{align*}
Then $R\in\R^{b\times n}$ is a AMS sketch matrix if we set $R_{i,j} = g_i(j)$.
\end{definition}

Next, we present the definition of count sketch matrix, which is also generated by hash functions. 

\begin{definition}[Count-Sketch Matrix \cite{ccf02}]\label{def:cs}
Suppose that $h : [n] \rightarrow [b]$  is a random $2$-wise independent hash function 

Assume that $\sigma : [n] \rightarrow \{-1,+1\}$ is a random $4$-wise independent hash function. 

Then we say $R\in\R^{b\times n}$ is a count-sketch matrix if the matrix satisfy that $R_{h(i),i} = \sigma(i)$ for all $i\in[n]$ and zero everywhere else.
\end{definition}

Next, we present one definition of sparse embedding matrix.

\begin{definition}[Sparse Embedding Matrix I \cite{nn13}]\label{def:sparse_1}
Let $R\in\R^{b\times n}$ be a sparse embedding matrix with parameter $s$ if each column of $R$ has exactly $s$ non-zero elements being $\pm 1/\sqrt{s}$ uniformly at random. Note that those locations are picked uniformly at random without replacement (and independent across columns)
\footnote{The signs need only be $O(\log d)$-wise independent. Each column can be specified by a $O(\log d)$-wise independent permutation. The seeds specifying the permutations in different columns need only be $O(\log d)$-wise independent.}.
\end{definition}

Finally, we present another equivalent definition of sparse embedding matrix.

\begin{definition}[Sparse Embedding Matrix II \cite{nn13}]\label{def:sparse_2}
Suppose that  $h : [n]\times[s] \rightarrow [b/s]$ is a random 2-wise independent hash function.

Assume that $\sigma:[n]\times[s]\to \{-1,1\}$ be 4-wise independent. 

We use $R\in\R^{b\times n}$ to represent a sparse embedding matrix II with parameter $s$ if we set $R_{(j-1)b/s+h(i,j),i} = \sigma(i,j)/\sqrt{s}$ for all $(i,j)\in[n]\times [s]$ and zero everywhere else.  
\end{definition}

\subsection{Differential Privacy Background}\label{subsec:app:dp_know}

In this section,  we first present the definition of privacy~\cite{d06}. Then we also present the simple composition theorem from~\cite{dr14}, the standard advanced composition theorem from~\cite{drv10}, and amplification via sampling theorem from~\cite{bnsv15}. After that, we present the generalization guarantee on $(\eps, \delta)$-$\mathrm{DP}$ algorithms~\cite{dfh+15}, and the private median algorithm. 

Here, we present the definition of $(\eps,\delta)$-differential privacy. The intuition of this definition is that any particular row of the dataset cannot have large impact on the output of the algorithm. 

\begin{definition}[Differential Privacy]
We say a randomized algorithm ${\cal A}$ is $(\epsilon,\delta)$-differentially private if for any two databases $S$ and $S'$ that differ only on one row and any subset of outputs $T$, the following
\begin{align*}
    \Pr[ {\cal A}(S) \in T ] \leq e^{\epsilon} \cdot \Pr[ {\cal A}(S') \in T ] + \delta ,
\end{align*}
holds. 
Note that, here the probability is over the randomness of ${\cal A}$.
\end{definition}

Next, we present the simple composition theorem, where the combination of differentially-private output has a privacy guarantee as the sum of their privacy guarantee. 
\begin{theorem} [Simple Composition (Corollary 3.15 of \cite{dr14})] \label{thm:na_c}
Let $\eps_1, \ldots, \eps_k \in (0,1]$, if each ${\cal A}_i$ is $\eps_i$-differentially private, then their combination, defined to be ${\cal A}_{[k]} = {\cal A}_k\circ\ldots \circ {\cal A}_1$ is $\sum_{s=1}^{k}\eps_s$-differentially private.

\end{theorem}

The above composition theorem gives a linear growth on the privacy guarantee. Next, the following advanced composition tool (Theorem~\ref{thm:ada_c}) demonstrates that the privacy parameter $\epsilon_0$ need not grow linearly in $k$. However it only requires roughly $ \sqrt{k}$.

\begin{theorem}[Advanced Composition, see \cite{drv10}] \label{thm:ada_c}
Given three parameters $\epsilon \in (0,1]$, $\delta_0 \in (0,1]$ and $\delta \in [0,1]$. If ${\cal A}_1, \cdots, {\cal A}_k$ are each $(\epsilon,\delta)$-$\mathrm{DP}$ algorithms, then the $k$-fold adaptive composition ${\cal A}_k \circ \cdots \circ {\cal A}_1$ is $(\epsilon_0, \delta_0 + k \delta)$-$\mathrm{DP}$ where
\begin{align*}
\epsilon_0 = \sqrt{2k \ln (1/\delta_0)} \cdot \epsilon + 2k \epsilon^2
\end{align*}
\end{theorem}

Next, we present the amplification theorem, where we can boost the privacy guarantee by subsampling a subset of the database of the original DP algorithm as the input. 

\begin{theorem}[Amplification via sampling (Lemma 4.12 of \cite{bnsv15}\footnote{\cite{bnsv15} gives a more general bound, and uses $(\epsilon, \delta)$-DP.})]\label{thm:amp}
Suppose that $\epsilon \in (0,1]$ is an accuracy parameter. 
Let ${\cal A}$ denote an $\epsilon$-$\mathrm{DP}$ algorithm. Let $S$ denote a dataset with size $|S|$.

Suppose that ${\cal A}'$ is the algorithm that, 
\begin{itemize}
    \item constructs a database $T \subset S$ by sub-sampling with repetition $k \leq n/2$ rows from $S$,
    \item returns ${\cal A}(T) $.
\end{itemize}
Finally, we have
\begin{align*} 
{\cal A}' \mathrm{~~~is~~~} ( \frac{6k}{n} \epsilon )\textnormal{-DP}.
\end{align*} 
\end{theorem}

Next, we present the generalization theorem (see \cite{dfh+15}, \cite{bns+21}) which gives the accuracy guarantee of our DP algorithm on \emph{adaptive} inputs. 
\begin{theorem}[Generalization of Differential Privacy ($\mathrm{DP}$)] \label{thm:general}
Given two accuracy parameters $\epsilon \in (0,1/3)$ and $\delta \in (0,\epsilon/4)$. Suppose that the parameter $t$ satisfy that $t \geq \epsilon^{-2} \log(2\epsilon/\delta)$. 

We ${\cal D}$ to represent a distribution on a domain $X$. Suppose $S\sim {\cal D}^t$ is a database containing $t$ elements sampled independently from ${\cal D}$. Let ${\cal A}$ be an algorithm that, given any database $S$ of size $t$, outputs a predicate $h:X\rightarrow \{0,1\}$.

If ${\cal A}$ is $(\epsilon,\delta)$-$\mathrm{DP}$, then the empirical average of $h$ on sample $S$, i.e., 
\begin{align*} 
h(S)=\frac{1}{|S|}\sum_{x\in S}h(x),
\end{align*}
and $h$'s expectation is taken over underlying distribution ${\cal D}$, i.e.,
\begin{align*} 
h({\cal D})=\E_{x\sim {\cal D}}[h(x)]
\end{align*} 
are within $10\epsilon$ with probability at least $1-{\delta}/{\epsilon}$:
\begin{align*}
    \Pr_{S\sim {\cal D}^t, h\leftarrow {\cal A}(S)} \Big[ \big| h(S) - h({\cal D}) \big| \geq 10\epsilon \Big] \leq {\delta}/{\epsilon}.
\end{align*}
\end{theorem}

Finally, we present the private median algorithm which has a differentially private output that is close to the median of the database. 

\begin{theorem}[Private Median]\label{thm:app:pm}
Given two accuracy parameter $\epsilon \in (0,1)$ and $\beta \in (0,1)$.
We use $X$ to represent a finite domain with total order. Let $\Gamma = O(\epsilon^{-1} \log(|X|/\beta) )$. 

Then there is an $(\epsilon,0)$-$\mathrm{DP}$ algorithm $\textsc{PrivateMedian}_{\epsilon,\beta}$ that, given a database $S \in X^*$ in 
\begin{align*}
O( |S| \cdot \epsilon^{-1} \log^3( |X| / \beta )  \cdot \poly\log|S|)
\end{align*}
time outputs an element $x \in X$ (possibly $x \notin S$) such that, with probability $1-\beta$, there are 
\begin{itemize}
    \item $\geq |S|/2- \Gamma$ elements in $S$ that are  $\geq x$,
    \item $ \geq |S|/2-\Gamma$ elements in $S$ that are $\leq x$. \footnote{The runtime dependency on domain size can be improved by other papers, if we have relatively small domain, we can use this theorem.}
\end{itemize}

\end{theorem}

\subsection{Data Structure with Norm Guarantee}\label{subsec:app:norm_result}
In this section, we present the definition of $\gamma$-approximation and the guarantee of the norm estimation algorithm against an adaptive adversary.
\begin{definition}
Given matrix $G\in \R^{n\times n}$ and $h\in \R^n$, we define function $f: \R^{n\times n}\times \R^n \rightarrow \R$ as 
\begin{align*}
    f(G, h) = & ~ \|G R^\top R h\|_2^2,
\end{align*}
where $R\in \R^{b\times n}$ satisfies $(\alpha,\beta,\delta)$-coordinate-wise embedding property (Def.~\ref{def:coordinate_embed}). We say $f(G, h)$ is a $\gamma$-approximation of $\|G h\|_2^2$ if
\begin{align*}
    \|Gh\|_2^2-\gamma \|G\|_F^2\|h\|_2^2 \leq f(G,h) \leq  \|Gh\|_2^2+\gamma \|G\|^2_F \|h\|_2^2.
\end{align*}
\end{definition}

The goal of this section is to prove the norm estimation guarantee(Theorem~\ref{thm:app:ada_Norm}) that, when given an approximation algorithm against an \emph{oblivious} adversary, we can adapt it to an approximation algorithm against an \emph{adaptive} adversary with slightly worse approximation guarantee.

\begin{theorem}[Reduction to Adaptive Adversary: Norm Estimation.]\label{thm:app:ada_Norm}
Given two parameters $\delta >0, \alpha >0$.
Suppose $\mathcal{U} := [-U,-\frac{1}{U}] \cup \{0\} \cup [\frac{1}{U}, U]$ for $U>1$.
We define function $f$ such that maps elements from domain $G \times H$ to an element in $\mathcal{U}$.

Assume there is a dynamic algorithm $\cal A$ against an oblivious adversary that, given an initial data point $x_0 \in X$ and $T$ updates, the following conditions are holding:
\begin{itemize}
    \item The preprocessing time is $\T_{\mathrm{prep}}$.
    \item The update time per round is $\T_{\mathrm{update}}$.
    \item The query time is $\T_{\mathrm{query}}$ and, with probability $\geq 9/10$, the answer $f(G_t, h_t)$ is a $\gamma$-approximation of $\|G_t h_t\|_2^2$ for every $t$, i.e.,
    \begin{align*} \|G_t h_t\|_2^2 - \gamma \|G_t\|_F^2 \|h_t\|_2^2 \leq f(G_t, h_t) \leq \|G_t h_t\|_2^2 + \gamma \|G_t\|_F^2 \|h_t\|_2^2 \end{align*}
\end{itemize}
Then, there is a dynamic algorithm $\cal B$ against an adaptive adversary, with probability at least $1-\delta$, obtains an $(\alpha +\gamma+\alpha\gamma)$-approximation of $\|G_t h_t\|_2^2$, guarantees the following:
\begin{itemize}
    \item The preprocessing time is $\wt{O}(\sqrt{T}\log(\frac{\log U}{\alpha \delta})\T_{\mathrm{prep}})$.
    \item The update time per round is $\wt{O}(\sqrt{T} \log(\frac{\log U}{\alpha \delta}) \T_{\mathrm{update}})$.
    \item The per round query time is $\wt{O}(\log(\frac{\log U}{\alpha \delta}) \T_{\mathrm{query}})$ and, with probability $\geq 9/10$, the answer $u_t$ is an $(\alpha+ \gamma+\alpha \gamma )$-norm approximation of $\|G_t h_t\|_2^2$ for every $t$, i.e.
    \begin{align*} \|G_t h_t\|_2^2 - (\alpha+ \gamma+\alpha \gamma ) \|G_t\|_F^2 \|h_t\|_2^2 \leq u_t \leq \|G_t h_t\|_2^2 + (\alpha+ \gamma+\alpha \gamma ) \|G_t\|_F^2 \|h_t\|_2^2 \end{align*}
    where $\| G \|_F$ denote the Frobenius norm of matrix $G$. 
\end{itemize}

Moreover, $\cal B$ undergoes $T$ updates in
\begin{align*}
\wt{O}({\sqrt{T} \log (\frac{\log U}{\alpha \delta})\cdot \T_{\mathrm{prep}} + T^{3/2} \log (\frac{\log U}{\alpha \delta})\cdot \T_{\mathrm{update}} + T \log (\frac{\log U}{\alpha \delta})\cdot \T_{\mathrm{query}}})
\end{align*}
total time, and hence $\cal B$ has an amortized running time of
\begin{align*}
\wt{O} (\log (\frac{\log U}{\alpha \delta})\T_{\mathrm{prep}}/\sqrt{T} + \sqrt{T}\log (\frac{\log U}{\alpha \delta})\T_{\mathrm{update}} + \log (\frac{\log U}{\alpha \delta})\T_{\mathrm{query}})
\end{align*}
per round. The $\wt{O}$ hides $\poly\log(T\log U/(\alpha\delta))$ factors.
\end{theorem}

\begin{proof}
\textbf{Algorithm $\cal B$}. We first describe the algorithm $\mathcal{B}$.
\begin{itemize}
    \item Suppose $L= \wt{O} (\sqrt{T} \log(\frac{\log U}{\alpha \delta}))$. Let us initialize $L $ copies of $\cal A$. Let us call them $\mathcal{A}^{(1)}, \cdots, \mathcal{A}^{(L)}$. Suppose the initial data point $x_0$.
    \item For time step $t =1,\ldots,T$:
    \begin{itemize}
        \item  We update each copy of $\cal A$ by $(G_t, h_t)$.
        \item  We independently uniformly sample $q = \wt{O} (\log (\frac{\log U}{\alpha \delta}))$ indices and we denote this index set as $S_t$. 
        \item  For every $l \in S_t$, we query $\mathcal{A}^{(l)}$ and let $\hat{f}_t^{(l)}$ denote its output for current update. For these nonzero output, we round them to the nearest power of $(1+\alpha)$, and denote it by $\wt{f}_t^{(l)}$. To be specific, $\wt{f}_t^{(l)}$ satisfies the following: 

        \begin{align*} \wt{f}_{t}^{(l)} = \frac{\hat{f}_{t}^{(l)}}{|\hat{f}_{t}^{(l)}|} (1+\alpha)^{ \lceil \log_{(1+\alpha)} |\hat{f}_{t}^{(l)}| \rceil} \end{align*}

        \item Finally, we aggregate the rounded output $\wt{f}_t^{(l)}$ by \textsc{PrivateMedian}  
        in Lemma~\ref{thm:app:pm}, 
        and then output the differentially private norm estimate $u_t$.
        
    \end{itemize}
where we use $\wt{O}$ to hide $\poly \log (T \log U/(\alpha\delta))$ factors.
   
\end{itemize}

Next, let us present the formal algorithm of the above statement:
\begin{algorithm*}[!ht]\caption{Our Norm Estimation Algorithm.  
}\label{alg:app:norm_est}
\begin{algorithmic}[1]
\Procedure{ReductionAlgorithm}{$T, U, \alpha, \delta$} \Comment{Theorem \ref{thm:app:ada_Norm}  
}
    \State $\delta_0 \leftarrow \delta/(4T)$
    \State $L \leftarrow \wt{O} (\sqrt{T}  \log(\frac{\log U}{\alpha \delta_0}))$ \label{line:setup_for_c}
    \For{$l \in [L]$}
        \State Initialize $\mathcal{A}^{(l)}$ with the initial data point $x_0$
    \EndFor
    \For{$ t = 1 \to T$}
        \For{$l \in [L]$}
            \State${\cal A}^{(l)}.\textsc{Update}(G_t, h_t)$.
        \EndFor
        \State $q \gets  \wt{O} (\log (\frac{\log U}{\alpha \delta_0}))$ \label{line:setup_for_k}
        \State We independently uniformly sample $q$ indices as the index set $S_t \subset [L]$.
        \For{$l \in S_t$}
            \State $\hat{f}_t^{(l)} \gets \mathcal{A}^{(l)}$.\textsc{Query}$()$
        \EndFor
        \For{$l \in S_t$}
            \State $ \wt{f}_{t}^{(l)} \gets \frac{\hat{f}_{t}^{(l)}}{|\hat{f}_{t}^{(l)}|} (1+\alpha)^{ \lceil \log_{(1+\alpha)} |\hat{f}_{t}^{(l)}| \rceil} $
        \EndFor
        \State $u_t \gets \textsc{PrivateMedian}(\wt{f}_t^{(l)})$ \label{line:out}
    \EndFor
\EndProcedure
\end{algorithmic}
\end{algorithm*}

\subsubsection{Proof Overview}
In this section, we present the choice
of our parameters, the informal and formal presentation of the proposed algorithm, and the intuition behind our implementation of differential privacy. 
\paragraph{Parameters}
Here, we choose the parameters of the algorithm as follows: 

\begin{align}\label{eq:app:choice_param}
\eps_{\pmedian} = \frac{1}{4} , ~~~ \delta_0 = \delta/(4T) , ~~~  q = \wt{O} (\log (\frac{\log U}{\alpha \delta_0})) ~~ \text{~~~and~~~} L = \wt{O} (\sqrt{T} \log(\frac{\log U}{\alpha \delta_0})).
\end{align}

\paragraph{Accuracy Guarantee} In the following sections, we argue that $\cal B$ maintains an accurate approximation of $\|G h\|_2^2$ against an adaptive adversary. At first, we prove that the transcript $\cal T$ between the \texttt{Adversary} and the algorithm $\cal B$ is differentially private with respect to the database $\cal R$, where $\cal R$ is a matrix generated by the randomness of $\cal B$. Then, we prove that for all $t$,  the aggregated output $u_t$ is indeed an $(\alpha + \gamma + \alpha \gamma)$-approximation of $\|G h\|_2^2$ with probability $1-\delta$ by Chernoff bound (Lemma~\ref{lem:chernoff_bound}). 

\paragraph{Privacy Guarantee} Let $r^{1},\ldots, r^{L}\in \{0,1\}^*$ denote the random strings used by copies of the oblivious algorithm $\cal A$ as \footnote{In our application, the random string is used to generate the random sketching matrices.} $\mathcal{A}^{1},\ldots, \mathcal{A}^{L}$ during the $T$ updates. We further denote $\mathcal{R} = \{ r^{1},\ldots, r^{L} \}$, and we view every $r^{l}$ as a row of the database $\cal R$. In the following paragraphs, we will show that the transcript between the \texttt{Adversary} and the above algorithm $\cal B$ is differentially private with respect to $\cal R$.

To proceed, for each step $t$, fixing the random strings $\cal R$, we define $u_t(\cal R)$\footnote{$\hat{f}_t(\cal R)$ is still a random variable due to private median step.} as the output of algorithm $\mathcal{B}$, and $\mathcal{T}_t({\cal R}) = ((G_t, h_t), u_t(\cal R))$ as the transcript between the \texttt{Adversary} and algorithm $\mathcal{B}$ at time step $t$. Furthermore, we denote $\mathcal{T} ({\cal R}) = \{ x_0, \mathcal{T}_1({\cal R}),\ldots, \mathcal{T}_T ({\cal R})\}$ as the transcript. We view $\mathcal{T}_t$ and $\cal T$ as algorithms that return the transcripts given a database ${\cal R}$. In this light, we prove in Section~\ref{subsubsec:privacy} that the transcript $\cal T$ is differentially private with respect to $\cal R$. 
\end{proof}

\paragraph{Runtime Analysis}

Here, we present the calculation of the total runtime of our algorithm.

\begin{lemma}[Runtime]\label{prop:run_time} The total runtime of $\cal B$ is at most 
\begin{align*} 
\Tilde{O} (\sqrt{T}\log(\frac{\log U}{\alpha \delta})\T_{\mathrm{prep}} + T^{3/2} \log(\frac{\log U}{\alpha \delta}) \T_{\mathrm{update}} + T \log(\frac{\log U}{\alpha \delta}) \T_{\mathrm{query}} ), 
\end{align*}

where $\wt{O}$ hides $\poly \log (T \log U/(\alpha\delta))$ factors.

\end{lemma}

\begin{proof}
We can calculate the update time in the following:
\begin{itemize}
    \item Preprocess $L$ copies of $\cal A$: $L \cdot \T_{\mathrm{prep}}$.
    \item Handle $T$ updates: $L T\cdot \T_{\mathrm{update}}$.
    \item For each step $t$,
    \begin{itemize}
        \item Query $q$ many copies of $\cal A$ cost: $q \cdot \T_{\mathrm{query}}$
        \item By binary search, rounding every output $\hat{f}_t$ to the nearest power of 
        $(1+\alpha)$ takes 
        \begin{align*} 
            O(q \cdot \log\frac{\log U}{\alpha})
        \end{align*}
        time.
        \item By Lemma~\ref{thm:app:pm}, computing $\textsc{PrivateMedian}$ with privacy guarantee $\eps_{\textsf{pm}}$ takes \begin{align*} 
            \wt{O}(q\cdot \poly \log (\frac{\log U}{\alpha \delta_0}))
        \end{align*}
        time.
    \end{itemize}
\end{itemize}

Therefore, we conclude that the total update time of $\mathcal{B}$ is at most 
\begin{align*}
\Tilde{O} (\sqrt{T}\log(\frac{\log U}{\alpha \delta})\T_{\mathrm{prep}} + T^{3/2} \log(\frac{\log U}{\alpha \delta}) \T_{\mathrm{update}} + T \log(\frac{\log U}{\alpha \delta}) \T_{\mathrm{query}} ).
\end{align*}

We can upper bound the $t_{\mathrm{total}}$ as follows:
\begin{align*}
    t_{\mathrm{total}} & = L \cdot \T_{\mathrm{prep}} + L T \cdot \T_{\mathrm{update}} + T ( q\cdot \T_{\mathrm{query}} + t_{\pmedian} + O(q \log \frac{\log U}{\alpha}))\\
    & = O(\T_{\mathrm{prep}} \cdot \sqrt{T} \log(\frac{T \log U}{\alpha \delta})\cdot \sqrt{\log \frac{T}{\delta}}) + O(T \cdot \T_{\mathrm{update}} \cdot \sqrt{T} \log(\frac{T \log U}{\alpha \delta})\cdot \sqrt{\log \frac{T}{\delta}})\\
   & +   O( T \log(\frac{T \log U}{\alpha \delta}) \T_{\mathrm{query}} + T \cdot \frac{1}{\eps_{\pmedian}} \log^3 (|X_{\pmedian}|/\beta) \cdot \poly \log |X_{\pmedian}|) \\
   & = \Tilde{O} (\sqrt{T}\log(\frac{\log U}{\alpha \delta})\T_{\mathrm{prep}} + T^{3/2} \log(\frac{\log U}{\alpha \delta}) \T_{\mathrm{update}} + T \log(\frac{\log U}{\alpha \delta}) \T_{\mathrm{query}} )
\end{align*}
where the first step follows from plugging in the running time of query $\T_{\mathrm{prep}}$, update $\T_{\mathrm{update}}$ and private median $t_{\pmedian}$, the second step follows from the choice of $q$~from Eq.(\ref{eq:app:choice_param}), and the last step follows from hiding the log factors into $\wt{O}(\cdot)$. 
\end{proof}

\subsubsection{Privacy Guarantee}\label{subsubsec:privacy}

We start by presenting the privacy guarantee for the transcript ${\cal T}_t$. 
\begin{lemma}\label{lemma:DP_i}
For every time step $t$, $\mathcal{T}_t$ is $(\frac{6q}{L}\cdot \eps_{\pmedian}, 0)$-$\mathrm{DP}$ with respect to ${\cal R}$.
\end{lemma}

\begin{proof}
For a given step $t$, the only way that the transcript $\T_t ({\cal R}) = (G_t,h_t, u_t({\cal R}))$ could leak information about ${\cal R}$ is by revealing the output $u_t({\cal R})$. In this light, we analyze the differential privacy guarantee of the algorithm $\cal B$ by analyzing the privacy of the output $u_t({\cal R})$. 

If we run $\textsc{PrivateMedian}$ with $\eps = \eps_{\pmedian}$ and $\beta= \delta_0$ on \emph{all} copies of $\cal A$, then from Theorem~\ref{thm:app:pm}, we get that the output $u_t({\cal R})$ would be $(\eps_{\pmedian}, 0)$-$\mathrm{DP}$. 

Instead, we run $\textsc{PrivateMedian}$ with $\eps = \eps_{\pmedian}$ and $\beta= \delta_0$ on the subsampled $q$ copies of $\cal A$ as in Line~\ref{line:out} of the Algorithm~\ref{alg:app:norm_est}. 
Then, from amplification theorem (Theorem~\ref{thm:amp}), by this subsampling, we can boost the privacy guarantee by $\frac{6q}{L}$, and hence $u_t({\cal R})$ is $(\frac{6q}{L} \cdot \eps_{\pmedian}, 0)$-$\mathrm{DP}$ with respect to ${\cal R}$.

\end{proof}

Next, we present the privacy guarantee for the composition of the transcripts as $\cal T$. 
\begin{corollary}\label{cor:app:privacy_whole}
$\cal T$ is $(\frac{1}{200}, \frac{\delta_0}{400})$-$\mathrm{DP}$ with respect to $R$.
\end{corollary}

\begin{proof}

Since our initialization of sketching matrices does not depend on the transcript $\T$, $x_0$ does not affect the privacy guarantee here. By Lemma~\ref{lemma:DP_i}, each $\T_t$ is $\frac{3q}{2L}$-$\mathrm{DP}$ with respect to ${\cal R}$.

Moreover, we can view $\T$ as a $T$-fold adaptive composition as follows:
\begin{align*}
\T_T \circ \T_{T-1} \circ \cdots \circ \T_2 \circ \T_1.
\end{align*}
In this light, we apply the advanced composition theorem (Theorem~\ref{thm:ada_c}) with per-round parameter $\eps = \frac{3q}{2L},~\delta_2 = \delta_0/400, \delta_1 = 0, k = T$, we have that $\T$ is $(\eps_1, \delta_1 k + \delta_2)$-$\mathrm{DP}$, where:
\begin{align*}
    \eps_1& = \sqrt{2 k \ln (1/\delta_2)} \cdot \eps + 2 k \eps^2  \\
    & = \sqrt{2T \ln ({400}/{\delta_0})}\cdot \frac{3q}{2L} + 2T \frac{9q^2}{4L^2} \\
    & \leq \frac{1}{400} + \frac{1}{400} = \frac{1}{200}
\end{align*}
for $L = 600 \cdot q \sqrt{4T \ln ({400}/{\delta_0})}$.
\end{proof}

Next, we prove that algorithm $\cal B$ has accuracy guarantee against an adaptive adversary. Let $x_{[t]} = (x_0, x_1, \ldots, x_t)$ denote the input sequence up to time $t$, where $x_t = (G_t, h_t)$. Let $\mathcal{A} (r, x_{[t]})$ denote the output of the algorithm $\cal A$ on input sequence $x_{[t]}$, given the random string $r$. Then, let $\mathbf{1}[x_{[t]}, r]$ denote the indicator whether $\mathcal{A} (r, x_{[t]})$ is an $(\gamma +\alpha+ \alpha\gamma)$-approximation of $\|G_t h_t\|_2^2$, i.e:
\begin{align*}   
\mathbf{1}[x_{[t]},r] = \mathbf{1} \{ \text{the~event~} \mathsf{E}_{t,r} \text{~holds}   \} 
\end{align*}
where event $\mathsf{E}_{t,r}$ is defined as 
\begin{align*}
   \|G_t h_t\|_2^2 - (\alpha + \gamma + \alpha \gamma) \|G_t\|_F^2\|h_t\|_2^2 \leq f(G_t, h_t) \leq \|G_t h_t\|_2^2 + (\alpha + \gamma + \alpha \gamma) \|G_t\|_F^2\|h_t\|_2^2 
\end{align*}

Now, we show that most instances of the copies of oblivious algorithm $\cal A$ maintains the $(\gamma + \alpha + \alpha\gamma)$-approximation of $\|G_t h_t\|_2^2$.

\subsubsection{Accuracy Guarantee}

In this section, we present the accuracy guarantee of our algorithm $\cal B$.

\begin{lemma}[Accuracy of Algorithm $\cal A$]\label{lem:app:rounded_A}
With probability 9/10, the output $\wt{u}_t$ of the algorithm $\cal A$ for every time step $t$ is an $(\alpha+\gamma + \gamma \alpha)$-approximation of $\|G_t h_t\|_2^2$, i.e., $\mathbb{E}[ \mathbf{1}[x_{[t]},r]] = 9/10$.
\end{lemma}

\begin{proof}
We know that the oblivious algorithm $\cal A$ will output an $\gamma$-norm approximation of $G h$ with probability $9/10$ as $\hat{f}$. For these $\hat{f}$, we proved that by rounding them up to the nearest power of $(1+\alpha)$, the resulting $\wt{u}$ remains to be an $(\gamma+\alpha + \alpha \gamma)$-approximation of $\|G h\|_2^2$. Hence, with probability $9/10$, the following two inequalities hold true simultaneously:
\begin{align*}
    \|G_t h_t\|_2^2 - \gamma \|G_t\|_F^2 \|h_t\|_2^2 \leq \hat{f}_t \leq \wt{u}_t
\end{align*}
where this step directly follows from the approximation guarantee of oblivious algorithm $\cal A$. 
\begin{align*}
    \wt{u}_t 
    \leq & ~ (1+\alpha) \hat{f}_t \\
    \leq & ~ (1+\alpha) ( \|G_t h_t\|_2^2 + \gamma\|G_t\|_F^2 \|h_t\|_2^2 ) \\
    \leq & ~ \|G_t h_t\|_2^2 + \alpha \|G_t h_t\|_2^2 + (1+\alpha) \gamma \|G_t\|_F^2 \|h_t\|_2^2\\
    \leq & ~ \|G_t h_t\|_2^2 + (\alpha + \gamma + \alpha \gamma) \|G_t\|_F^2 \|h_t\|_2^2
\end{align*}
where the first step 
follows from plugging in the approximation guarantee of oblivious algorithm $\cal A$ and our proposed rounding-up procedure, the second step follows from equation expansion, and the last step follows from applying Cauchy-Schwarz inequality that $\|G_t h_t\|_2 \leq \|G_t\|_F \|h_t\|_2$.
\end{proof}
Since every copies of $\cal A$ will output an $\wt{u}$ satisfies the above approximation result with probability $9/10$, we have that $\mathbb{E}[ \mathbf{1}[x_{[t]},r]] = 9/10$. 

Then, we present the accuracy guarantee of all $L$ copies of $\cal A$ at every time step $t$. 
\begin{lemma}[Accuracy of \emph{all} copies of $\cal A$]\label{lem:c_apx}
For every time step $t$, $\sum_{l=1}^{L} \mathbf{1}[x_{[t]}, r^{(l)}] \geq \frac{4}{5} L$ with probability at least $1-\delta_0$.

\end{lemma}

\begin{proof}
We view each $r$ as an i.i.d draw from a distribution $\cal D$, and we present the generalization guarantee on the database $\cal R$. From Lemma~\ref{lem:app:rounded_A}, we know that $\mathbb{E}[ \mathbf{1}[x_{[t]},r]] = 9/10$. By Corollary~\ref{cor:app:privacy_whole}, algorithm $\cal B$ is $(\eps_3, \delta_3)$-$\mathrm{DP}$ with $\eps_3 = \frac{1}{200}$ and $\delta_3= \frac{\delta_0}{400}$. 
Moreover, we can check that $L= \wt{O} (\sqrt{T} \log(\frac{\log U}{\alpha \delta_0})) \geq \eps_3^{-2} \log (2 \eps_3/\delta_3)$

Then, by applying generalization theorem~(Theorem~\ref{thm:general}) with $t = L$, we have that:
\begin{align*}
\Pr_{{\cal R} \sim \mathcal{D}^L, \mathbf{1}[x_{[t]}, \cdot] \leftarrow {\cal B}({\cal R})} \left[\frac{1}{|\cal R|} \sum_{l=1}^{L} \mathbf{1}[x_{[t]}, r^{(l)}] - \E_{r \sim \cal D}[\mathbf{1}[x_{[t]}, r]] \geq 10 \cdot \eps_3 \right] \leq \delta_3 / \eps_3
\end{align*}
and
\begin{align*}
\Pr_{{\cal R} \sim \mathcal{D}^L, \mathbf{1}[x_{[t]}, \cdot] \leftarrow \T({\cal R})} \left[\frac{1}{L} \sum_{l=1}^{L} \mathbf{1}[x_{[t]}, r^{(l)}] - \E_{r \sim \cal D}[\mathbf{1}[x_{[t]}, r]]  \geq \frac{1}{20} \right] \leq \frac{\delta_0}{2}
\end{align*}
where the second step follows from plugging in the value of parameters  $\eps_3$ and $\delta_3$. Then, it immediately follows that with probability at least $1-\delta_0/2$,
\begin{align*} 
\frac{1}{L} \sum_{l=1}^{L} \mathbf{1}[x_{[t]},r^{(l)}] \geq 9/10-1/20 = 0.85 .
\end{align*}
Thus, we complete the proof.
\end{proof}

Then, we prove that after the aggregation, $\textsc{PrivateMedian}$ outputs an $(\alpha +\gamma+ \alpha \gamma)$-approximation of $\|G_t h_t\|_2^2$ as $u_t$ with probability at least $1-3\delta_0$. Moreover, this statement holds true for all $t$ simultaneously with probability $1-\delta$.

\begin{corollary}[Accuracy of the final output]
With probability $1-\delta$, the following guarantee holds for all $t \in [T]$ simultaneously:

\begin{align*} 
\|G_t h_t\|_2^2 - (\alpha + \gamma + \alpha \gamma) \|G_t\|_F^2\|h_t\|_2^2 \leq u_t \leq \|G_t h_t\|_2^2 + (\alpha + \gamma + \alpha \gamma) \|G_t\|_F^2\|h_t\|_2^2
\end{align*}
\end{corollary}

\begin{proof}
Consider a fixed step $t$, we know that $\cal B$ independently samples $q$ indices as set $S_t$ and queries those copies of ${\cal A}$ with those indices. For ease of notation, we let $\mathbf{1}[l] = \mathbf{1}[x_{[t]},r^{(l)}]$ denote the indicator that whether ${\cal A}^{(l)}$ is accurate at time $t$. 

Since ${\cal A}^{(l)}$ are $i.i.d$ sampled, the $\mathbf{1}[l]$ is also i.i.d distributed. Then, from Lemma~\ref{lem:c_apx}, we know that with probability $1-\delta_0$,  $\E[\sum_{l= \in [L]} \mathbf{1}[l] \geq 0.85 L]$, which implies with probability $1-\delta_0$,  $\E[ \mathbf{1}[l] \geq 0.85 ]$. 

Then, for $l \in S_t$, with probability $1-\delta_0$, $\E[\sum_{l \in S_t} \mathbf{1}[l] \geq 0.85 q ]$. Moreover, from Lemma~\ref{thm:app:pm}, we know that with probability $1-\delta_0$, there are 49\% fraction of outputs that are at least $u_t$ as well as bigger than $u_t$ in set $S_t$. 

Then, by Hoeffding's bound (Lemma~\ref{lem:hoeffding}), we have the following:
\begin{align*} 
\Pr \Big[ | \sum_{l \in S_t} \mathbf{1}[l] - \E[\sum_{l=1}^{L} \mathbf{1}[l]| \geq 0.05q \Big]  \leq 2\exp(-\frac{1}{400}q)
\end{align*}
 Therefore, if the deviation in this bound is less than $0.05q$, then with probability $1-2\delta_0$,  $u_t$ is an $(\alpha +\gamma+ \alpha \gamma)$-approximation of $ \| G_t h_t\|_2^2$. Hence, with probability at most $\exp(-\Theta(q)) + 2\delta_0 \leq 3\delta_0$, $\textsc{PrivateMedian}_{\eps_{\pmedian}, \delta_0}$ returns $u_t$ without an $(\alpha +\gamma + \alpha\gamma)$- approximation guarantee.

Furthermore, by union bound, we have that with probability $1-3T\delta_0 \geq 1-\delta$, for all $t$, $u_t$ is an $(\alpha +\gamma+ \alpha\gamma)$-approximation of $\|G_t h_t\|_2^2$, where the last step follows from $\delta_0 = \delta/(4T)$.
\end{proof}

\section{Robust Set Query Data Structure}\label{sec:top_k}
This section is organized as follows: We present the definition of set query problem in Section~\ref{subsec:app:set_def}. We present our main results and the algorithm on the set query problem in Section~\ref{subsec:app:set_result}. We present the privacy guarantee for the transcript between the adversary and the algorithm of the $t$-th round in Section~\ref{subsec:app:privacy_i}, and for those of all rounds in Section~\ref{subsec:app:privacy_all}. We present the accuracy guarantee for the output of \emph{each} copy of oblivious algorithm $\cal A$ in Section~\ref{subsec:app:acc_i}, and for \emph{all} copies in Section~\ref{subsec:app:acc_all}. We present the accuracy guarantee of those outputs that aggregated by private median in Section~\ref{subsec:app:acc_pmedian}.

\subsection{Definition}\label{subsec:app:set_def}

At first, we present the definition of the set query problem and the associate $\epsilon$-approximation guarantee.

\begin{definition}[Set Query]
Let $G\in \R^{n\times n}$ and $h\in \R^n$. Given a set $Q\subseteq [n]$ and $|Q|=k$, the goal is to estimate the coordinates of $Gh$ in set $Q$. Given a precision parameter $\epsilon$, for each $j\in Q$, we want to design a function $f$ that is an $\epsilon$-approximation of $g_j^\top h$, i.e.,
\begin{align*}
    g_j^\top h - \epsilon \|g_j\|_2 \|h\|_2 \leq f(G, h)_j \leq g_j^\top h + \epsilon \|g_j\|_2 \|h\|_2
\end{align*}
where $g_j$ denotes the $j$-th row of $G$.
\end{definition}

In the remainder of this section, we denote $k$ as the number of elements defined in the set query problem, and we denote $q$ as the number of copies of algorithm $\cal A$ that we sample for each queried coordinate in each round.

\subsection{Main Results}\label{subsec:app:set_result}
 
The goal of this section is to prove the following norm estimation guarantee (Theorem~\ref{thm:app:top_k}) that, when given an approximation algorithm against an \emph{oblivious} adversary for the set query problem, we can adapt it to an approximation algorithm against an \emph{adaptive} adversary for the same problem with slightly worse approximation guarantee. 

\begin{theorem}[Reduction to Adaptive Adversary: Set query Estimation. Formal version of Theorem~\ref{thm:top_k}] \label{thm:app:top_k}
We define $\mathcal{U} := [-U,-\frac{1}{U}] \cup \{0\} \cup [\frac{1}{U}, U]$ for $U>1$.
Given two parameters $\delta, \alpha >0$. We define function $f$ to be a function that maps elements from domain $G \times H$ to an element in $\mathcal{U}^d$.

Suppose there is a dynamic algorithm $\cal A$ against an oblivious adversary that, given an initial data point $x_0 \in X$ and $T$ updates, the following conditions are holding:
\begin{itemize}
    \item The preprocessing time is $\T_{\mathrm{prep}}$.
    \item The update time per round is $\T_{\mathrm{update}}$.
    \item The query time is $\T_{\mathrm{query}}$ and given a set $Q_t\subset [n]$ with cardinality $k$, with probability $\geq 9/10$, the algorithm outputs $f(G_t, h_t)_j$ where $j \in Q_t$, and each $f(G_t,h_t)_j$ satisfies the following guarantee:
    \begin{align*} g_j^\top h_t-\gamma \|g_j\|_2\|h_t\|_2 \leq  f(G_t, h_t)_j \leq g_j^\top h_t + \gamma \|g_j\|_2 \| h_t\|_2 \end{align*}
    where $g_j$ denotes the $j$-th row of matrix $G_t$.
\end{itemize}
Then, there exists a dynamic algorithm $\cal B$ against an adaptive adversary, with probability at least $1-\delta$, obtains an $(\alpha +\gamma+\alpha\gamma)$-approximation of $g_j^\top h_t$ for every $j \in Q$, the following conditions are holding:
\begin{itemize}
    \item The preprocessing time is $\wt{O}(\sqrt{kT}\log(\frac{\log U}{\alpha \delta})\T_{\mathrm{prep}})$.
    \item The update time per round is $\wt{O}(\sqrt{kT} \log(\frac{\log U}{\alpha \delta}) \T_{\mathrm{update}})$.
    \item The per round query time is $\wt{O}(\min\{k,\sqrt{kT}\}\log(\frac{\log U}{\alpha \delta}) \T_{\mathrm{query}} + k \poly\log(\frac{\log U}{\alpha \delta}))$
    and, with probability $1-\delta$, for every $j \in Q$, the answer $(u_t)_j$ is an $(\alpha+ \gamma+\alpha \gamma )$-approximation of $g_j^\top h_t$ for every $t$, i.e.
    \begin{align*}
        &~g_j^\top h_t - (\gamma+\alpha + \gamma \alpha) \|g_j\|_2 \|h_t\|_2 \leq (u_t)_j \leq g_j^\top h_t + (\gamma+\alpha + \gamma \alpha) \| g_j\|_2 \| h_t \|_2.
    \end{align*}
\end{itemize}
Here $\wt{O}$ hides $\poly\log(kT\log U/(\alpha\delta))$ factors.
\end{theorem}

At first, we provide the algorithm $\cal B$ for set query problem as follows:

\begin{algorithm*}[!ht]\caption{Our Norm Estimation Algorithm for Set Query Problem}\label{alg:app:norm_est_set}
\begin{algorithmic}[1]
\Procedure{ReductionAlgorithm}{$T, U, \alpha, \delta$} \Comment{Theorem \ref{thm:top_k}}
    \State $L \leftarrow \wt{O} ( \sqrt{kT} \log(\frac{\log U}{\alpha \delta}))$\Comment{The total number of copies} 
    \State $q \gets  \wt{O} (\log (\frac{\log U}{\alpha \delta}))$ \Comment{The number of copies being used for each coordinate in each iteration}
    \For{$l \in [L]$} 
        \State Initialize $\mathcal{A}^{(l)}$ with the initial data point $x_0$
    \EndFor
    \For{$ t = 1 \to T$} 
        \State We receive $a$ query set $Q_t \subset [n]$
        \For{$l=1 \to L$} 
            \State${\cal A}^{(l)}.\textsc{Update}(G_t, h_t)$.
        \EndFor
        
        \For{$j \in Q_t$}
        \State We independently uniformly sample $q$ indices and denote this index set as $S_{t,j} \subset [L]$.
        \EndFor
        \For{$l \in \bigcup_{j \in Q_t} S_{t,j} $}
        \State $\hat{f}_t^{(l)} \gets \mathcal{A}^{(l)}$.\textsc{Query}$()$
        \EndFor

        \For{$j \in Q_t$}
        \For{$l \in S_{t,j} $}
        \State $ (\wt{f}_{t}^{(l)})_{j} \gets \frac{(\hat{f}_{t}^{(l)})_{j}}{|(\hat{f}_{t}^{(l)})_{j}|} (1+\alpha)^{ \lceil \log_{(1+\alpha)} |(\hat{f}_{t}^{(l)})_{j}| \rceil} $
        \EndFor
        \EndFor
        \For{ $j \in Q_t$ }
        \State $(u_t)_{j} \gets \textsc{PrivateMedian}(\{(\wt{f}_t^{(l)})_{j}\}_{l \in S_{t,j}})$.
        \EndFor
    \EndFor
\EndProcedure
\end{algorithmic}
\end{algorithm*}

\begin{proof}
\textbf{Algorithm $\cal B$}. We first describe the algorithm $\mathcal{B}$.
\begin{itemize}
    \item Let $L = \wt{O} (\sqrt{kT} \log(\frac{\log U}{\alpha \delta}))$. We initialize $L$ copies of $\cal A$. We call them $\mathcal{A}^{(1)}, \cdots, \mathcal{A}^{(L)}$. Suppose the initial data point is $x_0$.
    \item For time step $t =1,\ldots,T$:
    \begin{itemize}
        \item  We update each copy of $\cal A$ by $(G_t, h_t)$.
        \item
        For every $j \in Q_t$, we uniformly sample $q = \wt{O} ( \log (\frac{\log U}{\alpha \delta}))$ indices as set $S_{t,j}$, independently of everything else and of the samples for the other coordinates.
        \item For $l \in \bigcup_{j \in Q_t} S_{t,j}$, we query $\mathcal{A}^{(l)}$ and let $\hat{f}_t^{(l)}$ denote its output for current update. For these nonzero outputs, we round every entry $j \in Q_t$ with $l \in S_{t,j}$ to the nearest power of $(1+\alpha)$, and denote it by $\wt{f}_t^{(l)}$, i.e., for every $j \in Q_t$ and every $l \in S_{t,j}$:
        \begin{align*}
        (\wt{f}_{t}^{(l)})_{j} = \frac{(\hat{f}_{t}^{(l)})_{j}}{|(\hat{f}_{t}^{(l)})_{j}|} (1+\alpha)^{ \lceil \log_{(1+\alpha)} (|(\hat{f}_{t}^{(l)})_{j}|) \rceil}
        \end{align*}

        \item Finally, for every $j \in Q_t$, we aggregate the rounded values $(\wt{f}_t^{(l)})_{j}$ with $l \in S_{t,j}$ by \textsc{PrivateMedian} (Lemma~\ref{thm:app:pm}), and then output the differentially private estimate $(u_t)_{j}$.

    \end{itemize}
where $\wt{O}$ hides $\poly \log (kT \log U/(\alpha\delta))$ factors.
   
\end{itemize}

\paragraph{Parameters.}
Here, we choose the parameters of the algorithm as follows:  
\begin{align}\label{eq:app:param_set_query}
\eps_{\textsf{pm}} =\frac{1}{4}, ~~~ \beta = \delta/(4(k+2)T),
~~~ L = \wt{O} (\sqrt{kT} \log(\frac{\log U}{\alpha \delta})), ~~~ q = \wt{O} (\log(\frac{\log U}{\alpha \delta})) .
\end{align}
More precisely, we choose
\begin{align*}
q = C \log(\frac{\log U}{\alpha\beta}) ~~~\text{and}~~~ L = \lceil 1200 q \sqrt{2kT\log(800T/\beta)\log(800/\beta)} \rceil
\end{align*}
for a large enough absolute constant $C$. Since $\log(800T/\beta) = O(\log(kT/\delta))$, these values satisfy the bounds in Eq.~\eqref{eq:app:param_set_query}.

\paragraph{Update time.}
\begin{itemize}
    \item Preprocess $L$ copies of $\cal A$: $L \cdot \T_{\mathrm{prep}}$.
    \item Handle $T$ updates: $L T\cdot \T_{\mathrm{update}}$.
    \item For each step $t \in [T]$,
    \begin{itemize}
        \item Query the copies of $\cal A$ with indices in $\bigcup_{j \in Q_t} S_{t,j}$. There are at most $\min\{kq, L\}$ of them, so this costs at most $\min\{kq, L\} \cdot \T_{\mathrm{query}}$.
        \item By binary search, rounding up the $kq$ entries $(\hat{f}_t^{(l)})_j$ with $j \in Q_t$ and $l \in S_{t,j}$ to the nearest power of $(1+\alpha)$ takes $O(k q \cdot \log\frac{\log U}{\alpha})$ time.
        \item By Lemma~\ref{thm:app:pm}, computing $\textsc{PrivateMedian}$ for $q$ entries with $\eps_{\mathsf{pm}} = \frac{1}{4}$ takes $t_{\mathsf{pm}} =  \wt{O}(q\cdot \poly \log (\frac{\log U}{\alpha \beta}))$ time, and we compute $k$ of them.
    \end{itemize}
\end{itemize}

Therefore, we conclude that the total update time of algorithm $\mathcal{B}$ is at most $t_{\mathrm{total}}$, and we can upper bound $t_{\mathrm{total}}$ as follows:

\begin{align*}
     & ~ t_{\mathrm{total}} \\
    = & ~ L\cdot \T_{\mathrm{prep}} + LT \cdot \T_{\mathrm{update}} + T ( \min\{kq,L\} \cdot \T_{\mathrm{query}} + k t_{\pmedian} + O(kq \log \frac{\log U}{\alpha}))\\
    = & ~  O(\T_{\mathrm{prep}} \cdot q\sqrt{kT} \log\frac{kT}{\delta}) + O(T\cdot \T_{\mathrm{update}} \cdot q\sqrt{kT} \log\frac{kT}{\delta})\\
    + & ~   O( T q \min\{k, \sqrt{kT}\log\frac{kT}{\delta}\} \T_{\mathrm{query}}) \\
    + & ~   O( k T \cdot \frac{q}{\eps_{\pmedian}} \log^3 (|X_{\pmedian}|/\beta) \cdot \poly \log q + kTq \log \frac{\log U}{\alpha}) \\
   = & ~ \Tilde{O} (\sqrt{kT}\log(\frac{\log U}{\alpha \delta})\T_{\mathrm{prep}} +  \sqrt{k}T^{\frac{3}{2}} \log(\frac{\log U}{\alpha \delta}) \T_{\mathrm{update}}) \\
   + & ~ \Tilde{O} (T \min\{k,\sqrt{kT}\} \log(\frac{\log U}{\alpha \delta}) \T_{\mathrm{query}} + k T \poly \log (\frac{\log U}{\alpha \delta}) )
\end{align*}
where the first step follows from plugging in the running time of query $\T_{\mathrm{prep}}$, update $\T_{\mathrm{update}}$ and private median $t_{\pmedian}$ with privacy guarantee $\eps_{\pmedian} =\frac{1}{4}$, the second step follows from the choice of $L$ and $q$ in Eq.~\eqref{eq:app:param_set_query} together with $\log(800T/\beta)\log(800/\beta) = O(\log^2(kT/\delta))$ for $\beta = \delta/(4(k+2)T)$, and the last step follows from hiding the $\poly\log(kT\log U/(\alpha\delta))$ factors into $\wt{O}(\cdot)$. Dividing the last two terms by $T$ gives the per round query time in Theorem~\ref{thm:app:top_k}.

\paragraph{Privacy Guarantee.} In the following sections, we argue that $\cal B$ maintains an accurate approximation of $g_j^\top h$ for $j \in Q_t$ against an adaptive adversary. At first, we prove that the transcript $\cal T$ between the \texttt{Adversary} and the algorithm $\cal B$ is differentially private with respect to the database $\cal R$, where $\cal R$ is a matrix generated by the randomness of $\cal B$. Then, we prove that for every $j \in Q_t$, the $j$-th coordinate of aggregated output $u$ is indeed an $(\alpha + \gamma + \alpha \gamma)$-approximation of $g_j^\top h$ with probability $1-\delta$ by Chernoff–Hoeffding inequality.

Let $r^{1},\ldots, r^{L}  \in \{0,1\}^*$ denote the random strings used by the oblivious algorithms\footnote{In our application, the random string is used to generate the random sketching matrices.} $\mathcal{A}^{1}, \ldots, \mathcal{A}^{L}$ during the $T$ updates. We further denote $\mathcal{R} = \{ r^{1},\ldots, r^{L} \}$, and we view every $r^{l}$ as a row of the database $\cal R$. In the following paragraphs, we will show that the transcript between the \texttt{Adversary} and the above algorithm $\cal B$ is differentially private with respect to $\cal R$.

To proceed, for each time step $t$, fixing the random strings $\cal R$, we define $u_t(\cal R)$\footnote{$u_t(\cal R)$ is still a random variable due to private median step.} as the output of algorithm $\mathcal{B}$, and $\mathcal{T}_t({\cal R}) = ((G_t, h_t), u_t(\cal R))$ as the transcript between the \texttt{Adversary} and algorithm $\mathcal{B}$ at time step $t$. Furthermore, we denote 
\begin{align*}
\mathcal{T} ({\cal R}) = \{ x_0, \mathcal{T}_1({\cal R}),\ldots, \mathcal{T}_T ({\cal R})\}
\end{align*}
as the transcript. We view $\mathcal{T}_t$ and $\cal T$ as algorithms that return the transcripts given a database ${\cal R}$. 
In this light, we prove in the following that they are differentially private with respect to ${\cal R}$.
\end{proof}

\subsection{Privacy Guarantee for \texorpdfstring{$t$}{}-th Transcript 
}\label{subsec:app:privacy_i}
 
At first, we present the privacy guarantee for transcript ${\cal T}_t$ at time $t$. 

\begin{lemma}[Privacy guarantee for $\mathcal{T}_t$]\label{lemma:DP_i:C}
For every time step $t$, $\mathcal{T}_t$ is $(\frac{1}{400\sqrt{T\log(800/\beta)}}, \frac{\beta}{800T})$-$\mathrm{DP}$ with respect to $\cal R$.
\end{lemma}

\begin{proof}
For a given step $t$, the only way that a transcript $\T_t ({\cal R}) = (G_t,h_t, u_t({\cal R}))$ could possibly leak information about database ${\cal R}$ is by revealing $u_t({\cal R})$. Fix $j \in Q_t$. The coordinate $(u_t)_j$ is computed by $\textsc{PrivateMedian}_{\eps_{\pmedian}, \beta}$ from the values $(\wt{f}_t^{(l)})_j$ with $l \in S_{t,j}$, and each such value depends on ${\cal R}$ only through the row $r^{l}$. By Theorem~\ref{thm:app:pm}, we have that  $\textsc{PrivateMedian}_{\eps_{\pmedian}, \beta}$ gives an $(\eps_{\pmedian}, 0)$-$\mathrm{DP}$ output $(u_t)_j$ with respect to its $q$ inputs. Then, from amplification theorem (Theorem~\ref{thm:amp}), the subsampling of $S_{t,j}$ in our algorithm $\cal B$ boosts the privacy parameter by $\frac{6q}{L}$, and hence every queried coordinate of $u_{t}({\cal R})$ is $(\frac{6q}{L} \cdot \eps_{\pmedian}, 0)$-$\mathrm{DP}$ with respect to ${\cal R}$.

The samples $S_{t,j}$ for $j \in Q_t$ and the coins of the $k$ calls to $\textsc{PrivateMedian}$ are independent, so $u_t({\cal R})$ is the $k$-fold composition of these $k$ mechanisms, each of which uses its own randomness. Now, applying advanced composition theorem (Theorem~\ref{thm:ada_c}) to the $k$ coordinates of $u_t({\cal R})$ with $\delta_0 = \beta /(800T)$ gives us an $(\eps_0, \frac{\beta}{800T})$-DP guarantee of ${\cal T}_t$, for every $t$, where
\begin{align*}
    \eps_0 =&~ \sqrt{2k\log(800T/\beta)}\cdot \epsilon_{\pmedian} \cdot \frac{6q}{L}+2k\cdot \epsilon_{\pmedian}^2 \cdot (\frac{6q}{L})^2\\
    \leq & ~ \frac{1}{800\sqrt{T\log(800/\beta)}}+\frac{1}{800T\log(800/\beta)} \\
    \leq & ~ \frac{1}{400\sqrt{T\log(800/\beta)}},
\end{align*}
where the second step follows from $\eps_{\pmedian} \cdot \frac{6q}{L} = \frac{3q}{2L}$ and $L \geq 1200 q \sqrt{2kT\log(800T/\beta)\log(800/\beta)}$, and the last step follows from $T\log(800/\beta) \geq 1$. This concludes the proof.
\end{proof}

\subsection{Privacy Guarantee for All Transcripts}\label{subsec:app:privacy_all}
 
Next, we present the privacy guarantee of the whole transcript ${\cal T}$. 
\begin{corollary}[Privacy guarantee for $\mathcal{T}$]
$\cal T$ is $(\frac{1}{200}, \frac{\beta}{400})$-$\mathrm{DP}$ with respect to ${\cal R}$.
\end{corollary}

\begin{proof}
We can view $\T$ as an adaptive composition as:
\begin{align*}
\T_T \circ \T_{T-1} \circ \cdots \circ \T_2 \circ \T_1.
\end{align*}

Since our initialization of sketching matrices does not depend on the transcript $\T$, $x_0$ does not affect the privacy guarantee here. By Lemma~\ref{lemma:DP_i:C}, each $\T_t$ is $(\frac{1}{400\sqrt{T\log(800/\beta)}}, \frac{\beta}{800T})$-$\mathrm{DP}$ with respect to $R$. Then, we apply the advanced composition theorem (Theorem~\ref{thm:ada_c}) with $\delta_1 = \beta/800$, we have that $\T$ is  
$(\eps_1, \delta_0 T + \delta_1)$-$\mathrm{DP}$,  
where:
\begin{align*}
    \eps_1  = & ~ \sqrt{2T \log(1/\delta_1)} \cdot \frac{1}{400\sqrt{T\log(800/\beta)}} + 2T \frac{1}{400^2\cdot T\log(800/\beta)} \\
    = & ~ \frac{\sqrt{2}}{400} + \frac{1}{80000 \log(800/\beta)} \\
    \leq & ~\frac{1}{200}.
\end{align*}
Moreover, the $\delta$ guarantee of ${\cal T}$ is: 

\begin{align*}
    \delta_0 T + \delta_1 =& \frac{\beta}{800T} \cdot T + \frac{\beta}{800} \\
    = & \frac{\beta}{400}
\end{align*}
where the first step follows from plugging in the value of $\delta_0$ and $\delta_1$, and the final step follows from calculation.

Hence, our algorithm is $(\frac{1}{200},\frac{\beta}{400})$-DP.
\end{proof}

\subsection{Accuracy of \texorpdfstring{$\cal A$}{} on the  
\texorpdfstring{$t$}{}-th Output
}\label{subsec:app:acc_i}
 
Next, we prove that algorithm $\cal B$ has accuracy guarantee against an adaptive adversary. Let $x_{[t]} = (x_0, x_1, \ldots, x_t)$ denote the input sequence up to time $t$, where $x_t = (G_t, h_t)$. Let $\mathcal{A} (r, x_{[t]})$ and $\wt{u}_t$ denote the output of the algorithm $\cal A$ on input sequence $x_{[t]}$, given the random string $r$. Then, let $\mathbf{1}[x_{[t]}, r]$ denote the indicator whether for every $j \in Q_t$,  $\mathcal{A} (r, x_{[t]})_j$ is an $(\gamma +\alpha+ \alpha\gamma)$-norm approximation of $(G_t)_j^\top h_t$, i.e:
\begin{align*}   \mathbf{1}[x_{[t]},r] = {\bf 1} \{ \forall j \in Q_t, \text{the~event~} \mathsf{E}_{j,t,r} \text{~holds} \} \end{align*}
where $\mathsf{E}_{j,t,r}$ denotes the following event 
\begin{align*}
    g_j^\top h_t -(\alpha + \gamma + \alpha \gamma) \|g_j\|_2\|h_t\|_2 \leq \mathcal{A} (r, x_{[t]})_j \leq g_j^\top h_t + (\alpha + \gamma + \alpha \gamma) \|g_j\|_2\|h_t\|_2.
\end{align*}

Now, we show that most instances of the oblivious algorithm $\cal A$ are  $(\gamma + \alpha + \alpha\gamma)$-approximation of $g_j^\top h_t$.

\begin{lemma}[Accuracy of $\wt{u}_t$] \label{lem:ro:C}
For every time step $t$, with probability~9/10, the output $(\wt{u}_t)_j$ of the algorithm $\cal A$, is an $(\alpha+\gamma + \alpha\gamma)$-approximation of $g_j^\top h_t$ simultaneously for every $j \in Q$.
\end{lemma}

\begin{proof}

We know that the oblivious algorithm $\cal A$ will output an $\gamma$-approximation of $g_j^\top h_t$ with probability $9/10$ as $\hat{f}_j$,
for every $j \in Q$, i.e., $|\hat{f}_j - g_j^\top h_t| \leq \gamma \|g_j\|_2\|h_t\|_2$. For these $\hat{f}_j$, we prove that by rounding them to the nearest power of $(1+\alpha)$, they still remain to be an $(\gamma+\alpha + \alpha \gamma)$-approximation of $g_j^\top h_t$. The rounding keeps the sign of $\hat{f}_j$ and replaces $|\hat{f}_j|$ by $(1+\alpha)^{\lceil \log_{(1+\alpha)} |\hat{f}_j| \rceil} \in [|\hat{f}_j|, (1+\alpha)|\hat{f}_j|)$, and a zero output stays zero. Hence $|(\wt{u}_t)_j - \hat{f}_j| \leq \alpha |\hat{f}_j|$. Therefore, with probability $9/10$, the following holds simultaneously for every $j \in Q$:
\begin{align*}
    |(\wt{u}_t)_j - g_j^\top h_t| & \leq |(\wt{u}_t)_j - \hat{f}_j| + |\hat{f}_j - g_j^\top h_t| \\
    & \leq \alpha |\hat{f}_j| + \gamma \|g_j\|_2\|h_t\|_2\\
    & \leq \alpha (|g_j^\top h_t| + \gamma \|g_j\|_2\|h_t\|_2) + \gamma \|g_j\|_2\|h_t\|_2\\
    & \leq (\alpha + \gamma + \alpha \gamma) \|g_j\|_2\|h_t\|_2
\end{align*}
where the first step follows from the triangle inequality, the second step follows from the rounding up procedure and the guarantee of $\hat{f}_j$, the third step follows from the guarantee of $\hat{f}_j$, and the last step follows from Cauchy-Schwarz $|g_j^\top h_t| \leq \|g_j\|_2\|h_t\|_2$.
\end{proof}
Since every copy of $\cal A$ will have an output that satisfies the above approximation result with probability $9/10$, we have that $\mathbb{E}[ \mathbf{1}[x_{[t]},r]] = 9/10$.

\subsection{Accuracy of All Copies of \texorpdfstring{$\cal B$}{}}\label{subsec:app:acc_all}
In this section, we give the guarantee on the accuracy of \emph{all} copies that the algorithm $\cal B$ maintains. 

\begin{lemma}[Accuracy on all $L$ copies of Algorithm $\cal A$]\label{lem:c_apx:C}
For every time step $t$, $\sum_{l=1}^{L} \mathbf{1}[x_{[t]}, r^{(l)}] \geq \frac{4}{5} L$ with probability at least $1-\beta$.

\end{lemma}

\begin{proof}
We view each $r$ as an i.i.d draw from a distribution $\cal D$. By generalization theorem (Theorem~\ref{thm:general}), we have that:
\begin{align*}
\Pr_{R \sim \mathcal{D}^L, \mathbf{1}[x_{[t]}] \leftarrow \T({\cal R})} \Big[\frac{1}{L} \sum_{l=1}^{L} \mathbf{1}[x_{[t]},r^{(l)}] - \E_{r \sim \cal D}[\mathbf{1}[x_{[t]},r]]  \geq \frac{1}{20} \Big] \leq {\beta}/{2}
\end{align*}
By plugging in $\eps = \frac{1}{200}$, and $\delta = \frac{\beta}{400}$, we have that with probability at least $1-\beta/2$,
\begin{align*} \frac{1}{L} \sum_{l=1}^{L} \mathbf{1}[x_{[t]},r^{(l)}] \geq 9/10-1/20 = & ~  0.85. \qedhere
\end{align*}
\end{proof}

\subsection{Accuracy Guarantee of Private Median}\label{subsec:app:acc_pmedian}
In this section, we prove that after the aggregation, for every $j \in Q_t$,  $\textsc{PrivateMedian}$ outputs an $(\alpha +\gamma+ \alpha \gamma)$-approximation of $g_j^\top h_t$ with probability at least $1-3\beta$ by generalization theorem (Lemma~\ref{thm:general}). Moreover, this statement holds true for all $j$ and all $t$ simultaneously with probability $1-\delta$.

\begin{corollary}[Accuracy guarantee of private median]\label{cor:acc_pmedian}
With probability $1-\delta$, for every step $t$ and every $j \in Q_t$,
 the following statement holds true:
\begin{align*}
g_j^\top h_t - (\gamma + \alpha + \gamma \alpha) \|g_j\|_2 \| h_t\|_2 \leq (u_t)_j \leq g_j^\top h_t +  (\gamma+\alpha + \alpha\gamma)\|g_j\|_2 \|h_t\|_2
\end{align*}

\end{corollary}

\begin{proof}
Consider a fixed step $t$ and a fixed $j \in Q_t$. We know that $\cal B$ samples $q$ indices as set $S_{t,j}$ and computes $(u_t)_j$ from the $j$-th coordinates of the outputs of ${\cal A}^{(l)}$ for $l \in S_{t,j}$. For ease of notation, we let $\mathbf{1}[l] = \mathbf{1}[x_{[t]},r^{(l)}]$ denote the whether ${\cal A}^{l}$ is accurate at time $t$.

From Lemma \ref{lem:c_apx:C}, we know that with probability $1-\beta$, $\sum_{l=1}^{L} \mathbf{1}[l] \geq \frac{4}{5} L$. The sample $S_{t,j}$ is drawn after $x_{[t]}$ is fixed, independently of ${\cal R}$ and of all other random choices, so on this event $\E[\sum_{l \in S_{t,j}} \mathbf{1}[l]] \geq 0.8 q$. Then, by Hoeffding's bound (Lemma~\ref{lem:hoeffding}), we have the following:
\begin{align*}
\Pr \Big[ | \sum_{l \in S_{t,j}} \mathbf{1}[l] - \E[\sum_{l \in S_{t,j}} \mathbf{1}[l]]| \geq 0.05q \Big]  \leq 2\exp(- q/400)
\end{align*}
Hence, with probability at least $1-\exp(-\Theta(q)) \geq 1-\beta$, at least $0.75q$ of the $q$ sampled indices $l \in S_{t,j}$ have $\mathbf{1}[l] = 1$.

From Lemma~\ref{thm:app:pm}, we know that with probability $1-\beta$, at least 49\% of the values $(\wt{f}_t^{(l)})_j$ for $l \in S_{t,j}$ are at least $(u_t)_j$, and at least 49\% of them are at most $(u_t)_j$. At most $0.25q$ of these values come from inaccurate copies, so each of the two groups contains a value in the interval $[g_j^\top h_t - (\alpha+\gamma+\alpha\gamma)\|g_j\|_2\|h_t\|_2, g_j^\top h_t + (\alpha+\gamma+\alpha\gamma)\|g_j\|_2\|h_t\|_2]$, and hence $(u_t)_j$ lies in this interval as well. Therefore, with probability at least $1-3\beta$, $(u_t)_j$ is an $(\alpha +\gamma+ \alpha \gamma)$-approximation of $(G_t)_j^\top h_t$.

Furthermore, the event of Lemma~\ref{lem:c_apx:C} is shared by all $j \in Q_t$. So by union bound over the $k$ coordinates in $Q_t$ and the $T$ steps, we have that with probability at least $1-(2k+1)T\beta \geq 1-\delta$, for every $t$ and every $j \in Q_t$, $(u_t)_j$ is an $(\alpha +\gamma+ \alpha\gamma)$-approximation of $(G_t)_j^\top h_t$, where the last step follows from $\beta = \delta/(4(k+2)T)$.
\end{proof}

\end{document}